\documentclass[10pt,aps,prl, preprint, onecolumn,
superscriptaddress, 
notitlepage]{revtex4-1}

\usepackage{amssymb,amsmath}
\usepackage{natbib}
\usepackage{graphicx}
\usepackage{amsthm}
\usepackage{cancel}
\usepackage{color}
\usepackage{bbold}
\usepackage{wrapfig}
\usepackage{booktabs}
\usepackage{float}
\usepackage{enumitem}
\usepackage{hyperref}
\usepackage{tikz}
\usepackage{comment}
\usepackage{longtable}
\usepackage{multirow}
\usepackage{hyperref}

\usepackage[normalem]{ulem}

\usepackage[normalem]{ulem}

\begin{document}

\title{Anomalous supply shortages from dynamic pricing \\
in on-demand mobility}
			
\author{Malte Schr\"oder$^*$}
\affiliation{Chair for Network Dynamics, Center for Advancing Electronics Dresden (cfaed), Technical University of Dresden, 01062 Dresden, Germany}
\affiliation{Institute for Theoretical Physics, Technical University of Dresden, 01062 Dresden, Germany}
\author{David-Maximilian Storch$^*$}
\affiliation{Institute for Theoretical Physics, Technical University of Dresden, 01062 Dresden, Germany}
\author{Philip Marszal$^*$}
\affiliation{Chair for Network Dynamics, Center for Advancing Electronics Dresden (cfaed), Technical University of Dresden, 01062 Dresden, Germany}
\affiliation{Institute for Theoretical Physics, Technical University of Dresden, 01062 Dresden, Germany}
\author{Marc Timme}
\affiliation{Chair for Network Dynamics, Center for Advancing Electronics Dresden (cfaed), Technical University of Dresden, 01062 Dresden, Germany}
\affiliation{Institute for Theoretical Physics, Technical University of Dresden, 01062 Dresden, Germany}
\affiliation{Lakeside Labs, Lakeside B04b, 9020 Klagenfurt, Austria}

\date{\today, $^*$these authors contributed equally to the manuscript}

\maketitle

\textbf{Dynamic pricing schemes are increasingly employed across industries to maintain a self-organized balance of demand and supply. However, throughout complex dynamical systems, unintended collective states exist that may compromise their function. Here we reveal how dynamic pricing may induce demand-supply imbalances instead of preventing them. Combining game theory and time series analysis of dynamic pricing data from on-demand ride-hailing services, we explain this apparent contradiction. We derive a phase diagram demonstrating how and under which conditions dynamic pricing incentivizes collective action of ride-hailing drivers to induce anomalous supply shortages. By disentangling different timescales in price time series of ride-hailing services at 137 locations across the globe, we identify characteristic patterns in the price dynamics reflecting these anomalous supply shortages. Our results provide systemic insights for the regulation of dynamic pricing, in particular in publicly accessible mobility systems, by unraveling under which conditions dynamic pricing schemes promote anomalous supply shortages.}

\newpage

\section*{Introduction}
Complex engineered systems are known to exhibit unintended states in their collective dynamics that often disrupt their function \cite{Strogatz2005, buldyrev2010catastrophic, Havlin2012, Helbing2013, Schroder2018}. In complex mobility systems, examples include the emergence of congestion \cite{loder2019understanding, Helbing2000b}, anomalous random walks in human travel patterns \cite{brockmann2006scaling} and cascading failures of mobility networks \cite{bashan2013extreme, li2015percolation, zeng2019switch}. As urban mobility becomes increasingly self-organized and digitized, mobility services increasingly employ dynamic pricing schemes \cite{dynamicPricingGeneral, AmazonDynamicPricing, schafer2015decentral, LyftPricing2019, UberSurge2019}. Dynamic pricing in general serves two main purposes (Fig.~\ref{fig:Fig1}a). First, it adjusts the price of a product or service to compensate for changes in its intrinsic base cost. 
Second, it creates incentives for all market participants to equilibrate demand-supply imbalances by increasing the price if demand exceeds supply and vice versa. 
A higher price both imposes higher costs to customers incentivizing them to decrease their demand and, at the same time, offers higher profit for identical service to suppliers, in turn motivating them to increase their supply.

However, recent reports on on-demand ride-hailing \cite{ABC7News2019a, ABC7News2019b, Mohlmann2017} indicate that dynamic pricing may have the opposite effect and instead \emph{cause} demand-supply imbalances.
Here we quantitatively demonstrate the existence of these imbalances by comparing price time series and demand estimates for ride-hailing services. 
In a game theoretic analysis we reveal the incentive structure for drivers to induce anomalous supply shortages as a generic feature of dynamic pricing.
This observation suggests that similar dynamics should emerge independent of the location or industry. 
Comparing price time series for 137 locations in 59 urban areas across six continents we indeed find price dynamics reflecting anomalous supply shortages in several cities around the world.

\newpage

\section*{Results}
\begin{figure*}
    \centering
    \includegraphics{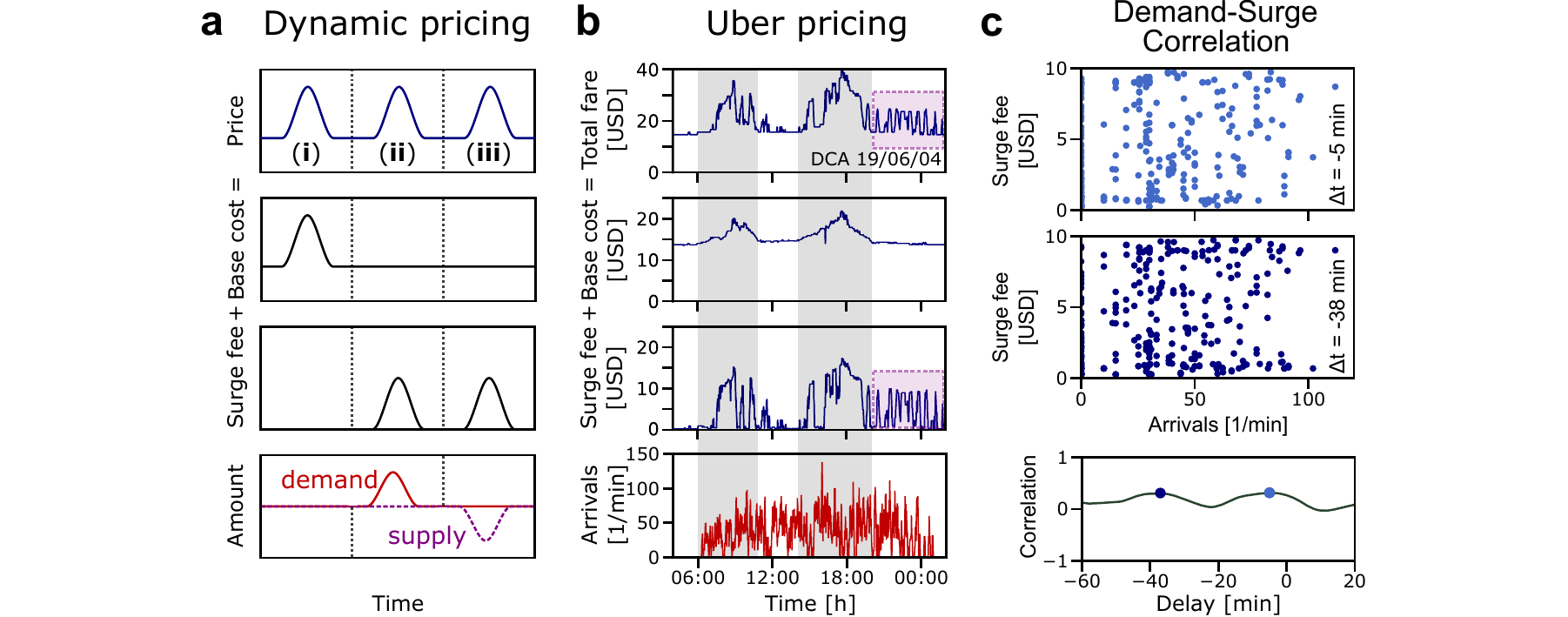}
    \caption{\textbf{Dynamic pricing in on-demand mobility}. \textbf{a}, Schematic illustration of dynamic pricing. The total price separates into the base cost of the product or service and a supply and demand dependent surge fee. Three fundamental mechanisms underlying price changes are (i) changes of the base cost, (ii) demand exceeding current supply levels and (iii) supply shortage compared to current demand. Price adaptations (ii) and (iii) are intended to drive the system back to a supply-demand equilibrium. \textbf{b}, The total fare for Uber ride-hailing services similarly decomposes into base cost and surge fee.  
    Base cost depend on trip duration and reflect current traffic conditions while surge fees result from supply-demand imbalances. Both effects are illustrated here for trips between Reagan National Airport (DCA) and Washington Union Station in Washington, D.C., USA. 
    During commuting hours (grey) base cost increase because of longer expected trip duration 
    during rush-hour. 
    The slower speed effectively reduces the supply of available drivers as they spend more time in traffic and naturally causes accompanying surge fees. During late evening and nighttime, the total fare exhibits repeated price surges triggered by supply-demand imbalances (dashed box) not reflected in the demand dynamics (passenger capacity of airplanes landing in DCA). \textbf{c}, Supporting the previous observation, no apparent correlation exists between the surge fee and the demand dynamics during the evening hours (20:00 - 02:00), even at five and 38 minute delays, the two local maxima of the correlation function (see Supplementary Material for a more detailed analysis).
    }
    \label{fig:Fig1}
\end{figure*}

Dynamic pricing schemes in mobility services are commonly applied by on-demand mobility service providers, such as Lyft and Uber \cite{LyftPricing2019, UberSurge2019}.
For Uber, the price of the service (the total fare for a ride) decomposes into the same two parts described above \cite{UberSurge2019}, base cost $p_\mathrm{base}$ and surge fee $p_\mathrm{surge}$,
\begin{equation}
    p = p_\mathrm{base} + p_\mathrm{surge}(D,S) \,,
\end{equation}
as illustrated in Figure~\ref{fig:Fig1}b for trips from Reagan National Airport (DCA) to Union Station in Washington, D.C. (see Methods and Supplementary Material for more information).

\newpage

\clearpage

The first component (base cost) are regular fees for a ride
\begin{equation}
    p_\mathrm{base} = p_0 + p_t \, \Delta t + p_l \, \Delta l \,,
\end{equation}
including one-off fees $p_0$ as well as trip fees $p_t$ and $p_l$, proportional to the duration $\Delta t$ and distance $\Delta l$ of the trip, similar to the fare for a typical taxi cab. 
These base cost increase, for example, during times of heavy traffic, such as morning and evening commuting hours (grey shading in Fig.~\ref{fig:Fig1}b) when the trip duration $\Delta t$ increases due to congestion. 

The second component (surge fee $p_\mathrm{surge}$) implements Uber's \emph{surge pricing} algorithm \cite{UberSurge2019, Garg2019} and reflects the time evolution of supply-demand imbalances.
The surge fee 
increases due to persistent supply-demand imbalance during commuting hours. 
Longer trip duration means that drivers spend more time in traffic serving the same number of customers which effectively reduces the supply of available drivers compared to the demand and causes an increase of the surge fee. 
These price surges are meant to incentivize customers to delay their request, reducing the current demand, as well as to incentivize drivers to offer their service in areas or at times with high demand, increasing the supply. 

As illustrated in Fig.~\ref{fig:Fig1}b, during the evening the system settles 
to constant base cost, reflecting constant trip duration in uncongested traffic.
Yet, 
even under these apparent equilibrium conditions, the surge fee exhibits a series of short, repeated price surges 
(dashed box in Fig.~\ref{fig:Fig1}b) that are not reflected in the demand dynamics (Fig.~\ref{fig:Fig1}c). 
Consistent with this observation, recent reports \cite{ABC7News2019a, ABC7News2019b, Mohlmann2017} suggest that Uber drivers at DCA and other locations cause artificial supply shortages on purpose to induce these price surges. 
This behavior enables drivers to increase their revenue by capitalizing on the increased total fare. 
Still, a couple of key questions remain open. 
First, what is the dynamic origin of these non-equilibrium dynamics and under which conditions do they emerge? 
Second, can this non-equilibrium state be identified from available data without direct observation of the supply dynamics?

\newpage

\begin{figure*}[h]
    \centering
    \vspace{5cm}
    \includegraphics{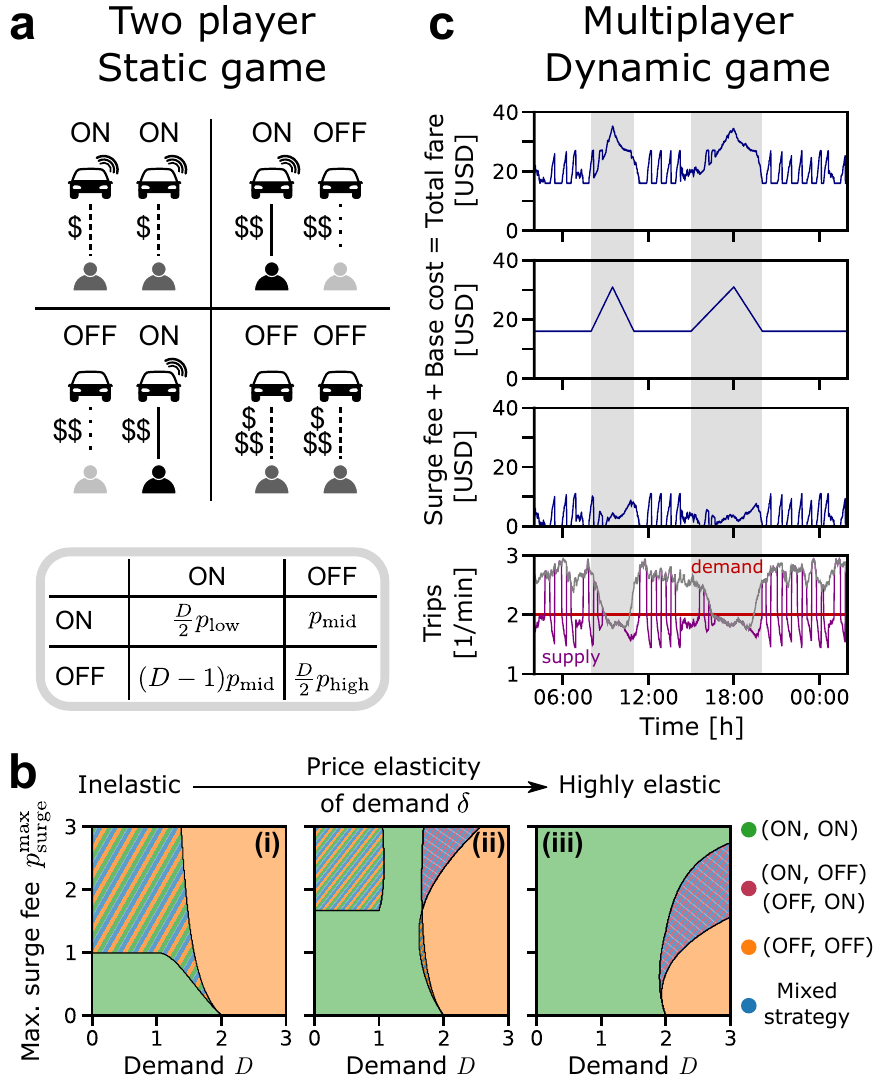}
    \vspace{3cm}
\end{figure*}
\newpage

\begin{figure*}
    \centering
    \caption{
    \textbf{
    Incentive structure in dynamic pricing}. \textbf{a}, A two player game captures the fundamental incentives for drivers. 
    Both drivers compete for a fixed average number of customers $1 \le D \le 2$. The drivers may choose to temporarily switch off their apps to induce an artificial supply shortage and additional surge fees (see Methods and Supplementary Material for details). (top left): If both drivers keep their apps on, both earn $p_\mathrm{low}$ (\$) with probability $D/2$. (top right and bottom left): If one driver switches their app off, the total fare increases to $p_\mathrm{mid}$ (\$\$). However, the other driver exploits their first-mover advantage to secure a customer, earning guaranteed $p_\mathrm{mid}$, while the offline driver only earns $(D-1) \, p_\mathrm{mid}$ from the remaining demand. (bottom right): If both drivers switch off their apps, they induce a larger supply deficit and thus a larger surge fee, resulting in the total fare $p_\mathrm{high}$ (\$\$\$). Both drivers again share the demand equally when they go back online. 
    \textbf{b}, Phase diagram of the resulting Nash equilibria in the two player game. (i): If the demand is sufficiently large, 
    the game is trivial and both drivers always switch off their app, triggering anomalous supply shortages (orange).
    At low demand 
    the game becomes a prisoner's dilemma \cite{PrisonersDilemma2019} or stag hung \cite{Osborne1994} and both drivers remain online (green). (ii) and (iii): As the demand becomes more elastic and decreases as the price increases, drivers switching off their app risk missing out on a customer completely and the parameter range promoting artificial price surges becomes smaller (orange). Drivers are more likely to both remain online (green). 
    \textbf{c}, A dynamic game with multiple drivers (see Methods and Supplementary Material for details) qualitatively reproduces the observed dynamics (compare DCA, Fig.~\ref{fig:Fig1}b): Sustained non-zero surge fees occur during commuting hours (grey). During non-commuting hours, drivers cooperate to induce artificial supply shortages and create price surges to optimize their collective profit. 
    }
    \label{fig:Fig2}
\end{figure*}

\clearpage

A first principles game theoretic description captures the fundamental incentives underlying the anomalous supply shortages: 
$S = 2$ drivers are competing for a fixed demand $D$ aiming to maximize their expected profit (Fig.~\ref{fig:Fig2}a). For illustration, we take a piecewise linear function such that drivers earn the total fare 
\begin{eqnarray*}
    p^\prime(S,D) = \begin{cases}
        p_\mathrm{base} \quad &
        \text{if}\quad S \ge D \\
        p_\mathrm{base} + p_\mathrm{surge}^\mathrm{max} \, \frac{D-S}{D}& \text{else} 
        \end{cases}
\end{eqnarray*}
when they serve a customer, where $p_\mathrm{base}$ denotes the (constant) base cost and $p_\mathrm{surge}^\mathrm{max}$ denotes the maximum possible surge fee when $S = 0$ (see Methods and Supplementary Material for details). 
Each driver has the option to temporarily not offer their service, contributing to an artificial supply shortage, $S < 2$. 
As drivers turn off their app, the fare increases from $p_\mathrm{low} = p^\prime(2,D)$ with both drivers online over $p_\mathrm{mid} = p^\prime(1,D) \ge p_\mathrm{low}$ as one driver goes offline to $p_\mathrm{high} = p^\prime(0,D) \ge p_\mathrm{mid}$ when both drivers withhold their service. 
While drivers who do not offer their service would typically miss out on a customer, the use of online mobile applications in most ride-sharing services enables them to quickly change their decision. 
Turning their app back on, they can capitalize on the additional surge fee and earn the higher total fare by quickly accepting a customer before the dynamic pricing algorithm reacts (Fig.~\ref{fig:Fig2}a, see Methods for details). 

Figure~\ref{fig:Fig2}b illustrates the phase diagram of the resulting Nash equilibria. 
When the demand is inelastic and does not change as the price increases [Fig.~\ref{fig:Fig2}b, panel (i)], the payoff structure of the game changes from a prisoner's dilemma \cite{PrisonersDilemma2019} over a stag hunt \cite{Osborne1994} to a trivial game as demand increases.
At low demand, the high risk of completely missing out on a customer if the other driver remains online disincentivizes switching off the app. 
At high demand, this risk disappears and drivers always profit from inducing artificial supply shortages to earn the additional surge fee. 
As the demand becomes more elastic, i.e. the demand decreases in response to an increase of the total fare as
\begin{align}
    D^\prime(p^\prime, D) = D \, (1 - \delta\, (p^\prime - p_\mathrm{base})) \,
\end{align}
where $\delta$ denotes the price elasticity of the demand,
the risk of missing out on a customer increases and the range of parameters where artificial price surges are incentivized becomes smaller [Fig.~\ref{fig:Fig2}b, panel (ii) and (iii)]. 

In general the specific conditions promoting artificial price surges depend on the details of the demand dynamics. Nonetheless, the supply-side incentives remain qualitatively unchanged and are a generic property of dynamic pricing schemes. 
To isolate the impact of these supply side incentives, we simulate a time-continuous game under constant conditions (constant demand, a constant number of drivers and a constant price elasticity of demand) where the ON-OFF-decisions of the drivers are the only remaining dynamics (Fig.~\ref{fig:Fig2}c). 
Drivers react to the current conditions and can choose to turn their app on or off at any time. 
They contribute to an artificial supply shortage if sufficiently many other idle drivers are willing to also participate, following their mean-field optimal strategy.
To avoid never making profit, however, individual drivers remain offline only for a short amount of time, explicitly fixing the timescale of potential artificial price surges (see Supplementary Material for details).
The simulations reproduce qualitatively the same non-equilibrium price dynamics as observed in the recorded price data (compare Fig.~\ref{fig:Fig1}b): 
Increases of the trip duration during commuting hours (grey shading in Fig.~\ref{fig:Fig2}c) are accompanied by a sustained supply-demand imbalance and surge fees without drivers turning off their app.
At other times, the drivers create short, artificial price surges to maximize their profit. 
This result demonstrates that the systemic incentives in dynamic pricing schemes alone are sufficient to cause anomalous supply dynamics.

The fact that these incentives are generic to dynamic pricing schemes suggests that artificial supply shortages and non-equilibrium surge dynamics emerge independent of the location. 
However, direct observation of the supply dynamics, e.g. of the number and location of online drivers, is typically impossible as this information is not publicly available. 
Even with the above results, a bottom-up prediction 
is practically not feasible since the exact conditions under which these dynamics are promoted depend on the specific details of the trip, the local demand dynamics, publicly unavailable details on the surge pricing algorithm as well as additional external influences such as local legislation. 

\begin{figure*}
    \centering
    \includegraphics[width=\textwidth]{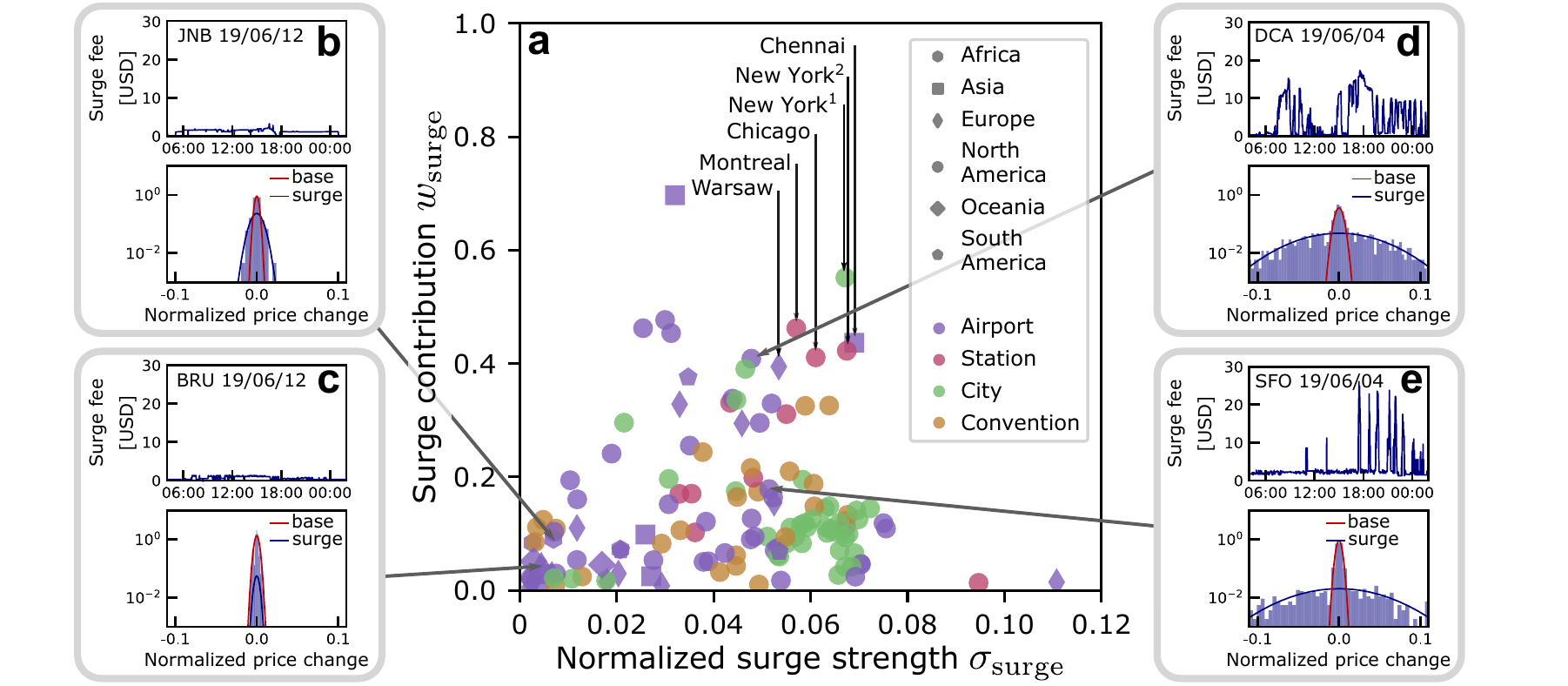}
    \caption{\textbf{Characterizing non-equilibrium surge dynamics}.
    The timescales of price changes characterize the surge dynamics at different locations. 
    These price changes separate into small, slow changes $P_\mathrm{base}$ corresponding to varying base costs and fast, strong changes $P_\mathrm{surge}$ corresponding to the surge fee (red and blue line in the histograms in \textbf{(b)}-\textbf{(e)}, respectively). 
    \textbf{a}, Characterizing locations by the total weight $w_\mathrm{surge}$ of the surge component of the price change distribution  
    and the magnitude $\sigma_\mathrm{surge}$ of the associated price changes 
    reveals several locations [e.g., Warsaw, Montreal, Chicago, New York City [city(1) and station(2)] and Chennai] with similar characteristics to DCA (see Fig.~\ref{fig:Fig4} and Methods and Supplementary Material for more details and additional examples). 
    \textbf{b} and \textbf{c}, Locations with low surge strength $\sigma_\mathrm{surge}$ exhibit no significant surge activity and no price changes on a fast time scale, shown here for Johannesburg (JNB, South Africa) and Brussels (BRU, Belgium).
    \textbf{d}, Locations with high surge strength $\sigma_\mathrm{surge}$ and small surge contribution $w_\mathrm{surge}$ exhibit relatively few price surges (San Francisco, USA).
    \textbf{e}, Locations with high surge strength $\sigma_\mathrm{surge}$ and high surge contribution $w_\mathrm{surge}$ exhibit a large number of fast price surges potentially driven by artificially induced supply shortages. Figure~\ref{fig:Fig4} confirms that the surge fee dynamics at these locations is indeed similar to the dynamics observed at DCA (Washington D.C., USA). 
    }
    \label{fig:Fig3}
\end{figure*}

We overcome these obstacles by exploiting the characteristic temporal structure of the surge dynamics (compare Fig.~\ref{fig:Fig1}b) to identify locations with similar dynamics that are 
characteristic for 
artificial supply shortages.
Based only on the price time series, without requiring further input on demand or supply or the specific dynamic pricing algorithm, we quantify the timescales of normalized price changes $\Delta p$ 
for 137 different routes in 59 urban areas across six continents (Fig.~\ref{fig:Fig3}a, see Methods for details).
The distribution of price changes separates into a slow and fast timescale and a contribution where the price does not change
\begin{equation}
    P\left(\Delta p\right) = w_\mathrm{base} \, P_\mathrm{base}\left(\Delta p; \sigma_\mathrm{base}\right) + w_\mathrm{surge} \, P_\mathrm{surge}\left(\Delta p; \sigma_\mathrm{surge}\right) + w_0 \, \delta(\Delta p) \,.
\end{equation}
The slow price changes $P_\mathrm{base}\left(\Delta p; \sigma_\mathrm{base}\right)$ describe changes of the base cost varying as slowly as traffic conditions change during the day. 
The fast price changes $P_\mathrm{surge}\left(\Delta p; \sigma_\mathrm{surge}\right)$ are associated with sudden changes of the surge fee. 
The last term $w_0 \, \delta(\Delta p)$ describes times when the price remains constant price and contributes only at $\Delta p = 0$, where $\delta$ represents the Dirac-Delta distribution and $w_0$ the remaining weight $w_0 = 1 - w_\mathrm{base} - w_\mathrm{surge}$.

Characterizing the contribution $w_\mathrm{surge}$ of the surge fee and the magnitude $\sigma_\mathrm{surge}$ of the associated price changes with a maximum likelihood Gaussian mixture model fit (see Methods for details) 
\begin{equation}
    P_x(\Delta p; \sigma_x) = \frac{1}{\sqrt{2 \pi \sigma_x^2}}\,e^{-\frac{\Delta p^2}{2\sigma_x^2}}
\end{equation}
with $x \in \left\{\mathrm{base}, \mathrm{surge}\right\}$ we find 
locations without surge activity (Fig.~\ref{fig:Fig3}b and c) as well as locations with strong but infrequent price surges (Fig.~\ref{fig:Fig3}e).
Importantly, we also identify several locations with price change characteristics similar to those observed at DCA, with a high magnitude and contribution of surge price changes, suggesting strong and frequent price surges potentially driven by anomalous supply dynamics (compare Fig.~\ref{fig:Fig3}d).
\newpage

\clearpage

Indeed, all of the identified locations exhibit qualitatively similar non-equilibrium surge fee dynamics with a large number of repeated price surges, in particular during evening hours, demonstrating that the phenomenon is ubiquitous (Fig.~\ref{fig:Fig4}, see Supplementary Material for additional examples). 
While these results do not directly imply that the price surges at these locations are artificially induced, the similarity to confirmed artificial price surges and the universality of the incentives for drivers makes it a likely conclusion.

\begin{figure*}
    \centering
    \includegraphics[width=\textwidth]{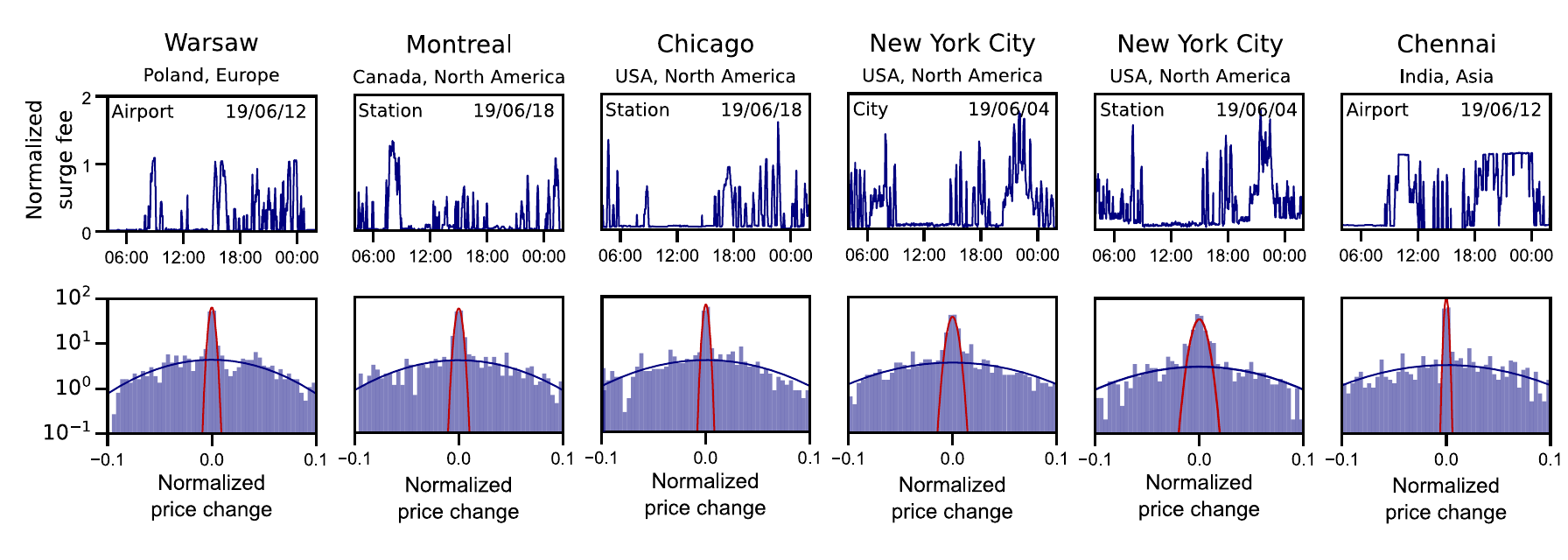}
    \caption{\textbf{Identifying non-equilibrium surge dynamics and anomalous supply shortages}. Repeated price surges similar to those observed at DCA (compare Fig.~\ref{fig:Fig3}e) emerge in locations across the globe (America, Asia and Europe) and independent of type of origin (airport, train station and other prominent locations). The surge dynamics at the six locations identified in Fig.~\ref{fig:Fig3}a is qualitatively and statistically similar to DCA. 
	The time scale separation indicates that all these observed surge dynamics are similar to characteristic dynamics 
    of anomalous supply shortages. 
    In particular, sustained periods with non-zero surge fee likely reflect a real supply-demand imbalance at that time while periods with repeated surge peaks are characteristic for price surges induced by artificial supply shortages (e.g. Warsaw evening, Montreal evening, Chicago evening, New York City afternoon and evening).
    }
    \label{fig:Fig4}
\end{figure*}

\clearpage

\section*{Discussion}
In summary, we 
quantitatively demonstrated the emergence of non-equilibrium price dynamics in on-demand mobility systems 
at various locations 
across the globe.
We 
showed that the fundamental incentives 
sufficient for 
promoting 
their 
emergence 
through supply anomalies
constitute a generic property of dynamic pricing schemes. 
In particular, these incentives are independent of the urban area, the specific route, or the exact type of mobility service and should even apply across industries. 
Our methodology to classify the price dynamics of 
on-demand mobility systems without explicit knowledge about the time-resolved demand and supply evolution
 
enables 
a direct, systematic search for 
supply 
anomalies based on price time series only. 
Furthermore, characterizing 
the incentives and the conditions that promote artificially induced price surges enables targeted action to suppress the emergence of 
such supply anomalies. 
Specific 
actions may include 
offering ride-sharing options \cite{Vazifeh2018, Tachet2017, Santi2014, Molkenthin2019} (effectively lowering the demand, compare Fig.~\ref{fig:Fig2}b) 
or providing more 
or 
alternative public transport options (effectively increasing the price elasticity of demand, compare Fig.~\ref{fig:Fig2}b). 
In particular, our results suggest that limiting the maximum surge fee, as 
done 
in response to the initial reports from DCA \cite{NPR2019} and frequently discussed as potential legislation \cite{Patel2019, ETTech2019}, is not an effective response and may even result in the opposite effect if the demand is highly elastic (compare Chennai, Fig.~\ref{fig:Fig4}). 

In general, 
with the emergence of digital platforms, sharing economies and autonomous vehicle fleets, mobility services are becoming increasingly self-organized and complex 
such that 
new, potentially unintended collective dynamics will emerge \cite{Millard-Ball2019, Helbing2000b, Strogatz2005, Havlin2012, Helbing2013, Schroder2018, zeng2019switch}. 
Our results provide conceptual insights into these dynamics to support the creation and regulation of fair, efficient and transparent publicly available 
mobility services \cite{Vazifeh2018, Tachet2017, Santi2014, auer2015dynamics, OKeeffe2019, Molkenthin2019}. 

\clearpage

\section*{Methods}

\textbf{Data sources and acquisition}.
In this work we recorded approximately 28 million ride-hailing price estimates for 137 routes 
of \textit{Uber} rides 
in 59 urban areas across six continents between 19-05-31 and 19-06-25. We distinguish between four types of routes based on the origin location: 
63 airport, 23 convention center, 12 train station and 39 city trips (see Supplementary Data for detailed information and precise GPS coordinates of the different routes). 

For each route, we prompted total fare requests with a fixed interval via Uber's \textit{price estimate} API endpoint recording the price estimates for each route every 2 to 30 seconds. Per request, the API returned lower and upper total fare estimates for all Uber products operating in the local area 
as well as estimated distance and duration of the trip
which we equipped with the request timestamp. 
Using Uber's \textit{products} API endpoint, we complemented the price estimate data with information on local booking fee, price per minute, price per mile, distance unit, minimum fees and the currency code parameter per product and location. We convert all price estimates to 
US Dollars 
based on currency exchange rates provided by the European Central Bank for the date of recording. 

In all our analyses, we work with the lower estimate of the local economy product (\textit{UberX}, \textit{UberGO} in India).\\ 

\textbf{Base cost}.
To determine the base cost (sum of pickup fee, trip fee and surcharges) of a trip we first compute the trip fee based on the price per mile, price per minute and the estimated trip length and duration. We add the pickup fee obtained from the Uber \textit{products} API. Since data on the surcharges (e.g. airport fees or tolls) of individual trips is not available, we take surcharges to be constant for each trip. We subtract the pickup fee and trip fee from the price estimate and take the minimum value of this remaining surge fee and surcharge cost as estimate of the surcharges, such that zero surge fee occurs at least once in the recorded price estimates.\\ 

\textbf{Surge fee}.
To estimate the surge fee time series we subtract the base cost of the respective product from the total fare estimate. Since the available price estimates are rounded to integer values, the recorded price estimate may not reflect all changes of the trip fare especially for shorter trips with lower absolute total fare). This leads to small fluctuations in the extracted surge fee that do not correspond to actual surge activity. \\

\textbf{Airport arrival data}. 
To estimate the demand for rides at airports, we record the number of arrivals at each of the 63 airports where we recorded price estimates. We collected aircraft landing times, call signs and type of aircraft using flightradar24's open API in the corresponding time frame, as well as information on the different aircraft's current seat configuration obtained via flightera.net. We disregard entries without call signs or real landing times. In rare cases where no seat configuration was available, we estimate the number of seats as the average of all recorded flights with the same aircraft model (or the average over all aircraft models if no other similar model was recorded).\\

\textbf{Airport demand}. 
We estimate 
the demand for ride-hailing services as 
proportional to the number of seats of all arriving airplanes (implying a constant fraction of potential Uber customers). 
To create a continuous time series from the discrete arrival events of individual airplanes we compute a five minute moving average to create equidistant records every minute. This also slightly reduces the strong variations between minutes with and without arrivals.

Because we have much more frequent but not equally spaced data for the Uber price estimates, 
we use the same procedure and compute a five minute moving average of the surge fee for every minute. This leaves us with the same granularity of the data as for the deplanements.

Using 
these 
data, we compute the cross-correlation between the Uber surge fee estimates and the deplanement data at the corresponding airport. In Fig.~1c we show the scatterplot at the timelag (deplanement earlier than surge) where this correlation is maximal for the illustrated window from 20:00 to 02:00 of the surge fee.\\

\newpage
\textbf{Comparison of surge dynamics}. To compare and characterize the surge dynamics for different trips we normalize the absolute surge fee time series by the base cost at that time, yielding an effective surge factor. For these normalized time series, we compute the per minute changes $\Delta p$ between consecutive time points (time $t$ in minutes),
\begin{eqnarray}
    \Delta p(t) &=& \frac{ \textrm{total~fare}(t) }{ \textrm{base~cost}(t) } - \frac{ \textrm{total~fare}(t-1) }{ \textrm{base~cost}(t-1) } \label{eq:supp_normalized_price} \\
    &=& \frac{ \textrm{surge~fee}(t) }{ \textrm{base~cost}(t) } - \frac{ \textrm{surge~fee}(t-1) }{ \textrm{base~cost}(t-1) } \,.\nonumber
\end{eqnarray}

To quantify and compare the statistical properties of the surge factor time series we split the price changes into three contributions. We take any data point with $\Delta p^2 < 10^{-7}$ to belong to a Dirac delta distribution at zero (not shown in the histograms) and fit a Gaussian mixture model with two Gaussian distributions to the remaining data.  
Taking both distributions 
to 
have a mean of zero (no price change on average) 
yields 
\begin{eqnarray*}
    \mathrm{Prob}\left(\Delta p\right) &=& w_0 \, \delta(\Delta p) \nonumber \\
    &+& w_\mathrm{base} \, \frac{1}{\sqrt{2 \pi \sigma_\mathrm{main}^2}} \, e^{-\frac{\Delta p^2}{2\,\sigma_\mathrm{base}^2}} \nonumber \\
    &+& w_\mathrm{surge} \, \frac{1}{\sqrt{2 \pi \sigma_\mathrm{surge}^2}} \, e^{-\frac{\Delta p^2}{2\,\sigma_\mathrm{surge}^2}}
\end{eqnarray*}
where the weight $w_\mathrm{surge}$ defines the \textit{surge contribution} and the standard deviation $\sigma_\mathrm{surge}$ is the \textit{normalized surge strength} used to characterize the surge dynamics. \\

\textbf{Two player game -- minimal theoretical model}.
The results presented in the manuscript (Fig.~\ref{fig:Fig2}B) are obtained with normalized parameters $p_\mathrm{base} = 1$ and $\delta \in \left\{0, 0.15, 0.30\right\}$, allowing up to $p_\mathrm{surge}^\mathrm{max} = 1/0.30 \approx 3.33$ before no customer orders a ride at the maximum surge fee. See Supplementary Material for a detailed description.\\

\textbf{Dynamic multiplayer game}.
For the dynamic multiplayer game, we consider a single origin location with $N=160$ drivers. Upon completing a trip, drivers return to the origin location after a total round-trip time $t_s$ uniformly distributed in $\left[\left<t_s\right> - 5, \left<t_s\right> + 5\right]$ minutes. We increase the round-trip time from the base value $\left<t_s\right> = 30$ minutes to $\left<t_s\right> = 60$ minutes in the morning and afternoon (starting at 08:00 and increasing linearly up to the maximum at 9:30 and back to the base value until 11:00. Similarly in the afternoon from 15:00 to the maximum at 18:00 and back until 20:00).

The base cost $p_\mathrm{base}$ depend linearly on the round-trip time as $p_\mathrm{base} = 1 + \left<t_s\right>/2 \in \left[16, 31\right]$ USD as the round-trip time changes during day. Similar to the two-player game, we 
take 
a linear price dependence for the surge pricing as 
\begin{equation}
    p^\prime(t) = \begin{cases}
                       p_\mathrm{base} \quad &\text{if} \quad N_\mathrm{idle}(t) \ge N_\mathrm{thresh} \\
                        p_\mathrm{base} + p_\mathrm{surge}^\mathrm{max} \left(1 - \frac{N_\mathrm{idle}(t)}{N_\mathrm{thres}}\right)    \quad &\text{else}
    \end{cases} \nonumber
\end{equation}
based on the number $N_\mathrm{idle}$ of online drivers at the trip origin and the number of drivers $N_\mathrm{thresh}$ before the surge fee becomes non-zero. We 
take 
$N_\mathrm{thresh} = \lambda \, \left<t_s\right>$, where $\lambda = 2$ requests per minute describes the demand modeled as a Poisson process in time. 
We model responses of the price to the current system state (number of available drivers and round-trip time) as instantaneous.

The behavior of customers and drivers is as follows: 
Each customer $i$ is assigned a uniformly random maximum price $p_\mathrm{max,i} \in [p_\mathrm{base}, p_\mathrm{max}]$ they are willing to pay, where we take the price of Uber Black as $p_\mathrm{max} = 54$ USD. When the customer makes a request, they check the current total fare. If the current total fare is smaller than $p_\mathrm{max,i}$, the customer orders the ride. If the total fare is higher or no drivers are online and idle, the customer waits and checks again every 2 minutes. 
After 10 minutes without ordering a ride, the customer leaves the system.

At every point in time the drivers decide whether to switch their app off or on. They make this decision based on the (mean field) optimal strategy to optimize their collective payoff. A driver switches off their app only if two conditions are fulfilled: first, if there are sufficiently many drivers available and willing to be offline to induce a non-zero surge fee. Second, if the price is less than the (mean field) optimal value for the drivers given the current system state. Each driver remains offline for at most 20 minutes. After this time, the driver only considers going offline again after serving a customer (drivers try to obtain similar individual profits whereas their optimal strategy based on maximizing their collective profit would be for some drivers to be always offline).

\newpage

\nocite{AmazonDynamicPricing, UberSurge2019, LyftPricing2019, UberCities2019, UberSurge2019, Garg2019, UberPriceEstimates2019, Garg2019, UberNewSurge2018, UberPriceEstimates2019, UberLaunchDates2014, ABC7News2019a, DCTaxiMarketShare2019, ABC7News2019a, ABC7News2019a, ABC7News2019a, UberNewSurge2018, Garg2019, Skyrms2003, PrisonersDilemma2019, UberStatistics2019, ABC7News2019a}

\bibliographystyle{unsrt}
\bibliography{uber_surge}

\begin{thebibliography}{10}

\bibitem{Strogatz2005}
Steven~H. Strogatz, Daniel~M. Abrams, Allan McRobie, Bruno Eckhardt, and Edward
  Ott.
\newblock {Crowd synchrony on the Millennium Bridge}.
\newblock {\em Nature}, 438(7064):43--44, 2005.

\bibitem{buldyrev2010catastrophic}
Sergey~V. Buldyrev, Roni Parshani, Gerald Paul, H.~Eugene Stanley, and Shlomo
  Havlin.
\newblock Catastrophic cascade of failures in interdependent networks.
\newblock {\em Nature}, 464(7291):1025--1028, 2010.

\bibitem{Havlin2012}
S.~Havlin, D.~Y. Kenett, E.~Ben-Jacob, A.~Bunde, R.~Cohen, H.~Hermann, J.~W.
  Kantelhardt, J.~Kert{\'{e}}sz, S.~Kirkpatrick, J.~Kurths, J.~Portugali, and
  S.~Solomon.
\newblock {Challenges in network science: Applications to infrastructures,
  climate, social systems and economics}.
\newblock {\em Europ. Phys. J. ST}, 214(1):273--293, 2012.

\bibitem{Helbing2013}
Dirk Helbing.
\newblock {Globally networked risks and how to respond}.
\newblock {\em Nature}, 497(7447):51--59, 2013.

\bibitem{Schroder2018}
Malte Schr{\"{o}}der, Jan Nagler, Marc Timme, and Dirk Witthaut.
\newblock {Hysteretic Percolation from Locally Optimal Individual Decisions}.
\newblock {\em Phys. Rev. Lett.}, 120(24):248302, 2018.

\bibitem{loder2019understanding}
Allister Loder, Lukas Amb{\"u}hl, Monica Menendez, and Kay~W. Axhausen.
\newblock Understanding traffic capacity of urban networks.
\newblock {\em Sci.. Rep.}, 9(1):1--10, 2019.

\bibitem{Helbing2000b}
Dirk Helbing, Ill{\'{e}}s Farkas, and Tam{\'{a}}s Vicsek.
\newblock {Simulating dynamical features of escape panic}.
\newblock {\em Nature}, 407(6803):487--490, 2000.

\bibitem{brockmann2006scaling}
Dirk Brockmann, Lars Hufnagel, and Theo Geisel.
\newblock The scaling laws of human travel.
\newblock {\em Nature}, 439(7075):462--465, 2006.

\bibitem{bashan2013extreme}
Amir Bashan, Yehiel Berezin, Sergey~V. Buldyrev, and Shlomo Havlin.
\newblock The extreme vulnerability of interdependent spatially embedded
  networks.
\newblock {\em Nat. Phys.}, 9(10):667--672, 2013.

\bibitem{li2015percolation}
Daqing Li, Bowen Fu, Yunpeng Wang, Guangquan Lu, Yehiel Berezin, H.~Eugene
  Stanley, and Shlomo Havlin.
\newblock Percolation transition in dynamical traffic network with evolving
  critical bottlenecks.
\newblock {\em Proc. Natl. Acad. Sci.}, 112(3):669--672, 2015.

\bibitem{zeng2019switch}
Guanwen Zeng, Daqing Li, Shengmin Guo, Liang Gao, Ziyou Gao, H~Eugene Stanley,
  and Shlomo Havlin.
\newblock Switch between critical percolation modes in city traffic dynamics.
\newblock {\em Proc. Natl. Acad. Sci.}, 116(1):23--28, 2019.

\bibitem{dynamicPricingGeneral}
Arvind Sahay.
\newblock {How dynamic pricing leads to higher profits}.
\newblock {\em MIT Sloan Management Review}, 48(4):53, 2007.

\bibitem{AmazonDynamicPricing}
Neel Mehta, Parth Detroja, and Aditya Agashe.
\newblock {How and why Amazon changes its prices so often}, 2018.
\newblock Accessed on 2019-07-05.

\bibitem{schafer2015decentral}
Benjamin Sch{\"a}fer, Moritz Matthiae, Marc Timme, and Dirk Witthaut.
\newblock Decentral smart grid control.
\newblock {\em New J. Phys.}, 17(1):015002, 2015.

\bibitem{LyftPricing2019}
{Lyft, Inc.}
\newblock {How to estimate a Lyft ride's cost}, 2019.
\newblock Accessed on 2019-07-05.

\bibitem{UberSurge2019}
{Uber Technologies, Inc.}
\newblock {How surge pricing works}, 2018.
\newblock Accessed on 2019-07-05.

\bibitem{ABC7News2019a}
Sam Sweeney.
\newblock {Uber, Lyft drivers manipulate fares at Reagan National causing
  artificial price surges}, 2019.
\newblock WJLA, Accessed on 2019-07-05.

\bibitem{ABC7News2019b}
Sam Sweeney.
\newblock {Uber drivers across the U.S. are manipulating fares to create
  artificial price surges}, 2019.
\newblock WJLA, Accessed on 2019-06-24.

\bibitem{Mohlmann2017}
Mareike M{\"{o}}hlmann and Lior Zalmanson.
\newblock Hands on the wheel: Navigating algorithmic management and drivers.
\newblock In {\em International Conference on Information Systems (ICIS 2017)},
  pages 1 -- 17, Seoul, South Korea, December 2017.

\bibitem{Garg2019}
Nikhil Garg and Hamid Nazerzadeh.
\newblock {Driver Surge Pricing}.
\newblock 2019.
\newblock ArXiv:1905.07544.

\bibitem{PrisonersDilemma2019}
Steven Kuhn.
\newblock Prisoner’s dilemma.
\newblock In Edward~N. Zalta, editor, {\em The Stanford Encyclopedia of
  Philosophy}. Metaphysics Research Lab, Stanford University, {Summer 2019}
  edition, 2019.

\bibitem{Osborne1994}
Martin~J. Osborne and Ariel. Rubinstein.
\newblock {\em {A course in game theory}}.
\newblock MIT Press, Cambridge, Massachusetts, 1994.

\bibitem{Vazifeh2018}
M.~M. Vazifeh, P.~Santi, G.~Resta, S.~H. Strogatz, and C.~Ratti.
\newblock {Addressing the minimum fleet problem in on-demand urban mobility}.
\newblock {\em Nature}, 557(7706):534--538, 2018.

\bibitem{Tachet2017}
R.~Tachet, O.~Sagarra, P.~Santi, G.~Resta, M.~Szell, S.~H. Strogatz, and
  C.~Ratti.
\newblock {Scaling Law of Urban Ride Sharing}.
\newblock {\em Sci. Rep.}, 7:42868, 2017.

\bibitem{Santi2014}
Paolo Santi, Giovanni Resta, Michael Szell, Stanislav Sobolevsky, Steven~H.
  Strogatz, and Carlo Ratti.
\newblock {Quantifying the benefits of vehicle pooling with shareability
  networks}.
\newblock {\em Proc. Natl. Acad. Sci.}, 111(37):13290--13294, 2014.

\bibitem{Molkenthin2019}
Nora Molkenthin, Malte Schr{\"{o}}der, and Marc Timme.
\newblock {Topological universality of on-demand ride-sharing efficiency}.
\newblock 2019.
\newblock ArXiV:1908.05929.

\bibitem{NPR2019}
{Karen Duffin}, {Sarah Johnston}, and {Nick Fountain}.
\newblock {Planet Money, Episode 921: Tales From The Parking Lot}, 2019.
\newblock Broadcasted on National Public Radio, 2019-06-21, 3:32 PM ET.

\bibitem{Patel2019}
{Anand Patel}.
\newblock {RSS economic wing wants Uber, Ola surge pricing capped, writes to
  Nitin Gadkari}, 2019.
\newblock India Today, Accessed 2019-10-01.

\bibitem{ETTech2019}
ETtech.com.
\newblock {Cap on surge pricing results in lower driver earnings: Uber}, 2019.
\newblock
  \url{https://tech.economictimes.indiatimes.com/news/mobile/cap-on-surge-pricing-results-in-lower-driver-earnings-uber/71205792}.
  Accessed 2019-10-01.

\bibitem{Millard-Ball2019}
Adam Millard-Ball.
\newblock {The autonomous vehicle parking problem}.
\newblock {\em Transp. Policy}, 75:99--108, 2019.

\bibitem{auer2015dynamics}
Sabine Auer, Jobst Heitzig, Ulrike Kornek, E~Sch{\"o}ll, and J{\"u}rgen Kurths.
\newblock The dynamics of coalition formation on complex networks.
\newblock {\em Sci. Rep.}, 5:13386, 2015.

\bibitem{OKeeffe2019}
Kevin~P. O'Keeffe, Amin Anjomshoaa, Steven~H Strogatz, Paolo Santi, and Carlo
  Ratti.
\newblock {Quantifying the sensing power of vehicle fleets}.
\newblock {\em Proc. Natl. Acad. Sci.}, 116(26):12752--12757, 2019.

\bibitem{UberCities2019}
{Uber Technologies, Inc.}
\newblock {Uber Cities Across the Globe }, 2019.
\newblock \url{https://www.uber.com/en/cities/}. Accessed on 2019-07-09.

\bibitem{UberPriceEstimates2019}
{Uber Technologies, Inc.}
\newblock {Uber Estimate - Get a Price Estimate in Your City}, 2019.
\newblock \url{https://www.uber.com/de/en/price-estimate/}. Accessed on
  2019-06-24.

\bibitem{UberNewSurge2018}
{Uber Technologies, Inc.}
\newblock {Your questions about the new surge, answered}, 2018.
\newblock
  \url{https://www.uber.com/blog/your-questions-about-the-new-surge-answered/}.
  Accessed on 2019-07-05.

\bibitem{UberLaunchDates2014}
Frank Bi.
\newblock {Uber launch cities and dates}, 2014.
\newblock
  \url{https://github.com/voxmedia/data-projects/tree/master/verge-uber-launch-dates}.
  Accessed on 2019-07-05.

\bibitem{DCTaxiMarketShare2019}
{Department of For-Hire Vehicles, Washington D.C.}
\newblock {Surcharge generated by Taxicabs, Private Sedans and Black Cars},
  2019.
\newblock
  \url{https://dfhv.dc.gov/page/dfhv-dashboard-and-statistical-data-sets}.
  Accessed on 2019-07-05.

\bibitem{Skyrms2003}
Brian Skyrms.
\newblock {\em {The Stag Hunt and the Evolution of Social Structure}}.
\newblock Cambridge University Press, Cambridge, 2003.

\bibitem{UberStatistics2019}
{Uber Technologies, Inc.}
\newblock {Facts \& Figures as of December 2018}, 2019.
\newblock \url{https://www.uber.com/en-PK/newsroom/company-info/}. Accessed on
  2019-07-05.

\bibitem{Note1}
This information on Uber's price model is based on Ref. \cite
  {UberPriceEstimates2019}, accessed on 19/06/24, and may be subject to change.

\bibitem{Note2}
According to the US Federal Communications Commission,.

\bibitem{Note3}
This is not accurate in individual cases where airport pickup fees may change
  over time or tolls may be in effect during the day but not at night.

\end{thebibliography}

\newpage

\section*{Acknowledgements}
We thank Nora Molkenthin and Jan Nagler for helpful discussions. M.T. acknowledges support from the German Research Foundation (Deutsche Forschungsgemeinschaft, DFG) through the Center for Advancing Electronics Dresden (cfaed).

\section*{Author contribution}
M.S. initiated the research with help from D.S. and P.M. All authors conceived the research. M.S. and M.T. planned the research, supported by D.S. and P.M. P.M. and D.S. collected and analyzed the empirical data. M.S. designed and analyzed the game theoretic models. All authors contributed to interpreting the results and writing the manuscript.

\section*{Competing interest}
The authors declare no competing interests.

\clearpage

\renewcommand{\thesection}{S\arabic{section}}
\renewcommand{\thesubsection}{\thesection.\arabic{subsection}}
\renewcommand{\thesubsubsection}{\thesubsection.\arabic{subsubsection}}

\makeatletter
\renewcommand{\p@subsection}{}
\renewcommand{\p@subsubsection}{}
\makeatother

\makeatletter
\def\l@subsubsection#1#2{}
\makeatother

\renewcommand{\thefigure}{S\arabic{figure}}
\renewcommand{\thetable}{S\arabic{table}}
\renewcommand{\theequation}{S\arabic{equation}}

{\centering
\large{Appendix}\\
{\small{accompanying the manuscript}}\\
Anomalous supply shortages from dynamic pricing in on-demand mobility
}\\[0.2cm]

In the main manuscript we illustrate how inherent incentives of dynamic pricing induce anomalous supply dynamics and artificial price surges in ride-hailing services. We show that the incentives driving these dynamics are generic to dynamic pricing schemes and analyze the specific conditions that promote them. Based on confirmed reports of these non-equilibrium dynamics at Reagan National Airport (DCA) in Washington, D.C. \cite{ABC7News2019a}, we identify locations with similar surge fee dynamics from Uber price estimates for trips originating from 63 airports around the globe as well as 23 convention centers, 12 train stations and 39 city trips in North America (largely US). 

In this Supplementary Material we provide additional details on all analyses and results presented in the main manuscript. This document is structured as follows:\\

\paragraph*{S1. Dynamic pricing in ride-hailing}\quad\\
\noindent We explain Uber's dynamic pricing model and isolate the surge fee component influenced by supply and demand imbalances. We give a detailed comparison of the surge activity to the estimated demand at DCA and other airports.\\

\paragraph*{S2. Statistical properties of surge pricing}\quad\\
\noindent We compare the surge fee dynamics of all recorded price estimates to identify locations with similar surge activity to DCA that may be driven by anomalous supply dynamics.\\

\paragraph*{S3. Incentive structure for drivers under dynamic pricing}\quad\\
\noindent We explain the incentives that determine the action of drivers under dynamic pricing in a simplified game-theoretic setting.\\

\paragraph*{S4. Data - methodology and details}\quad\\
\noindent We provide a detailed explanation of data collection and summarize our data processing approach.


\newpage

\section{\label{sec:uber_dynamic_pricing}Dynamic pricing in ride-haling}

Dynamic pricing is a general mechanism to 
adjust prices to time-varying conditions. Application of dynamic pricing schemes are widespread in online retail and used, for example, by Amazon \cite{AmazonDynamicPricing}, and is of increasing importance in transportation contexts, too. In particular for ride-hailing services, where service conditions may vary strongly over the day due to weather, special events or simply rush hour and congestion, many service providers apply dynamic pricing \cite{UberSurge2019, LyftPricing2019}.

\begin{figure*}[h]
\centering
    \includegraphics[width=\linewidth]{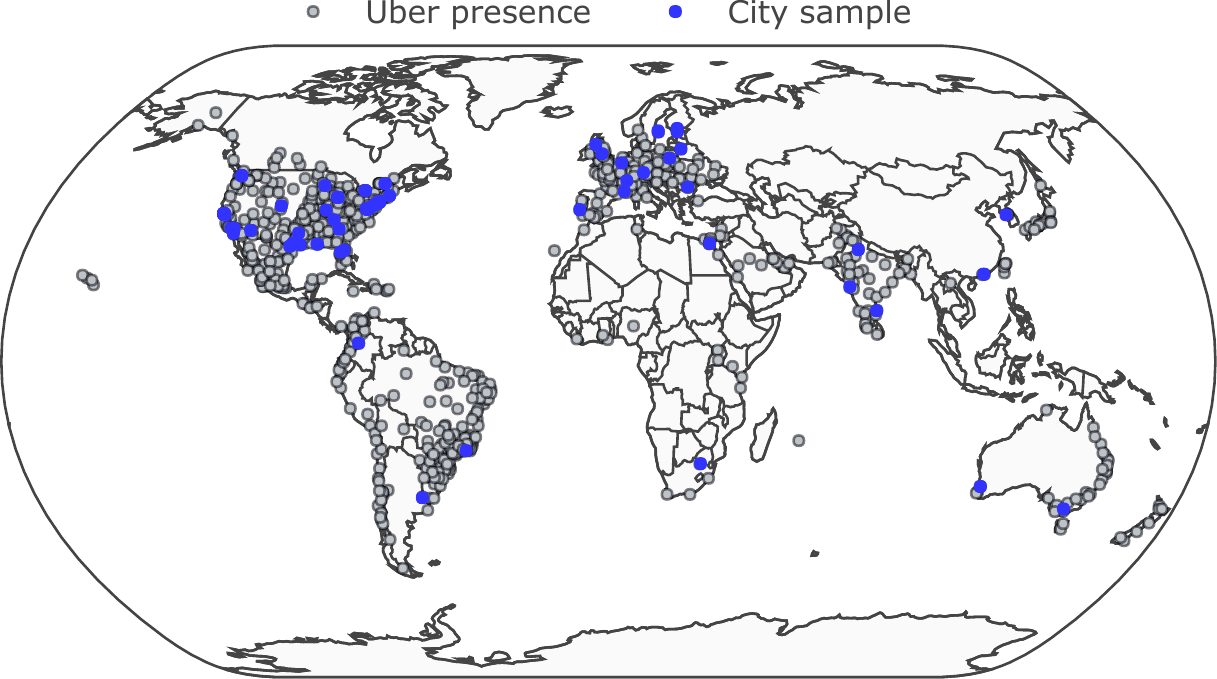}    
\caption{\textbf{Uber ride-hailing services around the world.} Blue markers indicate all cities where price estimate time series have been recorded out of all locations where Uber service is available (grey). In some locations, especially in the US, we recorded price estimates for multiple different trips in the same location (see Tab.~\ref{tab:CityList} in section~\ref{sec:data} for details).
}
\label{fig:WorldMap}
\end{figure*}

Uber Technologies Inc.~is a major ride-hailing platform operator that matches drivers with customers requesting transportation in 795 metropolitan areas worldwide (as of July 2019 \cite{UberCities2019}, Fig.~\ref{fig:WorldMap}). The company operates a digital marketplace for transportation services where riders voice their demand for a specific trip and drivers offer to deliver the service. 
The dynamic pricing scheme employed by Uber includes both an adaptive trip fee, based on local traffic conditions and similar parameters, as well as an additional component to balance the spatio-temporal distribution of demand and supply in the operating area, commonly denoted as \emph{surge pricing} \cite{UberSurge2019, Garg2019}.

In this section, we detail the different components of the pricing mechanism and isolate how they impact the evolution of the recorded price estimate time series.

\newpage
\subsection{Uber's price model}

Uber's dynamic pricing model consists of four different fee categories\footnote{This information on Uber's price model is based on Ref. \cite{UberPriceEstimates2019}, accessed on 19/06/24, and may be subject to change.}:
\begin{itemize}
    \item \textbf{Pickup fee}: cost for customer pick-up from the requested origin point of the trip. 
    
    Pickup fees contain a \textit{base fare} (flat fee per pick-up) and may be subject to additional \textit{long-pickup fees} if drivers and customers are far away from one another, e.g. when requesting a ride from a remote location. Long-pickup fees apply if the \textit{pickup duration} exceeds a pre-defined threshold value (e.g. 10 minutes in many US cities), and is charged for certain Uber products in selected operating areas only. 
    
    Long-pickup fees are calculated based on the distance and time the driver has to invest to pick up the customer, and is bound from above by a maximum pickup fee. \textit{Price per minute} and \textit{price per mile} parameters determine the absolute amount of the fee.
    
    \begin{center}
        Pickup fee = base fare + long-pickup fees (optional)
    \end{center}
    
    \textit{Input parameters}: base fare, price per minute (optional), pickup duration (optional), price per mile (optional), pickup distance (optional)
    
    \item \textbf{Trip fees}: cost for the passenger transportation component of the ride. 
    
    Trip fees consist of a fixed \textit{booking fee} as well as variable \textit{per minute} and \textit{per mile prices}. The trip fee is lower bounded by the \textit{minimum fare} parameter.
    
    \begin{center}
        Trip fee = booking fee + distance fee + time fee
    \end{center}
    
    \textit{Input parameters}: booking fare, price per minute, trip duration, price per mile, trip distance, minimum fare
    
    \item \textbf{Surcharges}: optional add-on fees for the specific type of ride.
    
    Surcharges may reflect additional fees such as tolls, airport pickup fees, tip, \dots, and thus depend on the trip specifics.
    
    \begin{center}
        Surcharges = tolls + airport pickup fees + \dots 
    \end{center}
    
    \textit{Input parameters}: tolls (optional), airport pickup fees (optional), tip (optional), \dots
        
    \item \textbf{Surge fee}: supply- and demand-based cost increment.
    
    \textit{Surge fees} consider the spatio-temporal distribution of ride requests and available drivers. They reflect a price increment on trip and base fares, either in terms of a \textit{surge multiplier} or as an \textit{additive surge} component that is intended to rebalance local demand and supply \cite{Garg2019, UberNewSurge2018}.
    
    \textit{Input parameters}: Unknown. Uber's surge pricing algorithm is not public. 
\end{itemize}
The total fare for the ride is the sum of these four components
\begin{eqnarray}
\mathrm{total~fare} = &\;\underbrace{\vphantom{\left(\frac{x_x^x}{x_x^x}\right)}\mathrm{pickup~fee}\;+\;\mathrm{trip~fee}\;+\;\mathrm{surcharges}}\;&+\; \mathrm{surge~fee} \nonumber\\
 = &\;\vphantom{\left(\frac{x_x^x}{x_x^x}\right)}\mathrm{base~cost}\;&+\;\mathrm{surge~fee} \,,
\label{eq:uber_price_components}
\end{eqnarray}
which we denote as trip dependent base cost and supply-demand dependent surge fee.

\newpage
Depending on the city, the values of the input parameters for the different price components may vary. Moreover, Uber offers different products that differ by their level of service, and have different input parameters to the pricing model. In our analysis, we focus on the standard service and, for comparison, a corresponding premium service available in the region:
\begin{itemize}
    \item \textbf{Standard service}: \textit{UberX}, \textit{UberGO} (India) 
    \item \textbf{Premium service}: \textit{Black}, \textit{Berline} (France), \textit{Exec} (Great Britain), \textit{Lux}, \textit{Premier} (India) or  \textit{Select} (Egypt, Argentina)
\end{itemize}
More details can be found in the data collection section at the end of this Supplementary Material. For all data presented in this section \textit{UberX} and \textit{Black} are available and results are based on price estimates for these services. Table~\ref{tab:UberPriceParameters} gives an exemplary overview of several parameter values serving as inputs to the Uber price mechanism for four US cities.\\

\begin{table}[!h]
\begin{tabular}{|l|l|cc|cc|cc|cc|}
\hline
                                      &                            & \multicolumn{2}{c|}{\textbf{Washington D.C.}}          & \multicolumn{2}{c|}{\textbf{Los Angeles}}            & \multicolumn{2}{c|}{\textbf{San Francisco}} &
                                      \multicolumn{2}{c|}{\textbf{Houston}} \\ \hline
                                      &          & Black & UberX & Black & UberX & Black & UberX  & Black & UberX \\ \hline
\multirow{5}{*}{\textbf{Pickup fees}} & Base fare {[}USD{]}        & 7.75                      & 1.21                      & 8.75                      & 0.00                      & 8.75                      & 2.20     & 7.75 & 1.00                 \\
                                      & Long-pickup fee {[}USD{]}  & -                         & Variable                  & -                         & Variable                  & -                         & Variable & -                         & Variable                      \\
                                      & - Price per minute {[}USD{]} & -                         & 0.30                      & -                         & 0.28                      & -                         & 0.39         & - & 0.17             \\
                                      & - Price per mile {[}USD{]}   & -                         & 0.80                      & -                         & 0.80                      & -                         & 0.91             & - & 0.80         \\ \hline
\multirow{4}{*}{\textbf{Trip fees}}   & Booking fee {[}USD{]}      & 0.00                      & 2.00                      & 0.00                      & 2.30                      & 0.00                      & 2.20       & 0.00 & 2.90               \\
                                      & Minimum fare {[}USD{]}     & 15.75                     & 7.00                      & 15.75                     & 5.80                      & 15.75                     & 7.20  & 15.75 & 5.95                    \\
                                      & Price per minute {[}USD{]} & 0.82                      & 0.30                      & 0.71                      & 0.28                      & 1.08                      & 0.39                & 1.03 & 0.17      \\
                                      & Price per mile {[}USD{]}   & 2.20                      & 0.80                      & 2.92                      & 0.80                      & 2.73                      & 0.91 & 2.20 & 0.80                      \\ \hline
\textbf{Surcharges}                  & Airport pickup fee           & 4.00                     & 4.00                      & 5.00                     & 4.00                      & N/A                     & N/A                   & 0.00 & 2.75   \\ \hline
\textbf{Surge fee}                   & Supply/demand balance           & N/A                     & N/A                      & N/A                     & N/A                      & N/A                     & N/A                      & N/A & N/A \\ \hline
\end{tabular}
\caption{\textbf{Input parameter values determining Uber prices differ per product and locality.} Price parameters for Uber's standard (\textit{UberX}) and premium (\textit{Black}) ride-hailing services in four exemplary US cities \cite{UberPriceEstimates2019}. 
}
\label{tab:UberPriceParameters}
\end{table}


\clearpage

\subsection{Time series of Uber price estimates}

\begin{figure*}
\centering
\includegraphics[width=1.0\linewidth]{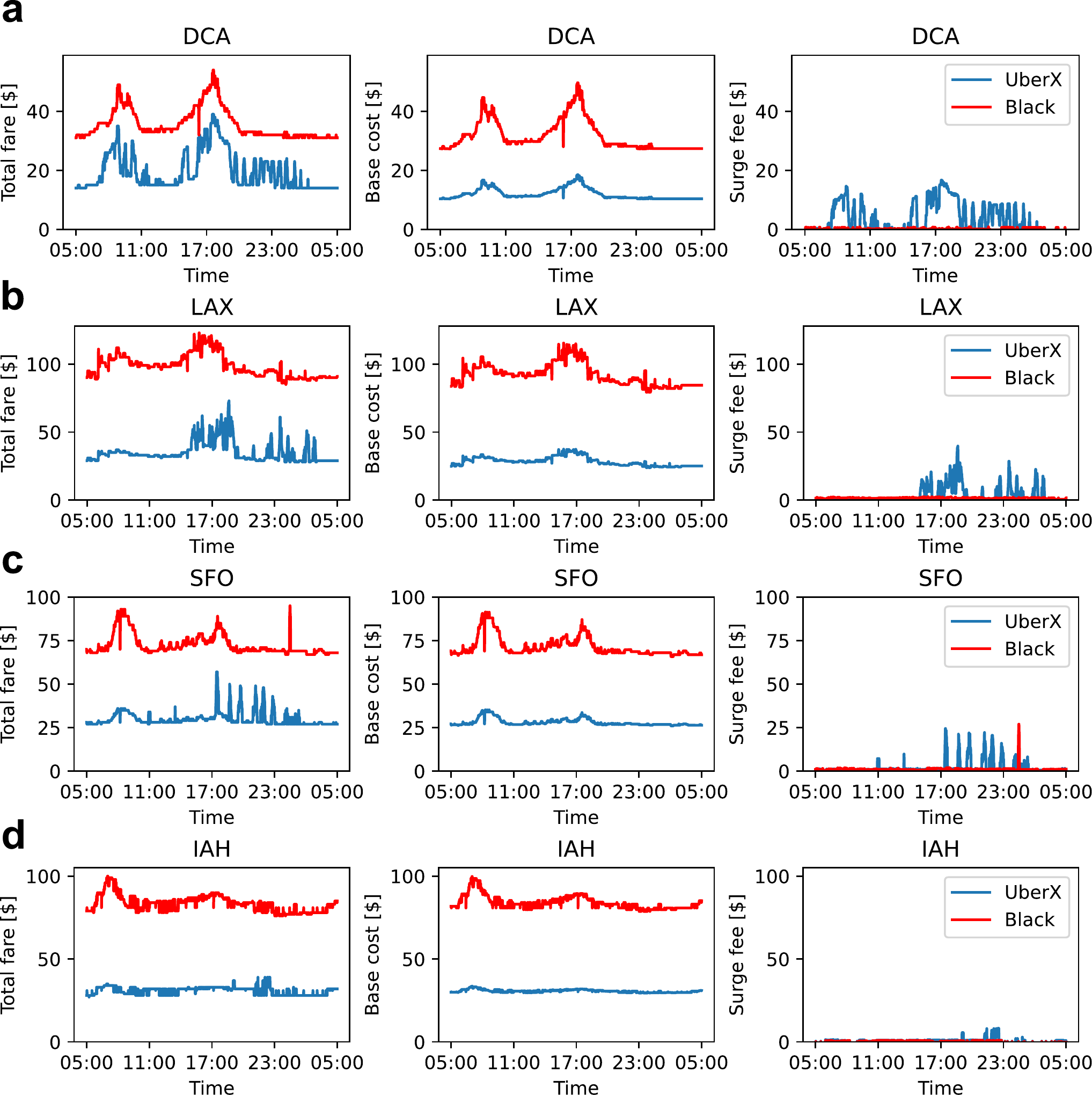}
\caption{\textbf{Uber \textit{Black} and \textit{UberX} trip fees evolve synchronously while the respective surge fees do not}. \textbf{a-d} Fare estimates for \textit{UberX} (blue) and Uber \textit{Black} (red) on 19/06/04 for trips originating from Reagan National Airport, Washington, D.C. (DCA), Los Angeles International Airport (LAX), San Francisco International Airport (SFO) and George Bush Intercontinental Airport, Houston (IAH). Columns show total fare, base cost and surge fee, respectively (from left to right).
}
\label{fig:PriceTimeSeries}
\end{figure*}

Figure~\ref{fig:PriceTimeSeries}a shows price estimate time series for four exemplary trips originating from airports in Washington, D.C. (DCA), San Francisco (SFO), Los Angeles (LAX) and Houston (IAH) for Uber \textit{Black} and \textit{UberX} over the time span of 24 hours. 
Trip fares change dynamically in all of the four cities and exhibit a slow and fast timescale of price volatility. While the slow dynamics modulates the price in timescales of several hours, the fast timescale adds price spikes in the order of ten minutes to half an hour. Here, we isolate the different contributions from pickup fees, trip fees, surge fee, and surcharges to the price evolution.

The price dynamics shown in Fig.~\ref{fig:PriceTimeSeries}a is driven by trip fees (see Fig.~\ref{fig:PriceTimeSeries}b) and surge fees (see Fig.~\ref{fig:PriceTimeSeries}c) which vary over time. We assume that pickup fees and surcharges are constant for all time series analyzed. Possible exceptions include tolls which apply during the day but not at night or are applicable only on alternative routes (e.g. for IAH). However, these surcharges are typically small and change over very long timescales ($\sim 12$ hour) and do not significantly alter the observed surge dynamics.

\begin{figure*}
    \centering
    \includegraphics[width=1.0\linewidth]{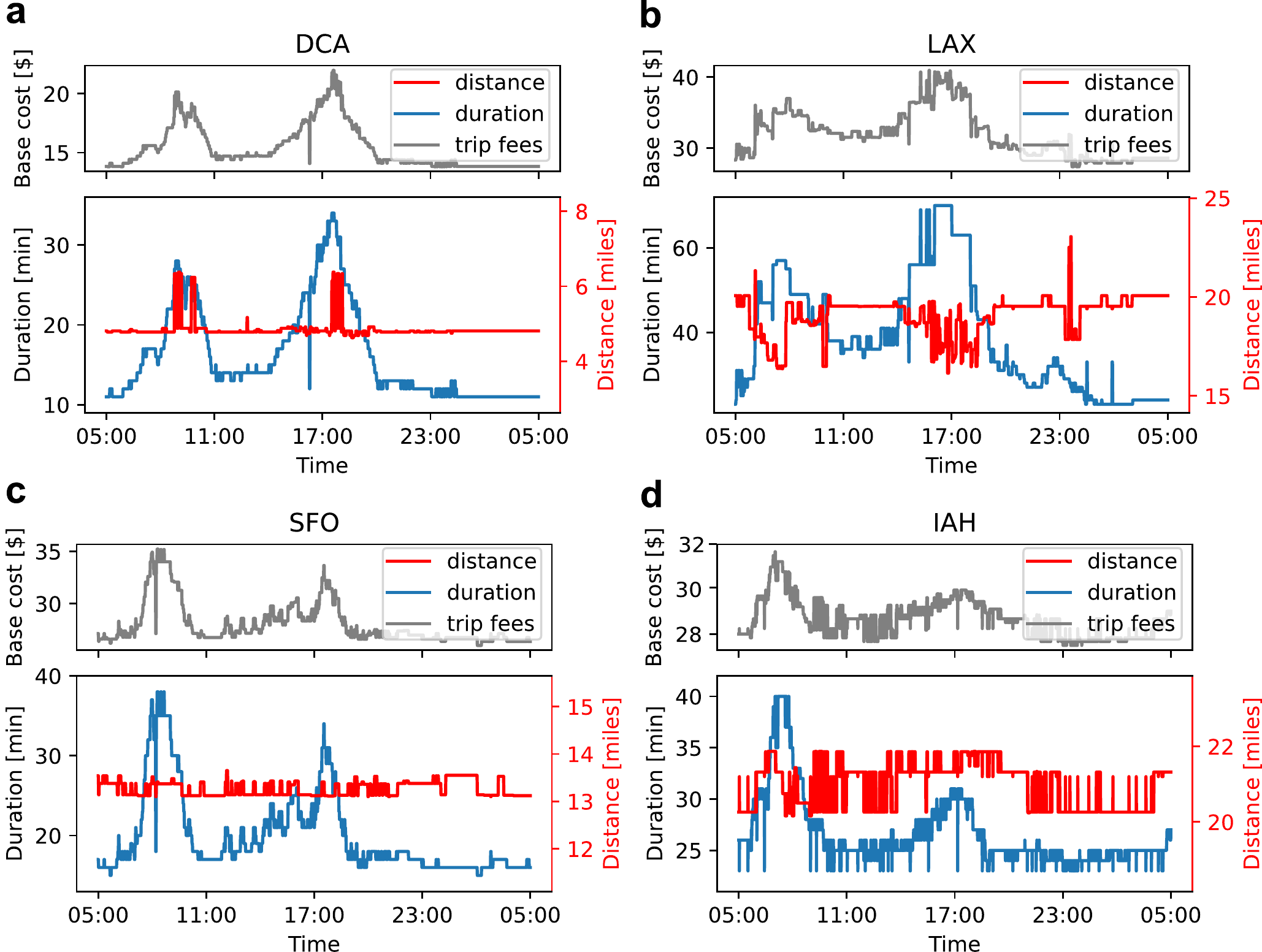}
    \caption{\textbf{Uber trip fees change proportional to city traffic flow conditions}. 
    \textit{UberX} base cost change intra-day at \textbf{a} DCA, \textbf{b} LAX, \textbf{c} SFO and \textbf{d} IAH (data from 19/06/04, compare Fig.~\ref{fig:PriceTimeSeries}) as trip fees adapt to latest estimates for trip duration (blue) and distance (red). Both estimates respond to changing traffic conditions, congestion and dynamic route choice. 
    During commuting hours rising trip duration estimates reflect congested streets, while changes in trip distance estimates correspond to alternative route choice advise as the fastest route option changes for given traffic conditions. Uber \textit{Black} trip fees behave qualitatively identically (compare Fig.~\ref{fig:PriceTimeSeries}) since they are based on the same duration and distance estimates.
    }
    \label{fig:TripDurationDistance}
\end{figure*}

Trip fees vary as a function of intra-day variation in local traffic conditions in each of the four cities. Two effects superimpose: On the one hand, trip duration estimates change as the streets from the different airports to the respective inner-city destinations become congested during commuting and business hours (see Fig.~\ref{fig:TripDurationDistance} for trip duration estimates). Hence, the time-dependent trip fee increases and decreases over the course of the day proportional to the street flow traffic conditions. On the other hand, route choice recommendations change as a function of the current traffic conditions. As traffic intensifies, alternative routes may become faster and thus more attractive to complete the trip. However, those trip duration-preferable routes might be longer compared to the shortest-distance path, implying higher distant-dependent trip fees (see Fig.~\ref{fig:TripDurationDistance}, note in particular DCA, where the trip distance increases only when the trip duration is large.). Together, both contributions define the time-dependent trip fee component of the total fare estimate. 
Trip duration and distance estimates are identical for both Uber \textit{Black} and \textit{UberX}. Hence, their trip fees evolve synchronously, though with different per-minute and per-mile charges (see Fig.~\ref{fig:PriceTimeSeries}b and Tab.~\ref{tab:UberPriceParameters}). \\

In Fig.~\ref{fig:PriceTimeSeries} ({right column}) we subtract pickup fees, trip fees and (estimated constant) surcharges from the price estimate time series to isolate the time evolution of the surge fee component (see section~\ref{sec:data} for more details). For Uber \textit{Black}, there is almost no surge activity at DCA, LAX or IAH over the illustrated time span of 24 hours. Similarly, SFO does not exhibit Uber \textit{Black} price surges for most of the day, but only a single distinctive surge at 00:30. Hence, we assume that the marketplace for Uber \textit{Black} is in equilibrium at all airports, and there are no stark spatio-temporal supply and demand imbalances. 

For \textit{UberX}, the surge fee dynamics differs substantially. 
At DCA, we observe substantial surge activity for most of the day. Only for four hours 
at night time the marketplace does not give rise to surge fees (consistent with the typically low demand for transportation during this time window. 
Longer price surges exist for approximately 1.5 hours during morning (07:30 to 09:00) and 3.5 hours during evening commuting hours (16:00 to 19:30). It is plausible that these price surges supplement the typical high commuting demand during rush hour. Additionally, the dynamic pricing mechanism induces a series of short, characteristic, almost periodic surges with approximately universal peak value of ten USD and duration of 20 minutes that appear between 19:00 and 03:00. 
Similar surge dynamics with short, repeated surges are clearly visible at SFO between 18:00 and 23:00, possibly between 23:30 and 01:30 at LAX, but not at IAH. 

Given the long-term presence of Uber in any of these cities (launch dates: San Francisco in May 2010, Los Angeles in March 2010, Washington, D.C. in December 2011 and Houston in February 2014 \cite{UberLaunchDates2014}) and the difficulty for the drivers to operate economically sustainable in a market with too many competitors for given demand, it is reasonable to assume that the different marketplaces are equilibrated with respect to long term fluctuations. In fact, the Uber \textit{Black} surge dynamics supports this hypothesis and even suggest equilibration down to small, intra-day timescales, with drivers having adjusted to when and where to work efficiently. In contrast, the repeated, sudden surges of \textit{UberX} indicate out-of-equilibrium dynamics. Counter-intuitively, the dynamic pricing mechanism seems to prevent the system from settling into an equilibrium.\\

\clearpage

\subsection{Demand model}
\label{sec:demand}
Possible origins for the surge dynamics are either local fluctuations in the demand (i.e. many customers requesting a ride at the same time) or changes in the supply (i.e. few available drivers). While driver induced price surges are confirmed to occur at DCA \cite{ABC7News2019a}, we do not know which, if any, of the observed peaks correspond to these artificial prices surges. To identify parts of the surge fee time series that are likely caused by supply-side action, we develop a demand model for Reagan National Airport (DCA) and assess to what extent it explains the price dynamics observed in Fig.~\ref{fig:PriceTimeSeries} ({top right}). In particular, we estimate the demand based on historic airport taxi departures 
and recorded aircraft deplanements. 
We find that a large part of the surge dynamics cannot be satisfactorily explained through the demand model, suggesting that in particular the short, repeated surges in the evening are primarily induced by changes of the supply.\\ 

\paragraph*{Historic taxi trip records:} Uber's ride-hailing service operates in the same niche as traditional taxi services. According to the Department For-Hire Vehicles, four of Washington D.C.'s large ride-hailing companies (HopSkipDrive, Lyft, Uber and Via) generated 45\% of the combined taxicab and ride-hailing tax revenues in 2016, 59\% in 2017 and 67\% in 2018 \cite{DCTaxiMarketShare2019}. As the absolute amount of those tax revenues remained approximately constant over the three years (2016: 6.75 million USD, 2017: 7.65 million USD, 2018: 7.19 million USD), ride-hailing companies kept on gaining market share from traditional taxi providers, underpinning that digital ride-hailing and traditional taxi services are substitutes. This suggests that taxis and ride-hailing services serve a similar demand and recorded taxi trips departing from DCA likely reflect the typical intra-day demand also for ride-hailing services.

\begin{figure*}[!h]
    \includegraphics[width=0.9\linewidth]{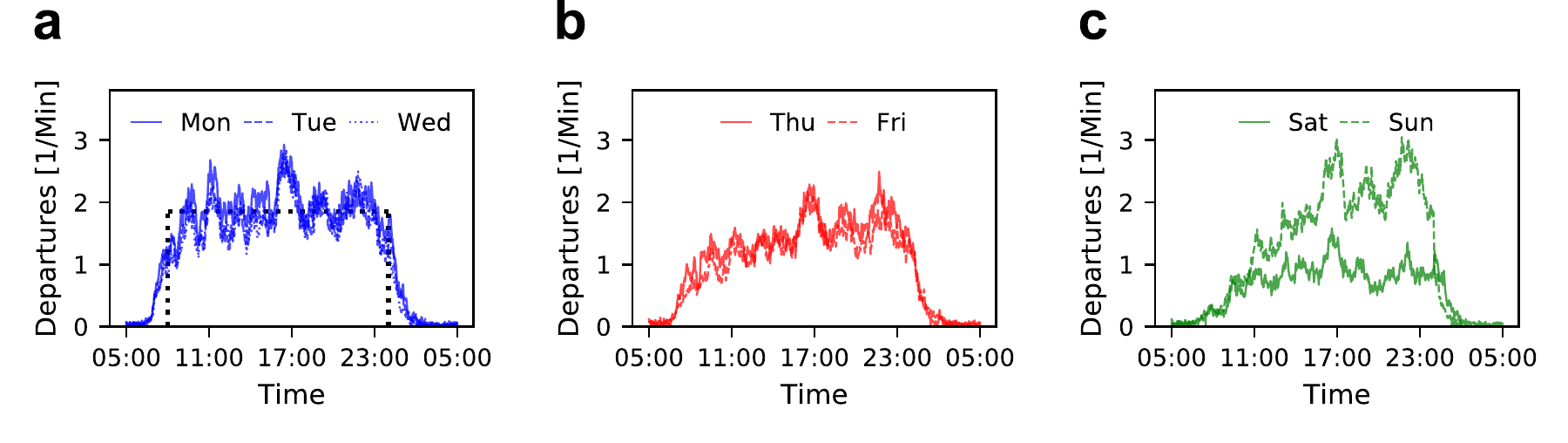}
     \caption{\textbf{Average demand for taxi trips from DCA is approximately constant during the day.} 
    Average number of taxi departures per minute from DCA recorded between 17/01/02-17/08/27. \textbf{a} From Monday to Wednesday, the average demand during business hours is approximately constant at $1.85$ taxi rides per minute (dashed black line). \textbf{b,c} Thursday through Sunday show a slightly higher demand in the evening. In general, larger fluctuations typically occur over one to two hours. Details on data processing can be found in section~\ref{sec:data}. 
    }
    \label{fig:DCATaxiDemand}
\end{figure*}

    Figure~\ref{fig:DCATaxiDemand} shows the average intra-day taxi demand evolution for trips originating from DCA airport, calculated for trips from January to August in 2017. 
    and, interestingly, do not reflect the rush hour traffic observed in Fig.~\ref{fig:TripDurationDistance}. In particular on Monday, Tuesday and Wednesday the average demand is approximately constant over the whole day with $1.85$ rides per minute (standard deviation $0.37$ rides per minute). Higher demand in the evening at other days may strengthen price surges during that time but no specific correspondence between taxi demand and the dynamics of the surge fee is visible.
    
    Overall, there seems to be no direct influence of the general demand evolution on the surge fees, consistent with our assumption of an equilibrated market on these timescales. However, the information about events of a specific day is lost by considering average historic demand data. Events of an individual day may reveal more direct correlations between the demand and the surge fee. In the following, we therefore attempt to match the surge fee with the demand (aircraft arrivals) of the respective day.

\newpage
\paragraph*{Aircraft passenger arrivals:} The taxi trips analyzed above only provide the average demand across many days. However, the surge dynamics of a specific day depend on the demand due to the specific arrival pattern of airplanes on that day. 
    
\begin{figure*}
    \centering
    \includegraphics[width=1.0\linewidth]{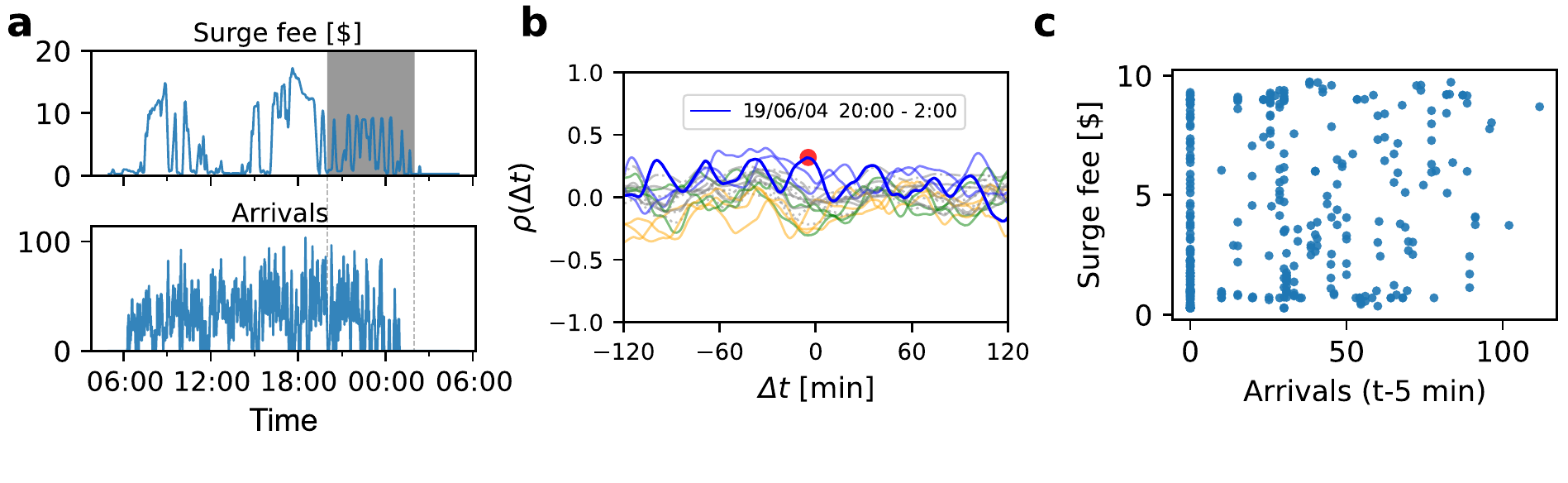}
    \caption{\textbf{Capacity of arriving aircraft is only weakly correlated with \textit{UberX} surge fees.}
    \textbf{a} \textit{UberX} surge fee (top) and capacity of arriving airplanes (bottom) at DCA on Tuesday, 19/06/04. 
    \textbf{b} Normalized cross-correlation $\rho(\Delta t)$ between fixed time windows of the surge fee and the capacity of arriving aicraft. The dark blue line corresponds to the shaded window from 20:00 to 02:00 in panel \textbf{a}, light blue lines to the same time windows on other days. Yellow lines show the correlation function for the 6-hour time window from 08:00 to 14:00, green lines for the 6-hour time windows from 14:00 to 20:00. Grey lines show 12- and 18-hour windows (dotted line: 08:00 to 20:00.; dotted-dashed line: 14:00 to 02:00; solid: 08:00 to 02:00).
    The maximum correlation $\rho(\Delta t^*) = 0.31$ for the highlighted time window is obtained at $\Delta t^* = -5\,\mathrm{minutes}$ (red point).
    \textbf{c} Surge fee and aircraft capacity at the time lag $\Delta t^*$ of maximum correlation (red point in panel \textbf{b}). There is no apparent relationship between the arrivals and the surge fee, suggesting a large supply-side influence on the observed surge dynamics.
       }
    \label{fig:AircraftCapacity}
    \end{figure*}

    Passengers arrive in DCA with each aircraft deplanement. Some of these passenger may continue on a connecting flight while the remainder likely travels in the direction of Washington, D.C., using for example ride-hailing services like Uber or taxi cabs. Taking a homogeneous load factor and a constant share of Uber customers across all arriving planes, we expect the demand for ride-hailing services to be proportional to the seat capacity of the arriving aircraft. 
    Figure~\ref{fig:AircraftCapacity}a shows the time series of the \textit{UberX} surge fee together with the corresponding capacity of arriving aircraft (compare DCA, Fig.~\ref{fig:PriceTimeSeries}). 
    
    The capacity of arriving aircraft follows a similar pattern to the average taxi demand above. At nighttime almost no deplanements are observed, in agreement with the DCA Nighttime Noise Rule.  
    Over the course of the day, DCA handles a near constant stream of aircraft landings, in line with the approximately constant demand for taxi service. Deplanements occur with a median interval of 2 minutes with aircraft equipped with with 50 to 213 seats. On average, 32 passengers arrive in DCA per minute, a fraction of which will use ride-hailing serivces.
    
    To estimate the influence of the arrivals on the surge dynamics, we compute the normalized cross-correlation function
    \begin{equation}
        \rho(\Delta t) = \frac{E\left[ (S(t)-\left<S(t)\right>) \, (A(t+\Delta t)-\left<A(t+\Delta t)\right> \right]}{\sigma_S(t) \sigma_A(t + \Delta t)} \label{eq:correlation}
    \end{equation}

    between fixed windows of the \textit{UberX} surge fee $S(t)$ and the capacity of arriving airplanes $A(t)$ with time lag $\Delta t$. 
    $E[\cdot]$ denotes the expectation value of the argument and $\sigma_S$ and $\sigma_A$ denote the standard deviation of the respective time series in the corresponding time window. For more details on the data preparation and processing see section~\ref{sec:data}.
    
    To avoid high correlations simply due to the significant night-day differences in both airplane arrivals and ride-hailing demand, we calculate the cross-correlation $\rho(\Delta t)$ only for time windows between 08:00 and 02:00 on the next morning 
    \begin{itemize}
        \item \textit{6-hour windows}: We analyze morning time windows between 08:00 to 14:00 (Fig.~\ref{fig:AircraftCapacity}b, yellow lines), afternoon time windows from 14:00 to 20:00 (green lines), and evening time windows from 20:00 to 02:00 (blue lines).
        \item \textit{12-hour windows}: We contrast the 6-hour time widows with half-day daytime windows between 08:00 to 20:00 (Fig.~\ref{fig:AircraftCapacity}b dotted gray lines) and late-afternoon to nighttime windows between 14:00 to 02:00 (dotted-dashed gray lines)
        \item \textit{18-hour windows}: Furthermore, we define an 18-hour time window between 08:00 to 02:00 extending over the full time span of surge activity (Fig.~\ref{fig:AircraftCapacity}b solid gray lines). 
    \end{itemize}
    In particular, we focus on the correlation of the 6-hour evening time window from 20:00 to 02:00 (see Fig.~\ref{fig:AircraftCapacity}, shaded in panel a, dark blue line in panel b), where the trip fee is constant and we expect no influence of traffic conditions on the surge fee. In this time window, we observe repeated peaks of the surge fee and supply-induced price surges are known to occur  \cite{ABC7News2019a}.
    
    \begin{figure*}
    \centering
    \includegraphics[width=1.0\linewidth]{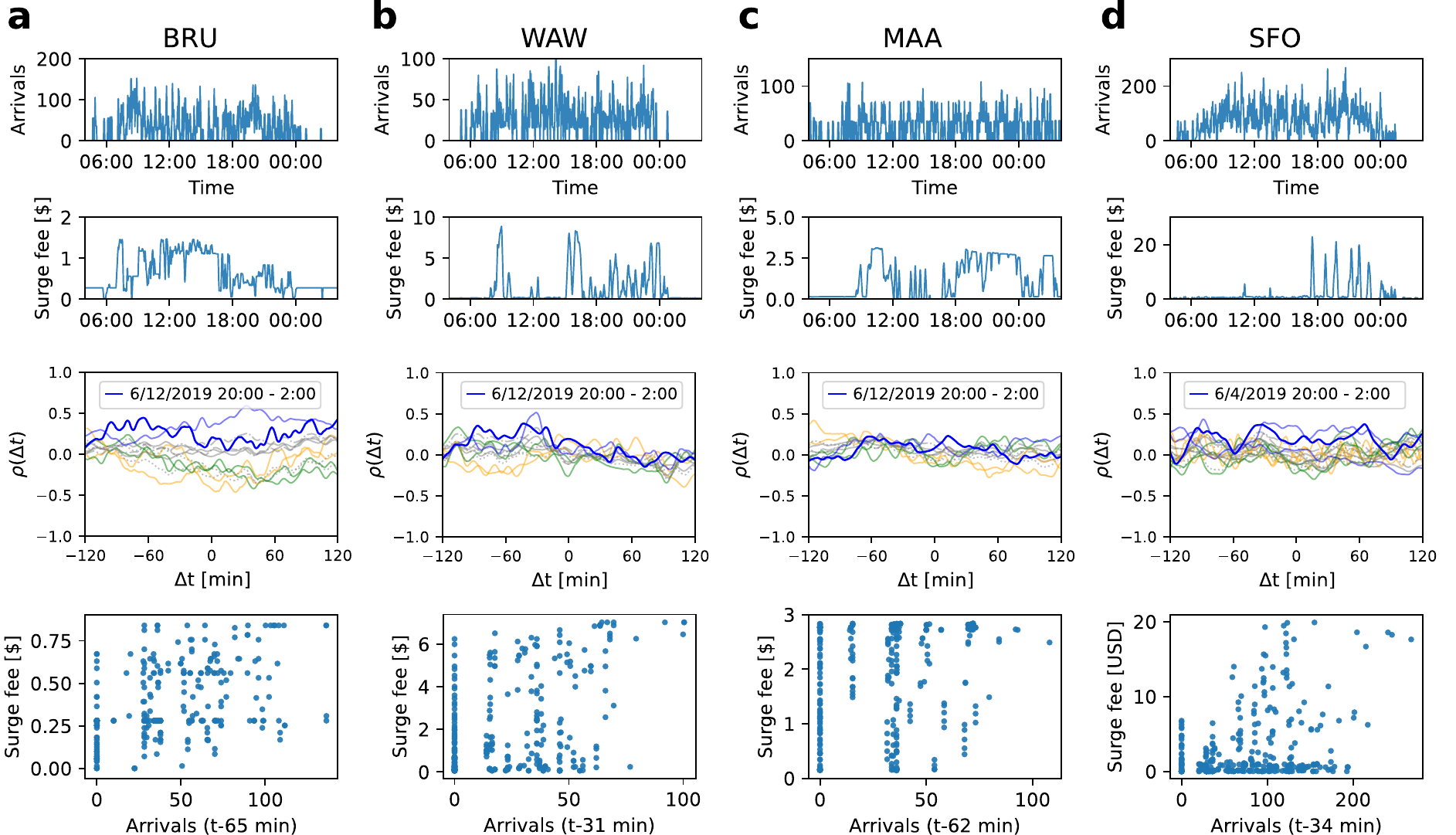}
    \caption{\textbf{Weak correlations between airport arrivals and surge fee across airports.}
    The time series of aircraft arrivals and \textit{UberX} surge fees for \textbf{a} Brussels, \textbf{b} Warsaw, \textbf{c} Chennai and \textbf{d} San Francisco (top two rows). The third row shows the cross-correlation (Pearson correlation coefficient) $\rho$ between the surge fee time series and the arrivals at different delays $\Delta t$ using the same approach as for DCA (compare Fig.~\ref{fig:AircraftCapacity}). The bottom row shows the surge fee vs. the number of arrivals plot at the delay $\Delta t^*$ of the largest cross-correlation. As in Fig.~\ref{fig:AircraftCapacity}c, the correlation between arrivals and surge fee is weak most of the time.
    }
    \label{fig:correlations_other}
    \end{figure*}
    
    The first maximum of the cross-correlation function is obtained for a time lag of $\Delta t^* = -5\,\mathrm{minutes}$ at a value of $\rho(\Delta t^*) = 0.31$, indicating only a weak direct influence of aircraft arrivals on surge dynamics. Several similar maxima of the correlation correspond to the 30 to 40 minute periodicity of the surges in this time window. In contrast, this periodicity is not reflected in the aircraft arrivals. 
    On other days, the correlation reaches up to $\rho(\Delta t^*) \approx 0.39$ in the same time window. 
    The correlation at $\Delta t^* = -5\,\mathrm{minutes}$ may be explained by passengers hailing a ride or checking prices very quickly after landing, possibly while taxiing to the gate or immediately after exiting the plane.\footnote{According to the US Federal Communications Commission, "[t]he use of cellular telephones while [..] aircraft [are] airborne is prohibited".} However, a scatter plot at the time lag of the maximum correlation reveals no clear relationship between the surge fee and aircraft arrivals (see Fig~\ref{fig:AircraftCapacity}c). Results for other time windows are similarly ambiguous, showing only weak correlation between the two time series and offering no clear explanation of the surge dynamics.     
   Therefore, the variability of aircraft arrivals only partially explains the surge dynamics. Other demand sources apart from aircraft arrivals seem unlikely at DCA. While we assumed a constant airplane load (fraction of occupied seats), we expect this factor to vary at most on the timescale of several hours with a typical daily pattern rather than on short timescales like the surge dynamics. Overall, we conclude that changes in ride-hailing demand do not offer a sufficient explanation for the observed surge dynamics. There must be additional, unobserved changes of supply affecting the surge fee.\\
    
\clearpage
\paragraph*{Conclusion.} The taxi and airport data analyzed above indicate an approximately constant demand during the day with no direct correspondence between demand changes and surge dynamics. Taxi data show a generally higher average demand during the evening, suggesting that price surges may be more likely during this time. Aircraft arrivals show weak correlation with the surge fee, offering only a partial explanation of the surge dynamics.

Unrelated to direct demand fluctuations, we consider other sources for demand and supply changes. Specifically, longer service times during rush hour (compare Fig.~\ref{fig:TripDurationDistance}) not only lead to higher trip fees but also to drivers spending more time serving each request. With a constant demand (compare taxi departures in Fig.~\ref{fig:DCATaxiDemand}), this means more drivers are busy and fewer drivers are available at a given time during this period. This general mechanism for supply shortages as a secondary effect of prolonged heavy traffic is consistent with the observed long duration price surges during main commuting times.

In summary, we only have a satisfactory explanation for the long rush hour peaks of the surge fee at DCA. The short, repeated, almost periodic surges with universal peak value and duration (in particular in the evening between 19:00 and 03:00, see Fig.~\ref{fig:PriceTimeSeries}, {top right}) cannot be explained by demand-side fluctuations alone or general demand and traffic conditions. We therefore consider these peaks as supply-side induced (or at least having a strong supply-side influence), consistent with the reported driver-induced price surges at DCA in the evening \cite{ABC7News2019a}. In the following, we attempt to find price time series with similar out-of-equilibrium surge dynamics to identify locations that may also be affected by artificially induced price surges.

\clearpage

\section{Statistical properties of surge pricing}

\begin{figure*}
\centering
    \includegraphics[width=1.0\linewidth]{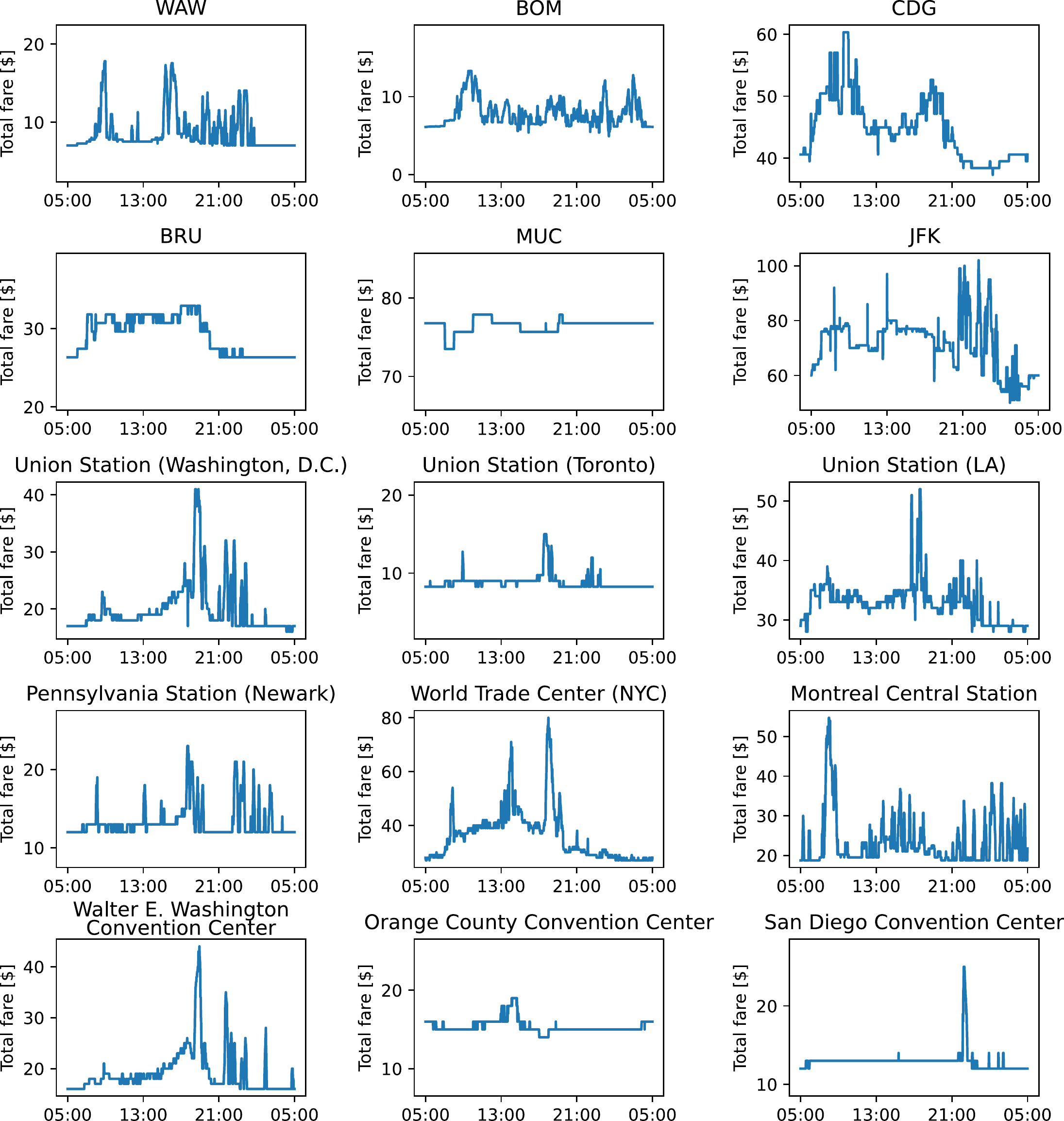}    
\caption{\textbf{The evolution of total fares for standard \textit{UberX} services differs qualitatively across locations.} Sample of total fare dynamics over a time span of 24 hours for rides originating from airports ({top two rows}), train stations ({third and fourth row}) and convention centers ({bottom row}) in different cities illustrating the range of possible price dynamics. Trip characteristics can be found in the data section and in Tab.~\ref{tab:CityList}. All currencies were converted to USD using the exchange rates for the respective day provided by the European Central Bank. 
}
\label{fig:Overview}
\end{figure*}

\begin{figure*}
\centering
    \includegraphics[width=1.0\linewidth]{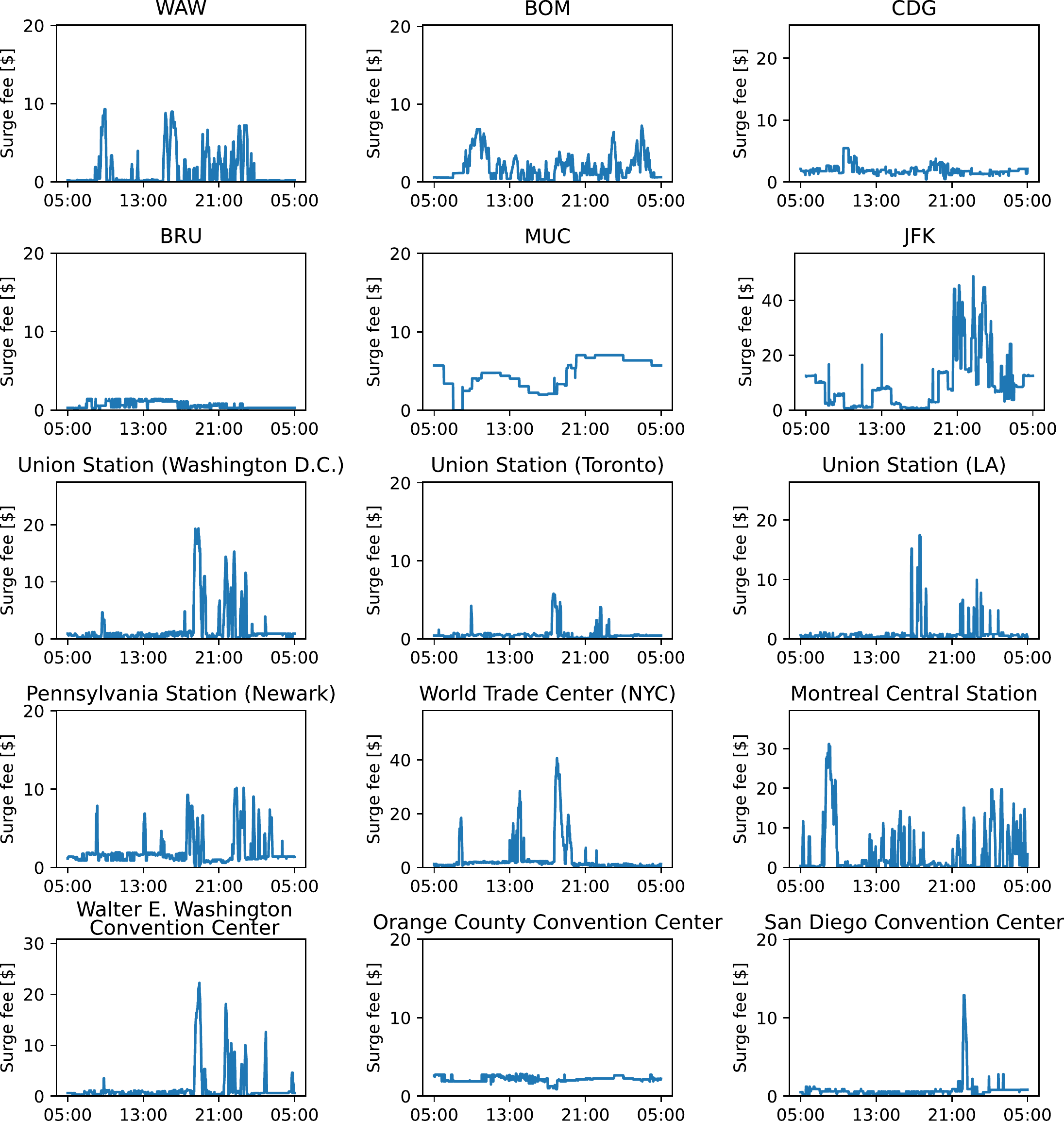}    
\caption{\textbf{The evolution of surge fees for standard \textit{UberX} services differs qualitatively across locations.} Estimated surge fee for the trip samples shown in Fig.~\ref{fig:Overview}. The surge fee shows a similarly broad range of dynamics as the total fare. All currencies were converted to USD using the exchange rates for the respective day provided by the European Central Bank.
}
\label{fig:Overview_surges}
\end{figure*}

Price surges reflect the imbalance of supply and demand at the origin location at the time of a given request. 
In the previous section, we have identified patterns of the surge dynamics in DCA that are likely related to artificial supply manipulation (compare DCA, Fig.~\ref{fig:PriceTimeSeries}) \cite{ABC7News2019a}. In the examples presented before, however, we also observe locations without surge activity (compare IAH, Fig.~\ref{fig:PriceTimeSeries}). Different locations around the globe exhibit different price dynamics, some without any price surges, some with single, distinct surge peaks and others with large, frequent surges similar to those observed in DCA (Fig.~\ref{fig:Overview} and \ref{fig:Overview_surges}). 

In this section, we attempt to characterize the out-of-equilibrium surge dynamics by quantifying properties of the price surges and the price time series at all observed locations (Fig.~\ref{fig:WorldMap}) and aim to identify cities with artificial price surges by comparing their surge dynamics to those observed at DCA.

We approach the problem of identifying price surges and quantifying the surge dynamics of locations from two different directions. First, 
we attempt to identify individual surge peaks directly. Second, we consider the distribution of the price changes across the whole price time series and identify two distinct contributions to the price dynamics, isolating changes corresponding to the surge fee component of the price changes in cities exhibiting price surges. While we find no clear distinction between locations with and without price surges or with and without potentially artificially induced price surges, we find several locations with similar or more volatile surge dynamics than DCA.

\begin{figure*}
\centering
    \includegraphics[width=1.0\linewidth]{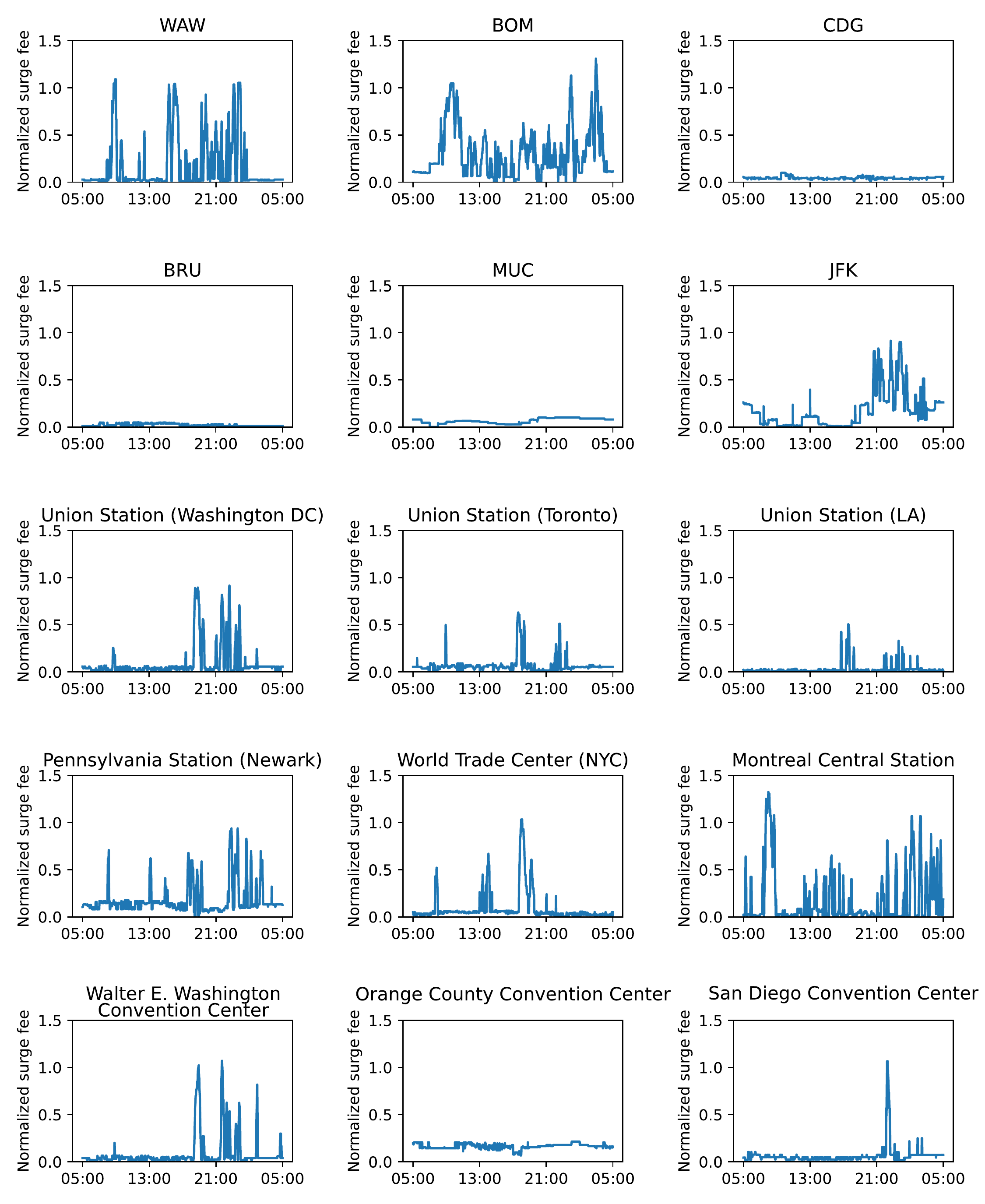}    
\caption{\textbf{The normalized surge fees for standard \textit{UberX} services enables comparison of the surge dynamics across locations.} Estimated normalized surge fee (surge fee divided by base cost) for the trip samples shown in Fig.~\ref{fig:Overview}. The normalized surge fee effectively represents a surge factor. Using this representation, the dynamics at different locations become comparable independent of the absolute cost of the trip (e.g. trip distance or local currency). In particular, differences between locations with and without price surges become clearer.
}
\label{fig:Overview_surge_factors}
\end{figure*}

\clearpage
\subsection{Statistics of individual surge peaks}
In order to identify individual surge peaks, we consider the surge fee component $S_\mathrm{X}(t)$ of the total \textit{UberX} fare, compare Fig.~\ref{fig:Overview_surges}. To compare different locations with varying total fare, we normalize the surge fee by the base cost at the current time, resulting in a surge factor $s_\mathrm{X}(t)$ describing the relative increase of the price due to price surges (compare Fig.~\ref{fig:Overview_surge_factors}). In order to filter surges due to overall demand increase, we subtract the corresponding premium service surge factor $s_\mathrm{prem}$ (e.g. of Uber \textit{Black}), resulting in a normalized surge factor 
\begin{equation}
    \tilde{s}_\mathrm{X} = \frac{ \textrm{surge~fee}_\textit{UberX}(t) }{\textrm{base~cost}_\textit{UberX}(t) } - \frac{ \textrm{surge~fee}_\textit{Black}(t) }{ \textrm{base~cost}_\textit{Black}(t) }\,. \label{eq:normalized_surge_factor_peak_identification}
\end{equation}
We use the premium products as a reference value since they tend to exhibit price surges only on rare occasions, presumably at times of generally high demand.

\begin{figure*}[!h]
\centering
\includegraphics[width=0.6\textwidth]{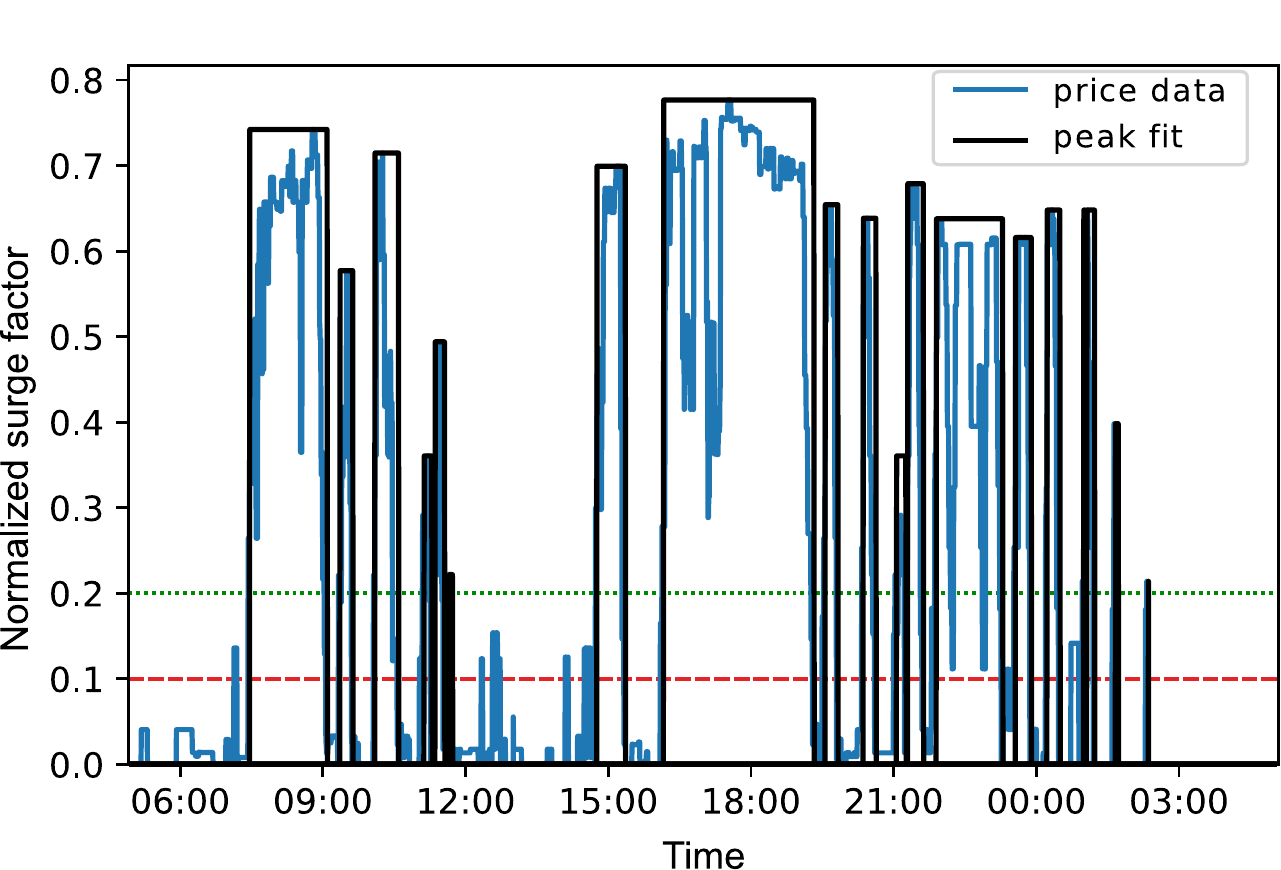}
\caption{\textbf{Surge peak detection.} Visualization of the surge peak identification for the normalized surge factor $\tilde{s}_\mathrm{X}$, Eq.~(\ref{eq:normalized_surge_factor_peak_identification}, in DCA. Peaks are defined to start when $\tilde{s}_\mathrm{X} > 0.2$ (dotted green line) and end when $\tilde{s}_\mathrm{X} < 0.1$ (dashed red line). Their amplitude is taken as the maximum value of $\tilde{s}_\mathrm{X}$ during the peak (solid black rectangles).
}
\label{fig:surge_box_fit}
\end{figure*}

Using this normalized surge factor, we define a surge by values $\tilde{s}_\mathrm{X} > 0.2$, that means a surge starts when $\tilde{s}_\mathrm{X}$ increases above $0.2$. To avoid peaks repeatedly ending and beginning due to small fluctuations, we define the end of a surge as the first time $\tilde{s}_\mathrm{X}$ decreases again to below $0.1$. Using different (reasonable) threshold values gives qualitatively similar results. These peaks are well described by boxes with start and end times defined by the peak identification conditions above and height given by the maximum value of the normalized surge factor during the peak. An example is illustrated in Fig.~\ref{fig:surge_box_fit}.

\begin{figure*}
\centering
\includegraphics[width=0.99\textwidth]{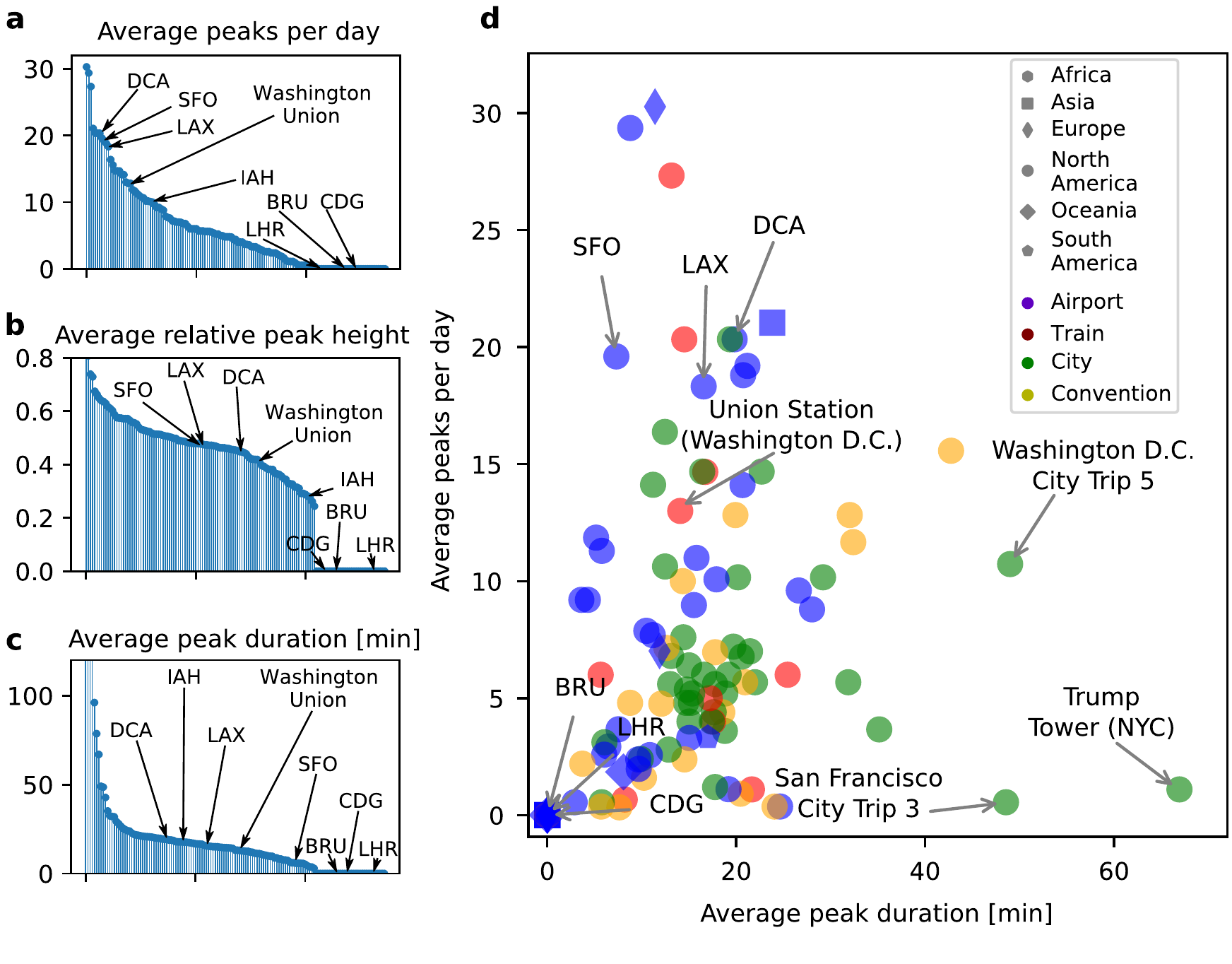}
\caption{\textbf{Single peak statistics do not separate locations with qualitatively different surge dynamics.}
None of the average peak characteristics (\textbf{a} number of peaks, \textbf{b} height, \textbf{c} duration] clearly classifies the cities with price surges into different categories, all measures gradually change from high to low values. The measures only distinguish between cities with peaks from those where no peaks where identified (e.g. panel \textbf{b} due to the threshold $\tilde{s}_\mathrm{X} = 0.2$ used for peak identification).
\textbf{d} Combining the average peak count per day and the average peak duration also reveals no clear pattern (results are qualitatively the same for other combinations of observables). Some locations with extremely long lasting surges (compare panel \textbf{c}) with an average peak duration upwards of 80 minutes are not visible in \textbf{d} (Orange County Convention Center, San Francisco City Trip 5, San Francisco City Trip 9, Minneapolis Convention Center). The long lasting price surges suggest a general supply-demand imbalance in these locations rather than artificially induced price surges, similar to the sustained rush-hour surge peaks in DCA (compare Fig.~\ref{fig:PriceTimeSeries}). No dependence on the trip type (airport, station, city or convention) is visible.
}
\label{fig:individual_surge_scatter}
\end{figure*}

We use the identified surges to compute aggregate characteristics of the surge dynamics for each location (see data section and Tab.~\ref{tab:CityList} for details). In particular we consider the mean duration and height of the peaks as well as the average number of peaks. Figure~\ref{fig:individual_surge_scatter}a-c shows the resulting statistics for each city by rank. The distinction between locations with and without peaks is clear only in the average peak height (Fig.~\ref{fig:individual_surge_scatter}b) due to the non-zero threshold used for peak identification. When peaks were identified at a location, neither the average duration, height nor number of peaks clearly separate locations with few or many or with strong or weak price surges. All measures change steadily across the different locations. Also the combination of these features (see Fig.~\ref{fig:individual_surge_scatter}d for an example) does not reveal distinct groups of locations.

\clearpage
\subsection{Statistics of surge time series}
In order to better characterize at which locations repeated price surges occur, we consider statistics of the entire price time series instead of individual price surge peaks. In particular, we consider per minute \emph{changes} in the normalized price to identify the typical timescales of the price dynamics.

For the analysis described in this section, we use a smoothed time series of normalized prices, i.e. the total fare divided by base cost, measuring the price in fractions of the time dependent base cost with equally spaced data points every minute (see also data processing section \ref{sec:data_processing:surge_time_series}). For these normalized time series, we compute the per minute changes $\Delta p$ between consecutive time points,
\begin{eqnarray}
    \Delta p(t) &=& \frac{ \textrm{total~fare}(t) }{ \textrm{base~cost}(t) } - \frac{ \textrm{total~fare}(t-1) }{ \textrm{base~cost}(t-1) } \label{eq:supp_normalized_price} \\
        &=& \frac{ \textrm{base~cost}(t) + \textrm{surge~fee}(t) }{ \textrm{base~cost}(t) } - \frac{ \textrm{base~cost}(t-1) + \textrm{surge~fee}(t-1) }{ \textrm{base~cost}(t-1) } \,.\nonumber \\
    &=& \frac{ \textrm{surge~fee}(t) }{ \textrm{base~cost}(t) } - \frac{ \textrm{surge~fee}(t-1) }{ \textrm{base~cost}(t-1) } \,.\nonumber
\end{eqnarray}
The normalized price can be interpreted as an effective surge factor (compare previous section). Although Uber began to transition to an additive surge computation \cite{UberNewSurge2018,Garg2019}, this normalization is necessary to compare price estimates across different locations regardless of base cost, currency or sampling frequency. 
Note that even in locations without surge activity, the estimated surge fee is not exactly zero, likely due to rounding of the price estimates. 
Taking into account these small fluctuations, we expect there to be three major contributions to the relative price changes: (i) Minutes without any change, e.g. during night where traffic and demand conditions do not change at all. (ii) Small changes due the rounding of the price estimate and slow changes of the base cost, mostly driven by changes of the trip fare as traffic conditions change over multiple hours. (iii) Fast changes of the surge fee component, increasing the price by up to 80\% in a matter of minutes. 

In order to quantify the contribution of these three parts, we fit an extended Gaussian mixture model to the data, consisting of one Dirac-delta distribution and two Gaussian distributions modelling the three contributions described above. We take the mean of both Gaussians to be zero (no price change on average), such that
\begin{equation}
    \mathrm{Prob}\left(\Delta p\right) = w_0 \, \delta(\Delta p) + w_\mathrm{base} \, \frac{1}{\sqrt{2 \pi \sigma_\mathrm{base}^2}} \, e^{-\frac{\Delta p^2}{2\,\sigma_\mathrm{base}^2}} + w_\mathrm{surge} \, \frac{1}{\sqrt{2 \pi \sigma_\mathrm{surge}^2}} \, e^{-\frac{\Delta p^2}{2\,\sigma_\mathrm{surge}^2}} \label{eq:gaussian_mixture}
\end{equation}
where we define the second Gaussian corresponding to the surge fee component to be broader, $\sigma_\mathrm{surge} > \sigma_\mathrm{base}$. We take all datapoints with $\Delta p^2 < 10^{-7}$ to belong to the Dirac-delta distribution indicating no price change and fit the two Gaussian distributions to the remaining data to determine the weights and standard deviations. A broad distribution corresponds to a fast changing behaviour (surge fee), while a narrow distribution describes a slowly changing price (base cost). For better visibility we only show the reduced data without the data points with no price change in the following, corresponding to the two Gaussian distributions without the Dirac delta distribution.\\

\begin{figure*}[!h]
\centering
\includegraphics[width=0.9\textwidth]{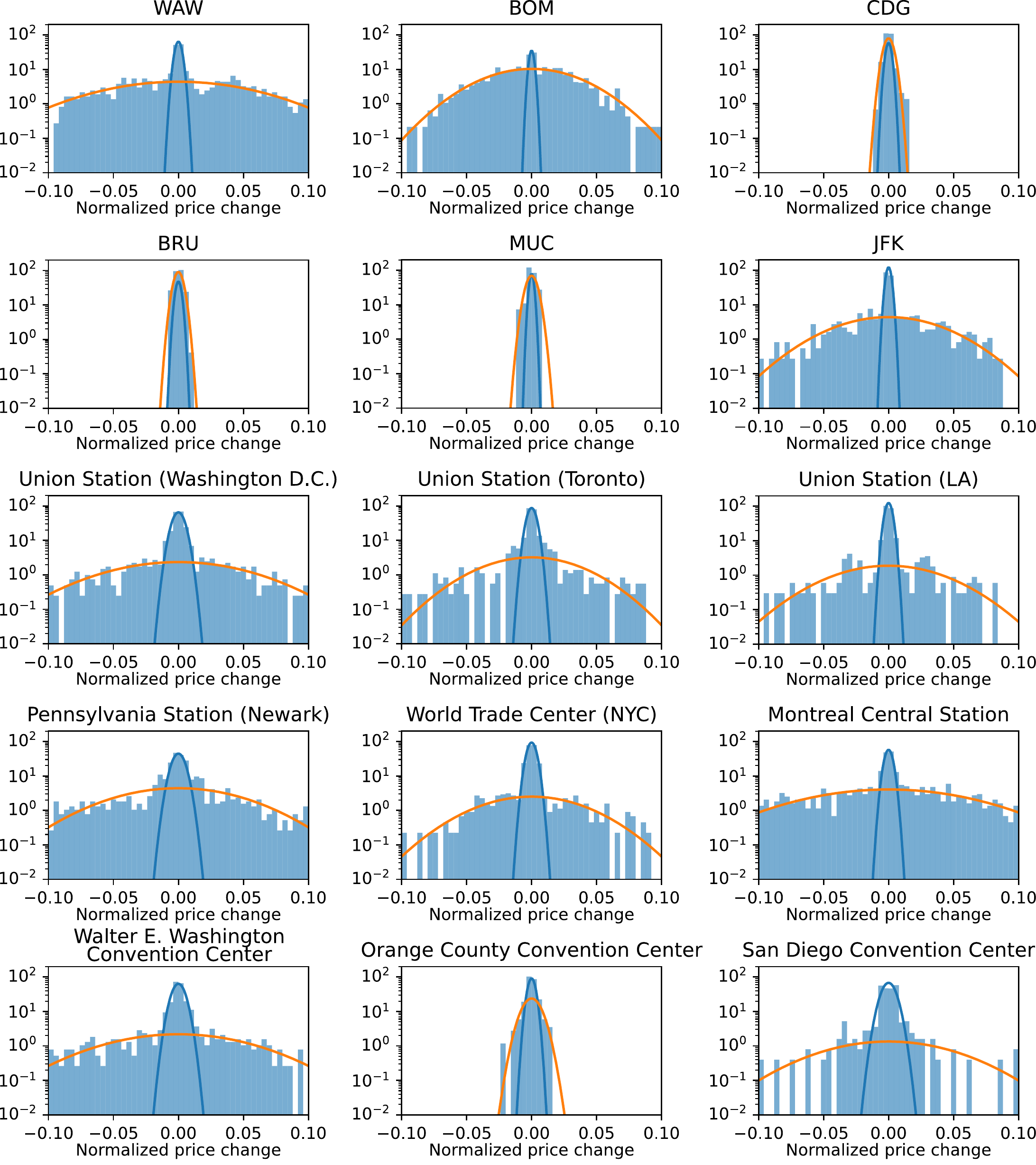}
\caption{\textbf{Relative price change distributions differ between locations with and without surge activity.} The relative price change distribution from the normalized \textit{UberX} price time series for different locations (blue histogram, see text). All locations show a narrow peak corresponding to slow changes of the base cost and small rounding errors of the price estimate (base cost fluctuations, blue line). Locations where price surges are prevalent (e.g. WAW, BOM in the top row) show a second part of the distribution, characterizing the larger and faster price changes during price surges (orange line). The orange and blue lines indicate the individual distributions of the Gaussian mixture model fit Eq.~\eqref{eq:gaussian_mixture}.
}
\label{fig:timescales}
\end{figure*}

Fig.~\ref{fig:timescales} shows the resulting distribution of the price changes at different locations without the Dirac-delta distribution. As expected, for all locations we find a large number of small changes, corresponding to small fluctuations of the base cost and rounding errors of the price estimates. Locations exhibiting price surges, like DCA, SFO and LAX, additionally have many larger price changes, corresponding to the dynamics of the surge fee. In contrast, locations without surge activity (LHR, CDG, BRU) have a very narrow distribution. 

For the data shown in Fig.~\ref{fig:timescales}, we find base cost fuctuations characterized by $\sigma_\mathrm{base} \approx 0.003$ whereas the relative price changes corresponding to the surge fee component (when they are different from the vase cost fluctuations) are characterized by $\sigma_\mathrm{surge} \approx 0.05$. When price surges do not exist, the relative price change distribution is typically well described by a single Gaussian such that the estimated values for $\sigma_\mathrm{base}$ and $\sigma_\mathrm{surge}$ are more similar.

\begin{figure*}
\centering
\includegraphics[width=0.7\textwidth]{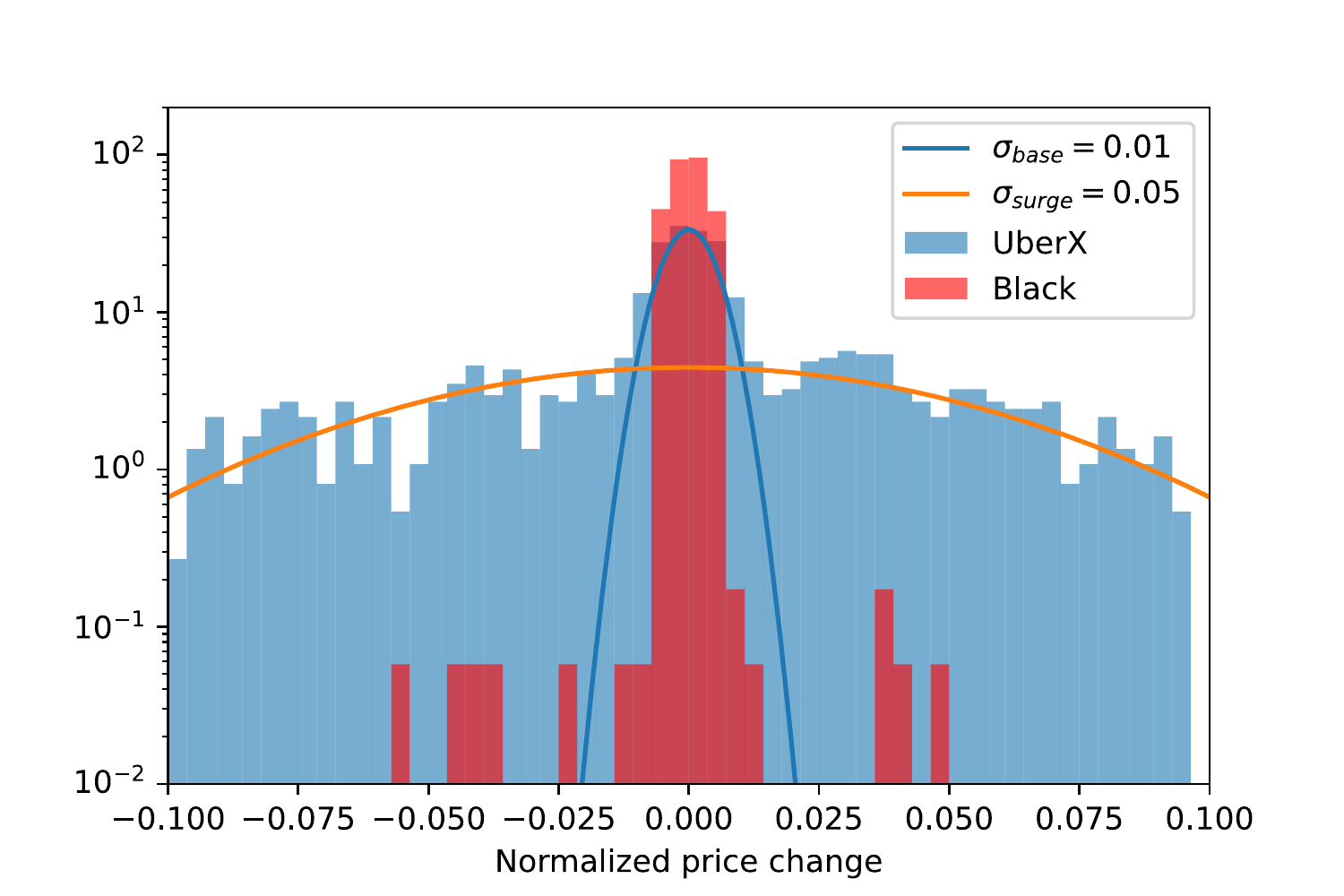}

\caption{\textbf{Uber \textit{Black} price changes are well described by only base cost fluctuations.} Distribution of the relative price changes at the Reagan National Airport (DCA) for \textit{UberX} (blue) and Uber \textit{Black} (red, higher values due to normalization of he probability distribution). The Gaussian distribution corresponding to the base cost fluctuations of \textit{UberX} approximately matches the price dynamics of Uber \textit{Black}, illustrating that these are base cost fluctuations unrelated to surge activity.
}
\label{fig:timescale_black}
\end{figure*}

To validate these observations, we compute the relative price changes also for Uber \textit{Black} at Reagan National Airport (DCA) in Washington D.C. Comparing the resulting distributions for \textit{UberX} and Uber \textit{Black} (see Fig.~\ref{fig:timescale_black}) shows that changes of the Uber \textit{Black} price are well described by only the base cost fluctuations. This is consistent with the observation that the trip fare evolves synchronously for \textit{UberX} and Uber \textit{Black} but only \textit{UberX} exhibits significant surge activity (compare Fig.~\ref{fig:PriceTimeSeries} above).

\clearpage
Based on these observations, we attempt to quantify the surge dynamics at different locations. If the estimated $\sigma_\mathrm{surge}$ is large, 
the price time series of that location likely exhibits price surges. The absolute value of $\sigma_\mathrm{surge}$ characterizes the strength of the surges, the weight $w_\mathrm{surge}$ quantifies the overall contribution of surge dynamics to the price behaviour, similar to the average number of surge peaks in the previous section. Note that in locations without surge activity, the estimated standard deviation of both distributions is similar and the weights are becoming more and more interchangeable. In the limit of $\sigma_\mathrm{main} = \sigma_{surge}$, all combinations of weights are indistinguishable.

\begin{figure*}
\centering
\includegraphics{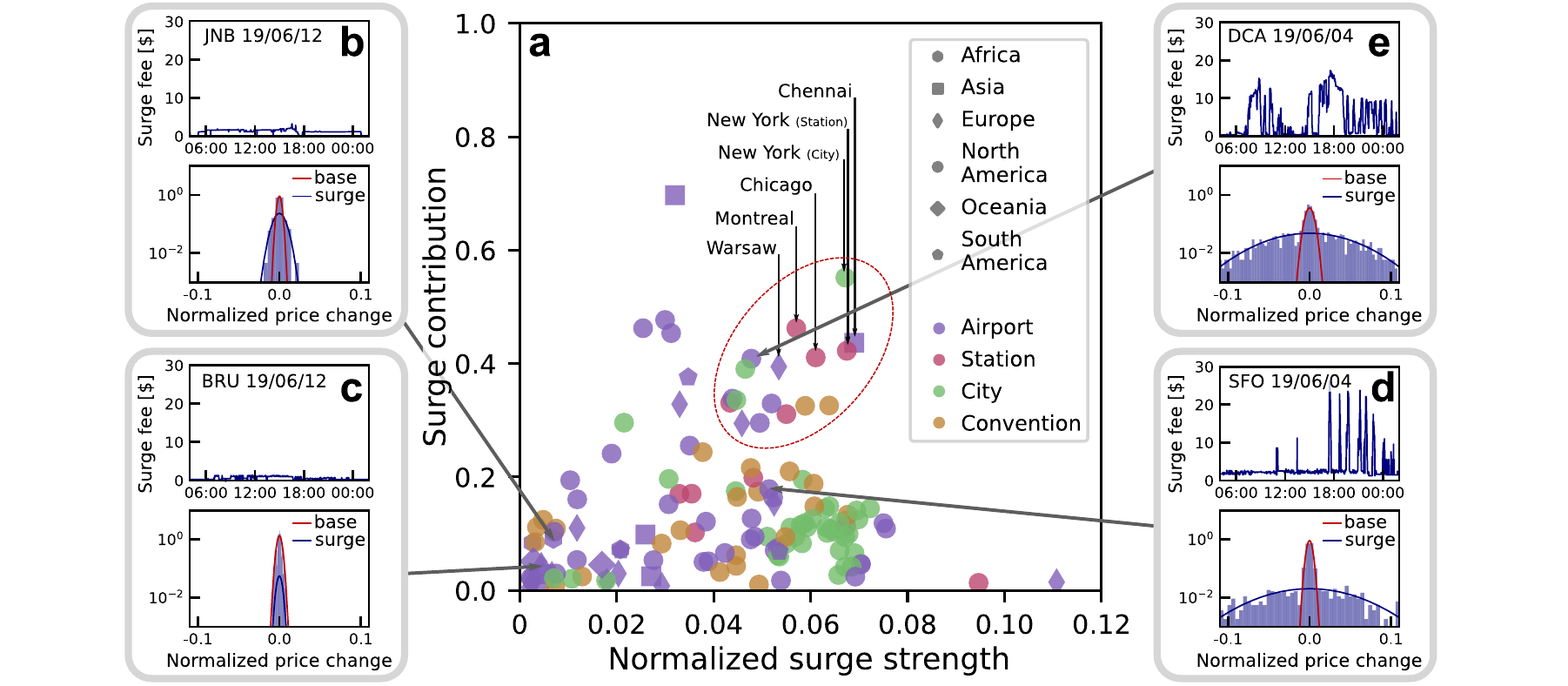}
\caption{\textbf{Price change distributions characterize the surge activity in different locations.}
\textbf{a} The main panel illustrates the weight $w_\mathrm{surge}$ (surge contribution) and the width $\sigma_\mathrm{surge}$ (normalized surge strength) of the broader distribution corresponding to the surge fee in the Gaussian mixture model fit [Eq.~\eqref{eq:gaussian_mixture}] for all observed locations. No clear classification of the locations is possible. However, we can distinguish three general regions. \textbf{b} and \textbf{c}, Small values of $\sigma_\mathrm{surge}$ indicate no surge activity as observed for example in JNB and BRU. Here, the distribution of relative price changes is well described by a single Gaussian corresponding to the base fare changes. On the other hand, large values of $\sigma_\mathrm{surge}$ signify fast and large price surges. \textbf{d}, If $w_\mathrm{surge}$ is small, the surges are infrequent (compare panel for SFO). \textbf{e}, If $w_\mathrm{surge}$ is also large, these price surges occur frequently, as shown for DCA. Several location (red ellipse) share similar characteristics of the price change distribution.
}
\label{fig:timescale_weight_std}
\end{figure*}

Figure~\ref{fig:timescale_weight_std} shows a scatter-plot of the time series statistics illustrating the magnitude $\sigma_{surge}$ (normalized surge strength, large values indicate strong price surges, small values indicate no surge activity) and the weight $w_\mathrm{surge}$ (surge contribution, large values of $w_\mathrm{surge}$ indicate many price surges, small values indicate few surges). Also in this representation, the data do not separate into distinct clusters. However, we clearly identify several locations where both $\sigma_\mathrm{surge}$ and $w_\mathrm{surge}$ are large (top right in Fig.~\ref{fig:timescale_weight_std}), including DCA (Washington D.C.), LAX (Los Angeles), WAW (Warsaw) and Pennsylvania Station (Manhattan, New York City). Prices in these locations show strong, highly volative surge activity similar to DCA, suggesting artificial causes for at least some of the price surges observed there. 

We also find a large number of locations with a broad distribution, $\sigma_\mathrm{surge} \ge 0.03$, but small weight $w_\mathrm{surge} \le 0.3$, including SFO (San Francisco) and most city trips in Manhattan, San Francisco and Washington, D.C.. These locations typically exhibit strong but infrequent price surges, some of which may be induced artificially. In particular for city trips, frequently organizing artificial surges is more difficult due to no central location where all drivers meet. Some of these locations may also exhibit a few long duration surge peaks caused by periods of high traffic and congestion or be affected by price surges at close by airports or other points of interest (e.g. city trips in the same city).

Interestingly, all 12 observed train station trips exhibit relatively large $\sigma_\mathrm{surge} \ge 0.03$. Here, it is plausible that arriving trains cause larger increases in demand, as trains can carry more passengers than planes and the time to leave the train and book a ride is likely more homogeneous than for airplane travelers. While this higher demand increases the incentive to induce a price surge (see game theoretic models below), it also offers a plausible demand-side explanation for the price surges in these locations.

Finally, locations with $\sigma_\mathrm{surge} \le 0.02$ either do not exhibit price surges at all or only show minor price changes due to surge pricing. At these locations, including BRU (Brussels), CDG (Paris) and LHR (London), no (artificial) price surges are induced. These include a larger fraction of non-US locations (relative to all observed locations), likely due to different local regulations legislating surge pricing and ride-hailing services. The value of $w_\mathrm{surge}$ is less meaningful for these locations, as in many cases $\sigma_\mathrm{surge} \approx \sigma_\mathrm{base}$ and both distributions in the mixture model are very similar.

\clearpage

\begin{figure}
    \centering
    \includegraphics[width = \textwidth]{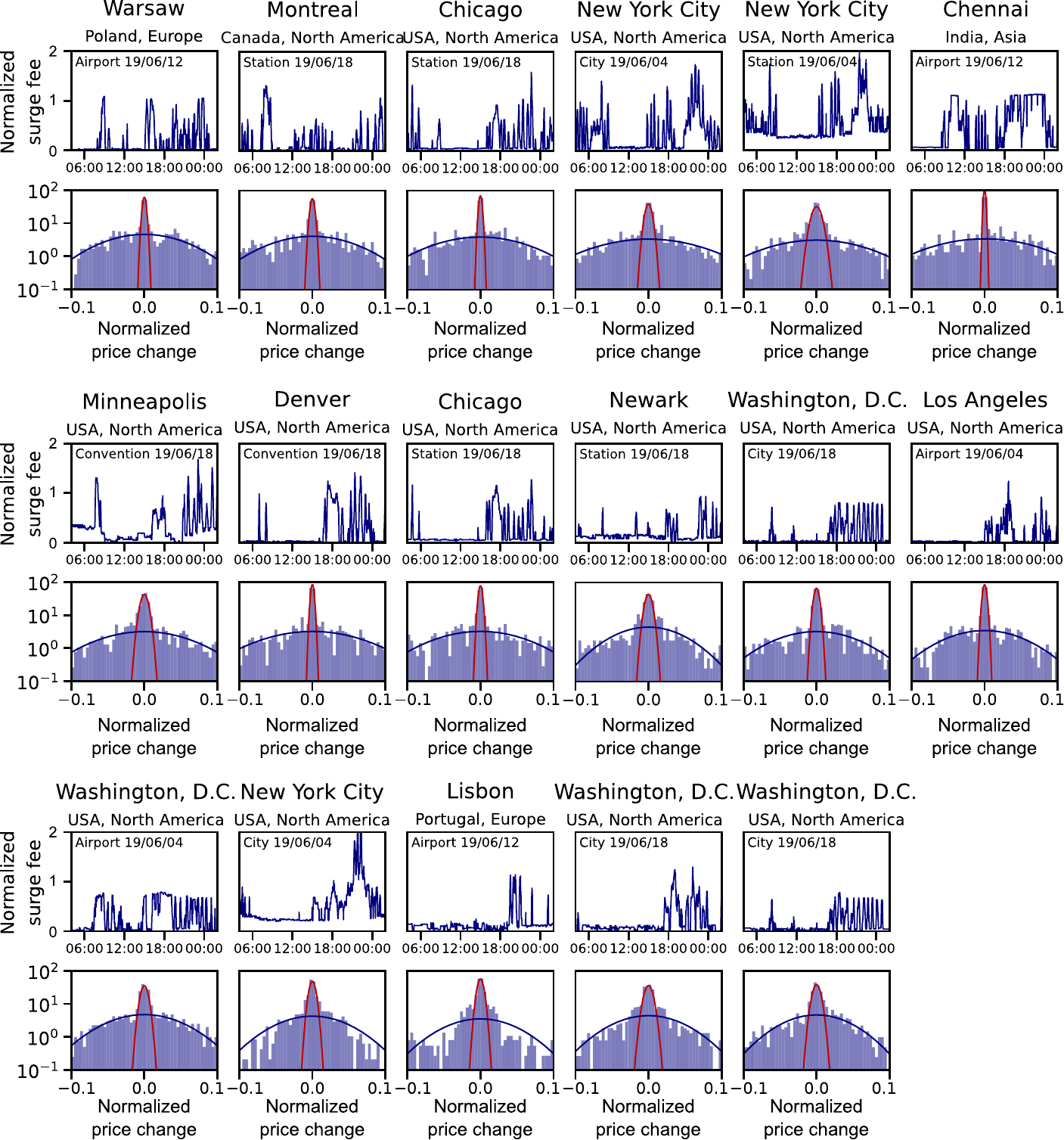}
    \caption{\textbf{Identifying non-equilibrium surge dynamics and anomalous supply dynamics}. Surge time series and price change statistics of all locations identified in Fig.~\ref{fig:timescale_weight_std} (red ellipse) that we expect to exhibit anomalous surge behaviour (compare also Fig.~4 in the main manuscript). Almost all identified locations show strong and frequent price surges. Exceptions are only Los Angeles (middle row, right) and Lisbon (bottom row, middle).}
    \label{fig:surge_city_normalized_surge}
\end{figure}

\clearpage

\section{Incentive structure for drivers under dynamic pricing}
By introducing dynamically adjustable prices, ride-hailing service providers create mobility services that, in principle, feature a self-organized equilibration of spatio-temporal demand and supply imbalances. Two groups of agents interact in such systems: customers and drivers. While customers have demand for mobility services, drivers offer such services against financial compensation. Price elasticities of demand and supply determine the equilibrium price point. Here, we illustrate in detail the fundamental incentive structure in dynamics pricing for drivers and how it may promote artificial supply shortages. First, we provide the simplest possible game-theoretical model revealing demand and supply constellations under which drivers collectively stimulate supply shortages. Second, we show that it is socially desirable for groups of drivers to coordinate in a broad regime of demand elasticities. Third, we reproduce the qualitative surge dynamics induced by collective action among drivers in a simple dynamic model with stochastic passenger arrivals, including the dynamics shown in Fig. 2 of the main manuscript.

\subsection{Two-player game}
\subsubsection{Two-player game with inelastic demand}

Consider a setting where the demand $D$ for rides is inelastic and fixed. Customers are insensitive to dynamic price changes and always agree to pay the current total fare to accommodate their demand. Such an assumption may be justified for constellations where business travelers have an urgent need to complete a trip in time, e.g. not to miss a business meeting. Typically, employers reimburse their employees for the cost of travel. Hence, business travelers have no incentive to prioritize their mobility demand based on the current price structure but aim for making it on time to their business meeting irrespective of the financial cost of mobility. 

Drivers may benefit from such constellations in terms of higher payoffs if they manage to stimulate dynamic price increases by creating an imbalance between demand and supply. Clearly, drivers can only affect the supply, i.e., increase the price by causing supply shortages, not the demand. Specifically, drivers can choose to switch OFF their ride-hailing mobile application to make their service temporarily unavailable. Thereby, they decrease the (observable) local supply $S$ while demand $D$ stays constant, making dynamic pricing algorithms increase the price per ride. If drivers switch their mobile application ON again and accept a customer's trip request before the dynamic pricing algorithm readjusts trip fares, drivers may secure higher payoffs. Note that the actually available supply of drivers (e.g. the total number of drivers idling at the airport) never changes, the dynamic pricing algorithm only reacts to an apparent supply shortage.

If $D \geq S$ all drivers benefit from this strategy. If $D < S$, drivers playing an OFF-strategy risk that other drivers remain ONline. These ON-players may exploit their first mover advantage in securing a ride and reduce the remaining demand for OFF-players. Hence, drivers play a game about cooperation and defection, where cooperation corresponds to the OFF-strategy and defection is the ON-strategy. Depending on price, supply and demand, the game undergoes a transition from stag hunt or prisoner's dilemma to a trivial game promoting the creation of artificial supply shortages.\\

We capture this idea in the following two-player game with strategies ON/OFF for which the drivers' payoff structure is given by
\begin{table}[H]
    \centering
    \begin{tabular}{l|lcccr|lcccr}
        & \multicolumn{5}{|c|}{ON} & \multicolumn{5}{c}{OFF} \\ \hline 
    ON &  $($&$\frac{D}{2}p_\mathrm{low}$&,& $\frac{D}{2}p_\mathrm{low}$&$)$ & $($&$p_\mathrm{mid}$&,& $(D-1)p_\mathrm{mid}$&$)$\\ \hline
    OFF & $($&$(D-1)p_\mathrm{mid}$ &,& $p_\mathrm{mid}$&$)$ & $($&$\frac{D}{2}p_\mathrm{high}$ &,& $\frac{D}{2}p_\mathrm{high}$&$)$ \\
    \end{tabular}
\end{table}
with $p_\mathrm{low} \le p_\mathrm{mid} \le p_\mathrm{high}$ modelling dynamic pricing as the apparent supply changes. To avoid case distinctions (each driver can serve a minimum of zero and a maximum of one customer), we assume that the demand $D$ is in the interval $1 \leq D \leq 2$. The expected payoffs shown in the payoff matrix assume a uniform probability to secure a ride from the currently available trip demand across currently available drivers. For example, if both drivers remain ON their respective payoff $\Pi_1$, $\Pi_2$ is
\begin{eqnarray}
    \Pi_1(\mathrm{ON}|\mathrm{ON}) &=& \frac{D}{2} \, p_\mathrm{low} \nonumber\\
    \Pi_2(\mathrm{ON}|\mathrm{ON}) &=& \frac{D}{2} \, p_\mathrm{low} \,,
\end{eqnarray}
where $D/2 = D/S$ is the probability of securing a ride at the respective instance in time. However, if one of the two switches their app OFF and the other remains ON, the probability of getting a ride shifts. The driver who remains ON has a first mover advantage because none of the demand have been served yet. The demand $D > 1$ meets a current supply of $S - 1 = 1$ and the dynamic pricing algorithm increases the price for a ride to $p_\mathrm{mid}$. At the same time, the ON-players probability of getting a ride increases because there is no competition from the OFF-player at the moment. With $D \ge 1$, a lone ON-player will always secure a ride. The OFF-player has to pick a ride from the left-over demand $D - 1 \le D/2$, reducing their chance to serve a customer. Consequently, the expected payoff is
\begin{eqnarray}
    \Pi_1(\mathrm{ON}|\mathrm{OFF}) &=& p_\mathrm{mid} \nonumber\\
    \Pi_2(\mathrm{OFF}|\mathrm{ON}) &=& (D-1) \, p_\mathrm{mid} \,.
\end{eqnarray}
If both players play OFF, they secure an even higher price $p_\mathrm{high} > p_\mathrm{mid}$ as the dynamic pricing algorithm reacts to the apparent supply $S - 2 = 0$. In this case, both players again have equal chance to secure a ride and achieve the expected payoff
\begin{eqnarray}
    \Pi_1(\mathrm{OFF}|\mathrm{OFF}) &=& \frac{D}{2} \, p_\mathrm{high} \nonumber\\
    \Pi_2(\mathrm{OFF}|\mathrm{OFF}) &=& \frac{D}{2} \, p_\mathrm{high} \,.
\end{eqnarray}

The Nash equilibria of the two-player game depend on the parameter values of $D, p_\mathrm{low}, p_\mathrm{mid}$ and $p_\mathrm{high}$. In particular, we distinguish two extreme cases:
\begin{itemize}
    \item $\mathbf{D=1}$: In the low demand limit $D=1$, the game is either a stag hunt or a prisoner's dilemma. 
            \begin{table}[H]
            \centering
            \begin{tabular}{l|c|c}
                & ON & OFF \\ \hline 
            ON &  $\left(\frac{1}{2}p_\mathrm{low},\frac{1}{2}p_\mathrm{low}\right)$ & $\left(p_\mathrm{mid},0\right)$\\ \hline
            OFF & $\left(0,p_\mathrm{mid}\right)$ & $\left(\frac{1}{2}p_\mathrm{high},\frac{1}{2}p_\mathrm{high}\right)$ 
            \end{tabular}
        \end{table}
    \begin{itemize}
        \item \textit{Stag Hunt}: For $p_\mathrm{mid} < \frac{p_\mathrm{high}}{2}$, the payoff matrix reduces to that of a stag hunt \cite{Skyrms2003} with $0 < \frac{1}{2}p_\mathrm{low} < p_\mathrm{mid} < \frac{p_\mathrm{high}}{2}$, where OFF is a payoff dominant, and ON a risk dominant Nash equilibrium.
        
        \item \textit{Prisoner's Dilemma}: For $p_\mathrm{mid}>\frac{p_\mathrm{high}}{2}$, the payoff structure changes to that of a prisoner's dilemma \cite{PrisonersDilemma2019}, $0 < \frac{1}{2}p_\mathrm{low} < \frac{p_\mathrm{high}}{2} < p_\mathrm{mid}$, where ON is the unique Nash equilibrium. 
    \end{itemize}
    
\item $\mathbf{D=2}$: In the high demand limit $D=2$, the payoff matrix is a trivial, fully symmetric game
\begin{table}[H]
    \centering
    \begin{tabular}{l|c|c}
        & ON & OFF \\ \hline 
    ON &  $\left(p_\mathrm{low},p_\mathrm{low}\right)$ & $\left(p_\mathrm{mid},p_\mathrm{mid}\right)$\\ \hline
    OFF & $\left(p_\mathrm{mid},p_\mathrm{mid}\right)$ & $\left(p_\mathrm{high},p_\mathrm{high}\right)$ 
    \end{tabular}
\end{table}
where OFF is the strictly dominant Nash equilibrium. Drivers always have an incentive to coordinate to cause a price surge in lack of competition for rides.
\end{itemize}
Hence, the dynamic price response and the demand to supply ratio govern the drivers' incentives under which constellations to switch OFF their mobile applications or when to remain ON.

\newpage
\subsubsection{Two-player game with elastic demand}

Based on the simplified game modelling the fundamental incentives for drivers in ride-hailing systems described in the previous section, we now consider in more detail under which conditions drivers induce artificial price surges. In particular, we relax the assumption of inelastic demand $D$, such that only some customers $D' \le D$ actually book a ride. 

\begin{figure*}[h]
    \centering
    \includegraphics[width=0.7\linewidth]{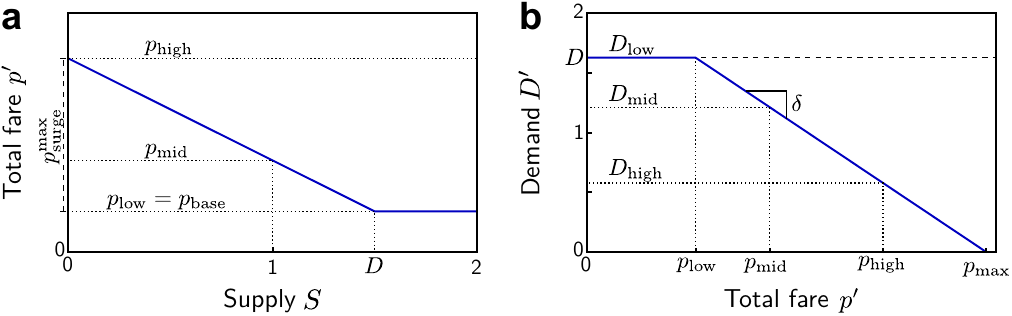}
    \caption{\textbf{Price structure and demand response in the two-player game.} \textbf{a}, The total fare payed by the customers and earned by the drivers is the base cost $p_\mathrm{base}$ and an additional surge fee. When the supply decreases below the demand $D$, the price increases linearly to a maximum value $p_\mathrm{high} = p_\mathrm{base} + p_\mathrm{surge}^\mathrm{max}$ [compare Eq.~\eqref{eq:total_fare_game}].
    \textbf{b}, The demand $D'$ of customers actually ordering a ride decreases as soon as the price increases above $p_\mathrm{base}$. It decreases to $D' = 0$ when the price reaches $p_\mathrm{max}$ [compare Eq.~\eqref{eq:two_player_demand}]. We assume $p_\mathrm{high} \le p_\mathrm{max}$ as otherwise the demand would vanish while there are still drivers available, indicating an unreasonably large value of $p_\mathrm{surge}^\mathrm{max}$.
    }
    \label{fig:two_player_schematic}
\end{figure*}

For the following analysis, we assume a linear increase of the surge fee as the supply decreases such that the total fare $p^\prime$ is given by
\begin{align}
    p^\prime(S < D) = p_\mathrm{base} + p_\mathrm{surge}^\mathrm{max} \, \frac{D-S}{D} \,.\label{eq:total_fare_game}
\end{align}
The price increases beyond the constant base cost $p_\mathrm{base}$ as soon as the supply $S$ falls below the total demand $D$ (all potential customers checking the app for the cost of a ride). Without loss of generality, we assume $p_\mathrm{base} = 1$ in all following calculations for the two-player game. The dependence of the total fare on the supply is illustrated in Fig.~\ref{fig:two_player_schematic}a.

Rather than paying high surge fees, however, customers may choose to wait until the price for their trip decreases or they may choose alternative modes of transportation such as (fixed price) taxi cabs or public transport to accommodate their demand. We model this price elasticity of the demand by assuming a willingness of customers to pay for a ride uniformly distributed in the interval $\left[p_\mathrm{base}, p_\mathrm{max}\right]$. As the price increases, the number of customers $D^\prime$ actually booking a ride thus decreases linearly as
\begin{align}
    D^\prime(p^\prime) = D_0 \left(1 - \frac{p^\prime-p_\mathrm{base}}{p_\mathrm{max}-p_\mathrm{base}} \right) = D_0 \left(1 - \frac{p^\prime-p_\mathrm{base}}{p_\mathrm{max}-p_\mathrm{base}} \right)=D_0(1-\delta (p^\prime - p_\mathrm{base})) \, \label{eq:two_player_demand}
\end{align}
where $\delta = \left(p_\mathrm{max}-p_\mathrm{base}\right)^{-1}$ describes the price elasticity of demand. Here, we assume that the base costs are set at the maximum possible value where all customers book a ride if there is no surge (Fig.~\ref{fig:two_player_schematic}b). Additionally, we assume that $p_\mathrm{high} = p_\mathrm{base} + p_\mathrm{surge}^\mathrm{max} \le p_\mathrm{max}$ such that there are some customers willing to pay the highest possible surge fee. Otherwise, the surge fee is chosen unreasonably high as it might completely nullify the demand, even if some supply of drivers is still available.

\newpage
In summary, for a total supply of $S_\mathrm{tot} = 2$ players, we have $S \in \left\{0,1,2\right\}$ depending on the drivers choice of strategy with corresponding values $p^\prime \in \{p_\mathrm{low},p_\mathrm{mid},p_\mathrm{high}\}$ for the total fare and $D^\prime \in \{D_\mathrm{low},D_\mathrm{mid},D_\mathrm{high}\}$ for the demand (Fig.~\ref{fig:two_player_schematic}). The payoff matrix is then given by
\begin{table}[H]
    \centering
    \begin{tabular}{l|lcccr|lcccr}
        & \multicolumn{5}{|c|}{ON} & \multicolumn{5}{c}{OFF} \\ \hline 
    ON &  $($&$\frac{D_\mathrm{low}}{2} \, p_\mathrm{low}$&,&$\frac{D_\mathrm{low}}{2} \, p_\mathrm{low}$&$)$ &  $($&$ p_\mathrm{mid}$&,&$(D_\mathrm{mid}-1) \, p_\mathrm{mid}$&$)$\\ \hline
    OFF & $($&$(D_\mathrm{mid}-1) \, p_\mathrm{mid}$&,&$ p_\mathrm{mid}$&$)$ &  $($&$\frac{D_\mathrm{high}}{2} \, p_\mathrm{high}$&,&$\frac{D_\mathrm{high}}{2} \, p_\mathrm{high}$&$)$ 
    \end{tabular}
\end{table}
\noindent with the total fare and demand depending on the drivers' decisions\newline
\begin{itemize}
\item (ON, ON) with supply $S = 2$
        \begin{eqnarray*}
            p_\mathrm{low} &=& p_\mathrm{base} \\
            D_\mathrm{low} &=& D
        \end{eqnarray*}
\item (ON, OFF) or (OFF, ON) with supply $S = 1$ 
        \begin{eqnarray*}
            p_\mathrm{mid} &=& p_\mathrm{base} + p_\mathrm{surge}^\mathrm{max} \frac{D - 1}{D} \\
            D_\mathrm{mid} &=& D \, \left( 1 - \delta \, p_\mathrm{surge}^\mathrm{max} \frac{D - 1}{D} \right)
        \end{eqnarray*}
\item (OFF, OFF) with supply $S = 0$ 
        \begin{eqnarray*}
            p_\mathrm{high} &=& p_\mathrm{base} + p_\mathrm{surge}^\mathrm{max} \\
            D_\mathrm{high} &=& D \, \left( 1 - \delta \, p_\mathrm{surge}^\mathrm{max} \right)
        \end{eqnarray*}
\end{itemize}
where, for simplicity of presentation, we do not explicitly note the case distinctions to ensure that each driver serves at most one and at least zero customers and the total fare only increases if $S < D$. The equations as presented are valid for $1 \le D \le 2$ and $\delta \, p_\mathrm{surge}^\mathrm{max} < 1$.

\begin{figure*}[h]
    \centering
    \includegraphics[width=\textwidth]{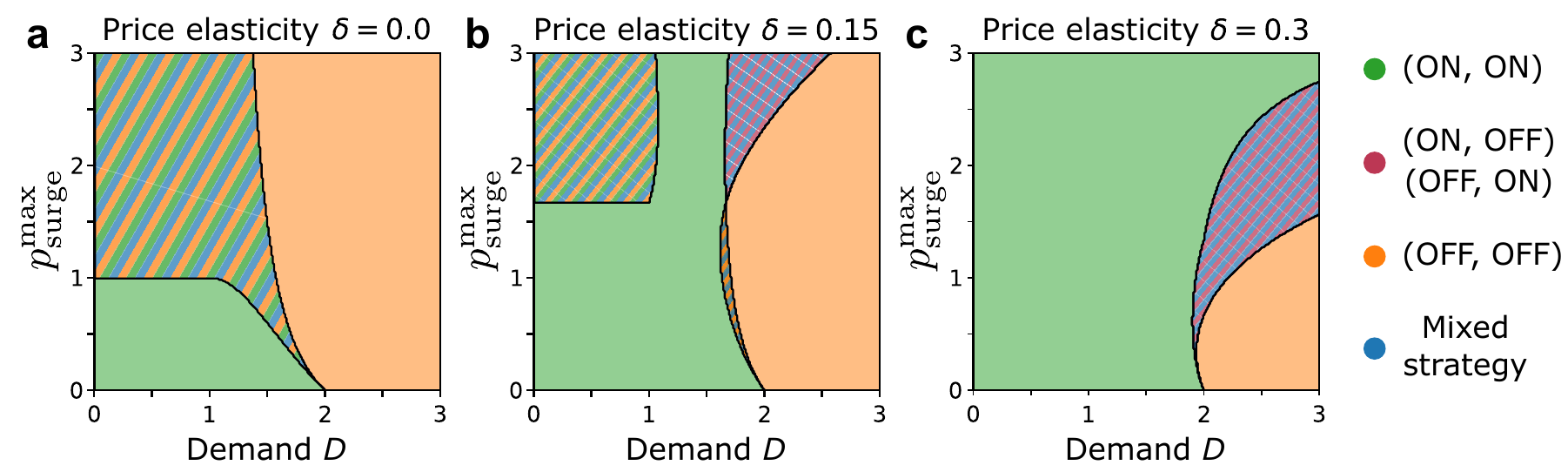}
    \caption{\textbf{Low price elasticity and high demand promote artificial supply shortages.} \textbf{a}, Nash equlibria of the two-player game (see text) with completely inelastic demand, $\delta = 0$. As discussed above, at $D=1$ the game transitions from a Prisoner's Dilemma at low $p_\mathrm{surge}^\mathrm{max} < 1$ with a single Nash equlibrium to a Stag Hunt with multiple Nash equilibria. For sufficiently high demand $D \ge 2$, the game becomes trivial with a single (OFF, OFF) socialy optimal Nash equlibrium. 
    \textbf{b}, Nash equlibria of the two-player game (see text) with completely elastic demand $\delta = 0.15$. 
    \textbf{c}, Nash equlibria of the two-player game (see text) with completely elastic demand with higher elasticity $\delta = 0.30$. 
    As the demand elasticity increases, ON becomes the only Nash equlibrium in a large part of the parameter space. Only for very high demand OFF remains a feasible strategy. 
    Overall, these results illustrate that high demand and low price elasticity of demand incentivize drivers to create artificial supply shortages.
    }
    \label{fig:two_player_phase}
\end{figure*}

\newpage
Figure~\ref{fig:two_player_phase} shows the resulting Nash equilibria for this two-player game with elastic demand for different demand $D$, maximum surge fee $p_\mathrm{surge}^\mathrm{max}$ and demand elasticities $\delta$ (compare Fig. 2b in the manuscript). 
\begin{itemize}
    \item $\delta = 0$: In the limit of inelastic demand we reproduce the findings discussed above. For demand $D \leq 2$ and $p_\mathrm{mid}>p_\mathrm{high}/2$ strong competition between players results in a Nash equilibrium in ON-strategies (green in Fig.~\ref{fig:two_player_phase}a). For sufficiently high surge fees the game changes to a stag hunt ($p_\mathrm{mid}>p_\mathrm{high}/2$) where both ON, OFF and mixed strategy Nash equilibria coexists (striped area in Fig.~\ref{fig:two_player_phase}a). For sufficiently high demand (guaranteed for $D \ge 2$), drivers are strongly incentivized to switch OFF their apps (orange in Fig.~\ref{fig:two_player_phase}a).
    \item $\delta = 0.15$: The drivers' incentive structure starts to shift when introducing demand elasticitiy (see Fig.~\ref{fig:two_player_phase}b). Inducing artificial price surges becomes unrealiable as the increase of the total fare (profit per ride) may be compensated by the reduced number of customers and drivers switching OFF their app have a higher risk missing out on potential customers. At low demand, larger $\delta$ intensifies the competition between drivers and forces them to stay ONline. The regime where remaining ONline is the Nash equilibrium extends to much higher surge fees than for inelastic demand since. However, in the low demand limit $D<1$ the coexistence of ON and OFF Nash equilibria persists for large possible surge fees as the risk becomes independent of the surge fee when there is less than one customer $D < 1$. Importantly, the regime where OFF is the pure strategy Nash equilibrium shrinks compared to the situation for $\delta=0$. 
    \item $\delta = 0.3$: For strongly elastic demand drivers are interacting in a highly competitive environment. For almost all $D<2$ and very large surge fees, the Nash equilibrium is in ON-strategies as strong demand response which lead to expected profit loss for drivers that go OFFline (green in Fig.~\ref{fig:two_player_phase}c). The coexistence regime of ON and OFF Nash equilibria for low demand disappears for high elasticity $\delta=0.3$. Only for low values $p_\mathrm{surge}^\mathrm{max}<1.5$ and sufficiently strong demand $D \ge 2$, playing OFFline strategies remains a Nash equilibrium (orange in Fig.~\ref{fig:two_player_phase}c). In the intermediate regime, the incentive structure forces drivers into a mixed strategy equilibrium (blue-red striped in Fig.~\ref{fig:two_player_phase}c).
\end{itemize}

Hence, there exist four well pronounced regimes of Nash equilibria in ride-hailing games. While strong competition compels drivers to always remain ONline in the limit of low overall demand, the reverse applies to situations with high demand where going OFFline is the dominant strategy. Demand elasticities and the maximum possible surge fee modulate this picture. These results illustrate the basic incentives for drivers: high, inelastic demand and sufficient additional profit due to surge fees promotes drivers to exploit their control over the market and induce artificial price surges. High price elasticity (e.g. due to other competitive public transport options) increases competition and reduces these incentives.

\clearpage

\newpage
\subsection{Multiplayer game with elastic demand}
In a more realistic setting, a larger number $N > 2$ of drivers interact, making a decision to switch their ride-hailing application ON or OFF. In this section, we consider a version of the game described above with $N > 2$ players.

For simplicity, we consider only the case of high demand, where the total demand $D$ is equal to the total supply $N$ drivers. When drivers switch OFF their ride-hailing application, the dynamic pricing algorithm reacts to the apparent supply $S$ and increases the price, as in the two-player game. We assume the same cost function Eq.~\eqref{eq:total_fare_game}, that simplifies to 
\begin{align}
    p^\prime(\rho_\mathrm{OFF}) = p_\mathrm{base} + p_\mathrm{surge}^\mathrm{max} \, \rho_\mathrm{OFF} \label{eq:multiplayer_price}
\end{align}
under the assumption of high demand, $D = N$. $\rho_\mathrm{OFF} = \frac{N-S}{N}$ is the fraction of drivers playing OFF and $p_\mathrm{surge}^\mathrm{max}$ is the maximal supply-driven surge fee when no drivers are available (e.g. when all drivers collectively switch to the OFF state). This dynamic pricing function implies that an individual driver turning his or her app OFF, $\rho_\mathrm{OFF} = 1/N$, changes the price by $p_\mathrm{surge}^\mathrm{max} / N$.

As before, customers react to the new price $p^\prime$. With the above assumption, equation~\eqref{eq:two_player_demand} simplifies to
\begin{align}
    D^\prime(p^\prime) = D_0 \left(1 - \frac{p^\prime-p_\mathrm{base}}{p_\mathrm{max}-p_\mathrm{base}} \right) = D_0 \left(1 - \delta \, p_\mathrm{surge}^\mathrm{max} \, \rho_\mathrm{OFF}\right) \, \label{eq:multiplayer_demand}
\end{align}
where $\delta = (p_\mathrm{max}-p_\mathrm{base})^{-1}$ again describes the price elasticity of the demand. \\

\begin{figure*}[h]
    \centering
    \includegraphics[width=0.66\linewidth]{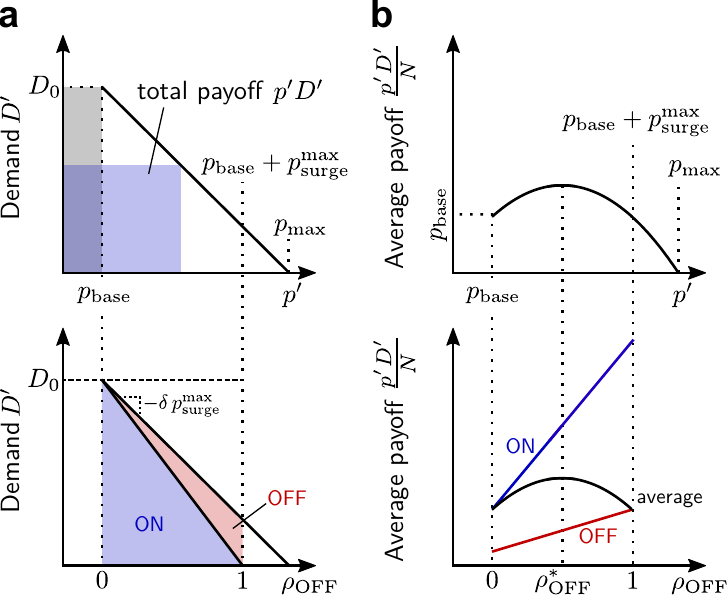}
    \caption{\textbf{Price elasticity of demand and optimal payoff with collective action.} \textbf{a} The demand decreases as the price increases with price elasticity $\delta$ [top panel, compare Eq.~(\ref{eq:multiplayer_demand})]. The increase in price is linearly related to the fraction $\rho_\mathrm{OFF}$ of OFF drivers [Eq.~(\ref{eq:multiplayer_price})] such that the demand also decreases linearly with $\rho_\mathrm{OFF}$ (bottom panel). For $p_\mathrm{max} > p_\mathrm{base} + p_\mathrm{surge}^\mathrm{max}$, the demand decreases such that every ON driver always gets a customer (bottom panel, shaded blue) and OFF drivers have to compete for the left-over customers (shaded red).
    \textbf{b} The average payoff per driver has a maximum at an intermediate price (top panel). This price corresponds to the optimal fraction $\rho_\mathrm{OFF}^*$ of OFF drivers Eq.~(\ref{eq:multiplayer_opt}). As a group, all drivers achieve a maximal collective (socially optimal) payoff. However, for each driver individually, ON is the better strategy (bottom panel). \\
    }
    \label{fig:multiplayerGame}
\end{figure*}

\paragraph*{Selfish action:} We first consider this game in the case where all drivers act selfishly and try to maximize their own payoff. Since we assume a total demand $D = N$ and $p_\mathrm{max} \geq p_\mathrm{base} + p_\mathrm{surge}^\mathrm{max}$, a driver choosing to remain ONline will always be able to secure a ride. Their expected payoff is then
\begin{align}
    \Pi[\mathrm{ON}|\rho_\mathrm{OFF}] = p^\prime = p_\mathrm{base} +  p_\mathrm{surge}^\mathrm{max} \, \rho_\mathrm{OFF}.
\end{align}
ON divers thus profit from the decision of other drivers to go OFFline.
On the other hand, a driver playing the OFF strategy receives
\begin{align}
    \Pi[\mathrm{OFF}|\rho_\mathrm{OFF}]= \frac{D^\prime-N \, (1 - \rho_\mathrm{OFF})}{N \rho_\mathrm{OFF}} \, p^\prime  = (1 - \delta \, p_\mathrm{surge}^\mathrm{max}) \, \left(p_\mathrm{base} +  p_\mathrm{surge}^\mathrm{max} \, \rho_\mathrm{OFF} \right), 
\end{align}
where the first term corresponds to the probability of an OFF-driver to get one of the remaining $D^\prime - N(1-\rho_\mathrm{OFF})$ customers after ON drivers have already secured their rides (illustrated in Fig.~\ref{fig:multiplayerGame}a).

Clearly, playing ON is always more beneficial for the individual player (see also Fig.~\ref{fig:multiplayerGame}b). Still, a driver may choose to play OFF if the expected payoff is larger than the payoff $\Pi[\mathrm{ON}|\rho_\mathrm{OFF} = 0] = p_\mathrm{base}$ in the case where everyone remains ONline, that means when $(1 - \delta \, p_\mathrm{surge}^\mathrm{max}) \, \left(p_\mathrm{base} +  p_\mathrm{surge}^\mathrm{max} \, \rho_\mathrm{OFF} \right) \ge p_\mathrm{base}$. For large $N$, $\rho_\mathrm{OFF} = 1/N$ is small for a single driver. Therefore, the inequality is only true (selfish action beneficial) if $N$ is sufficiently small, $N \le N^* = \frac{1 - \delta p_\mathrm{surge}^\mathrm{max}}{\delta p_\mathrm{base}}$.\\

\paragraph*{Collective action:} In contrast to selfish action, drivers may choose to cooperate to drive up the price and potentially increase their \emph{collective} payoff $p^\prime D^\prime$ even if this is not beneficial for an individual driver (see Fig.~\ref{fig:multiplayerGame}a). The expected collective payoff per driver is (see Fig.~\ref{fig:multiplayerGame}b)
\begin{eqnarray}
    \frac{\Pi_\mathrm{coll}[\rho_\mathrm{OFF}]}{N} &=& \frac{p^\prime \, D^\prime}{N} = \frac{ \left(p_\mathrm{base} +  p_\mathrm{surge}^\mathrm{max} \, \rho_\mathrm{OFF} \right) \, D_0 \left(1 - \delta \, p_\mathrm{surge}^\mathrm{max} \, \rho_\mathrm{OFF}\right) }{N} \nonumber\\
    &=& p_\mathrm{base} + p_\mathrm{surge}^\mathrm{max} \, \rho_\mathrm{OFF}  \left(1 - \delta  \, p_\mathrm{base} - \delta \, p_\mathrm{surge}^\mathrm{max} \, \rho_\mathrm{OFF}\right) \,.
\end{eqnarray}
In the limit $\rho_\mathrm{OFF} \rightarrow 0$, this reduces to the average payoff $\frac{\Pi_\mathrm{coll}[\rho_\mathrm{OFF} = 0]}{N} = \frac{D_0 \, p_\mathrm{base}}{N} = p_\mathrm{base}$ when everyone remains ONline. Maximizing this collective payoff gives the socially optimal fraction of OFF players
\begin{align}
    \rho_\mathrm{OFF}^* = \frac{1 - p_\mathrm{base} \, \delta}{2 \, \delta \, p_\mathrm{surge}^\mathrm{max}} \,, \label{eq:multiplayer_opt}
\end{align}
assuming $0 < \rho_\mathrm{OFF}^* < 1$, otherwise $\rho_\mathrm{OFF}^*$ takes the limiting value $0$ or $1$. In this socially optimal configuration, drivers only serve a fraction $1 - \frac{1 - p_\mathrm{base} \delta}{2}$ of potential customers (illustrated in Fig.~\ref{fig:multiplayerGame}) while all other customers choose a different mode of transport or wait for the price to drop.\\

Overall, cooperation between drivers allows them to achieve a higher collective payoff. In a repeated game, different drivers may rotate playing OFF for the benefit of the group, resulting in a game of cooperation (coordinated OFF) and defection (always ON) among the drivers. Note that the collective payoff depends only on the product $p_\mathrm{surge}^\mathrm{max} \, \rho_\mathrm{OFF}$. The qualitative behavior is thus independent of $p_\mathrm{surge}^\mathrm{max}$ and a general consequence of the dynamic pricing mechanism in this simplified model. If $p_\mathrm{surge}^\mathrm{max}$ is changed, the fraction of offline drivers $\rho_\mathrm{OFF}$ will change to compensate (as far as possible) and the overall price $p^\prime$, payoff and fraction of riders served remains unchanged. As for the two-player game, lower demand $D_0 < N$ or higher price elasticity of the demand increases the payoff gap between ON and OFF players, increasing the incentive to defect and making it more difficult to organize cooperation.

\newpage
\subsection{Dynamic multiplayer game under elastic demand}
Based on the previous two games, we now consider a repeated dynamic game, where drivers interact with the demand not only once. Instead they do so in a time-continuous system, making decisions whether to switch their app ON or OFF all the time and being busy for some time when they serve a customer. 

We consider a system where $N$ drivers serve customers that arrive following a stochastic process. For simplicity, we model the arrivals of customers as a Poisson process with a constant rate $\lambda$. In particular, we assume a uniform request rate over time (a valid approximation during most of the day as suggested by our demand analysis in section \ref{sec:demand}) and do not model, for example, the discrete arrivals of individual planes. This describes the limiting case where the time passengers need to cross the airport, claim baggage etc. is sufficiently heterogeneous to smooth out the discrete arrival events.

When a customer arrives, they check the current price of transportation and, depending on their willingness to pay, either book a ride, wait for some time or leave the system (e.g. choosing a different mode of transportation or a different ride-hailing service provider). As above, the maximum price a customer is willing to pay is distributed uniformly in the interval $\left[p_\mathrm{base}, p_\mathrm{max}\right]$. When the customer requests a ride, the oldest (longest waiting) ONline driver at the airport is selected to serve the ride. Here, we include the time the driver may have spent OFFline in this waiting time. The driver then serves the customer and returns to the airport after a round-trip time $t_s$ uniformly distributed in $\left[2 t_d - \Delta, 2 t_d + \Delta\right]$.

Drivers waiting idly at the airport may decide to switch OFF their app to induce a surge. We calculate the current price $p^\prime(t)$ at time $t$ based on the current number $N_\mathrm{idle}(t)$ of ONline drivers at the airport similar to the prevoulsy discussed games as
\begin{equation}
    p^\prime(t) = \begin{cases}
                       p_\mathrm{base} \quad &| \quad N_\mathrm{idle}(t) \ge N_\mathrm{thresh} \\
                        p_\mathrm{base} + p_\mathrm{surge}^\mathrm{max} \left(1 - \frac{N_\mathrm{idle}(t)}{N_\mathrm{thres}}\right)    \quad &| \quad N_\mathrm{idle}(t)  < N_\mathrm{thresh} \,. 
    \end{cases} \label{eq:dynamic_game_price} 
\end{equation}
Since there is not absolute demand in this case, we encode the demand in the number of drivers $N_\mathrm{thresh}$ below which the surge fee begins. We take $N_\mathrm{thresh} = \lambda \, (2 t_d)$ as the number of drivers required to serve all $\lambda \, (2 t_d)$ expected incoming requests before one of these drivers returns to the airport.

Note that we assume instantaneous updates of the price as a function of the current state of the system. There is no delay or dependence on the history of the system. As such, individual drivers can never profit themselves from the surge they induce (in contrast to the static games). When they go ONline, the price immediately decreases. Instead, other drivers may get higher payoffs and the drivers may profit as a group (as in the multiplayer game).

In the following we describe the parameters used in the model, give a mean-field calculation of the socially optimal strategy and describe the detailed dynamics of the time-continuous simulation here and in the main manuscript. Finally, we illustrate the robustness of the induced price surges across a range of model parameters.\\

\subsubsection{Parameters} Parameters used to illustrate the dynamic model here and in the main manuscript are loosely based on the observed price estimates from Washington D.C. (DCA). For the base cost and possible surge fee, we take $p_\mathrm{base} = 16$ and $p_\mathrm{surge}^\mathrm{max} = 20$. We take $p_\mathrm{max} = 54$ as the maximum observed total fare for an Uber Black ride.

Other parameters were chosen to (i) be in line with realistic values and (ii) be in the correct range to exhibit price surges in the simulations. In particular, we choose the round-trip time $\left<t_s\right> = 2 t_d = 30\,\mathrm{min}$, corresponding to the expected time of taxi rides from the airport of approximately $15\,\mathrm{min}$, 
distributed in the interval $\left[2 t_d - \Delta, 2 t_d + \Delta\right]$ with $\Delta = 5$ minutes. We choose the request rate $\lambda = 2\,\mathrm{min}^{-1}$, approximately in line with the average number of taxi trips recorded at DCA. 
Together with the number of drivers $N=160$, these parameters corresponds to drivers spending $50\,\mathrm{min}$ between each of their $30\,\mathrm{min}$ rides waiting at the airport and to an average number of $100$ drivers waiting at the airport at any given time. 

With the above parameters we require at least $N_\mathrm{thresh} = \lambda \, (2 t_d) = 60$ drivers at the origin to avoid surge fees. When all drivers are always ONline, we have on average $100$ drivers idle at the airport. In this setting, we thus expect a constant price $p_\mathrm{base}$ with only very rare fluctuations.

All simulations start with all drivers ONline and currently waiting for a request.

\newpage
\subsubsection{Mean field}
In order to determine the optimal strategy for the drivers, we first consider a mean field description of the system described above. For this calculation, we are only interested in the average steady state values of the price $p^\prime$, the request rate $\lambda^\prime$ and the number of ONline, OFFline and idle drivers $N_\mathrm{ON}$ , $N_\mathrm{OFF}$ and $N_\mathrm{idle}$. 

On the demand side, of all potential customers only a fraction $\frac{p_\mathrm{max} - p^\prime}{p_\mathrm{max} - p_\mathrm{base}}$ will actually request transport after checking the price. This effectively reduces the request rate to 
\begin{equation}
    \lambda^\prime = \frac{p_\mathrm{max} - p^\prime}{p_\mathrm{max} - p_\mathrm{base}} \lambda \,. \label{eq:dynamic_game_reqrate}
\end{equation}
Note that, as before, we assume that the demand relevant for calculating the surge fee does not change, i.e. there is no demand dependence in the price function Eq.~\eqref{eq:dynamic_game_price}. The total rate $\lambda$ of customers checking the price (though not necessarily requesting a ride) and $N_\mathrm{thres}$ remain constant.

On the supply side, we are interested in the number $N_\mathrm{idle}$ of idle drivers at the airport to calculate the total fare. As in the static game, we assume a constant fraction $\rho_\mathrm{OFF}$ of drivers to be OFFline. On average, the remaining $N_\mathrm{ON} = N - N_\mathrm{OFF} = N \, (1 - \rho_\mathrm{OFF})$ drivers spend $\left< t_s \right> = 2 t_d$ driving and $\left< t_w \right> = N_\mathrm{idle} / \lambda^\prime$ waiting at the airport for their next customer. Weighting the distribution of drivers with these times gives the number of drivers waiting idly as the airport as the fraction 
\begin{equation}
    N_\mathrm{idle} = \frac{\left< t_w \right>}{\left< t_s \right> + \left< t_w \right>} \, N_\mathrm{ON} \nonumber
\end{equation}
which results in
\begin{eqnarray}
    N_\mathrm{idle} &=& N_\mathrm{ON} - \lambda^\prime \, (2 \, t_d) \nonumber\\
                    &=& N_\mathrm{ON} - 2\, \lambda\, t_d\, \frac{p_\mathrm{max} - p^\prime}{p_\mathrm{max} - p_\mathrm{base}} \label{eq:dynamic_game_nidle}
\end{eqnarray}

\begin{figure*}[h]
    \centering
    \includegraphics[width=0.9\linewidth]{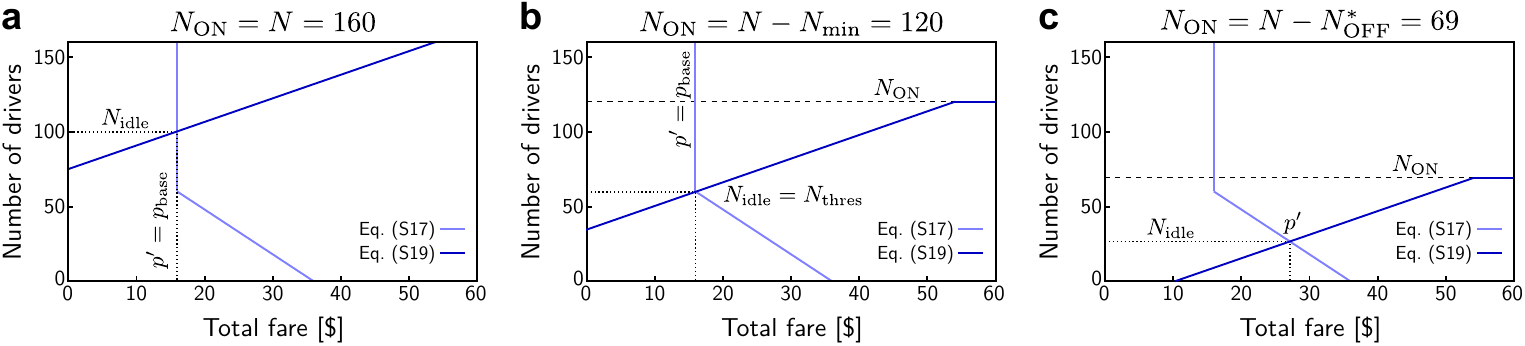}
    \caption{\textbf{Self-consistency condition for the price and the number of idle drivers.} The plots show the functions $p^\prime(N_\mathrm{idle})$ [Eq.~(\ref{eq:dynamic_game_price}), light blue] and $N_\mathrm{idle}(p^\prime)$ [Eq.~(\ref{eq:dynamic_game_nidle}), dark blue] when \textbf{a} all drivers are ONline, \textbf{b} $N_\mathrm{min} = 40$ drivers are OFFline at the boundary to price surges and \textbf{c} in the mean-field optimal case when $N_\mathrm{OFF}^* = 91$ drivers are OFFline. The price increases only when sufficiently many drivers $N_\mathrm{OFF} > N_\mathrm{min}$ go OFFline. 
    }
    \label{fig:dynamic_game_self_consistency_solution}
\end{figure*}

Together with the dynamic price function Eq.~\eqref{eq:dynamic_game_price}, we  now have a self-consitency condition for relating the price $p^\prime$ and the number of idle drivers $N_\mathrm{idle}$ [Eq.~(\ref{eq:dynamic_game_price}) and (\ref{eq:dynamic_game_nidle})]. The solution of this self-consisstency condition gives the equilibrium total fare $p^\prime$ and the optimal number of ONline drivers $N_\mathrm{ON}$. Due to the number of case distinctions arising from to the piece-wise definition of the price and to ensure, for example, $0 \le N_\mathrm{idle} \le N_\mathrm{ON}$ for all possible parameter choices, we do not give the full solution here. Instead, the solution is illustrated in Fig.~\ref{fig:dynamic_game_self_consistency_solution}.

With the price and the average request rate in equilibrium, the average earning rate of a single ONline driver follows as the profit from a ride divided by the total time for one ride (round-trip time and waiting for the next ride) as
\begin{equation}
    \Pi_\mathrm{ON} = \frac{p^\prime}{\left<t_s\right> + \left<t_w\right>} = \frac{p^\prime \lambda^\prime}{2 t_d \, \lambda^\prime + N_\mathrm{idle}} \label{eq:dynamic_game_on_earning}
\end{equation}
and the collective earning rate as the weighted average with the OFFline drivers, who earning nothing,
\begin{equation}
    \Pi_\mathrm{coll} = (1 - \rho_\mathrm{OFF}) \, \Pi_\mathrm{ON} = \frac{p^\prime \lambda^\prime \, (1 - \rho_\mathrm{OFF})}{2 t_d \, \lambda^\prime + N_\mathrm{idle}} \,. \label{eq:dynamic_game_social_earning}
\end{equation}

For the parameters described above ($\left<2 \, t_d\right> = 30\,\mathrm{min}$, $N = 160$), we obtain the following results: If all drivers are always ONline, they earn an average of $12$ USD per hour or $p = p_\mathrm{base} = 16$ USD per ride with $30 + 50$ minutes between rides. The drivers can earn more if they reduce the number of idle ONline drivers to at most $N_\mathrm{thresh} = 60$ drivers by switching OFF their app. They maximize their profit when only $N_\mathrm{idle}^* = 26$ drivers are available at the airport ($N_\mathrm{ON}^* = 69$), collectively earning $14.39$ USD per hour. Here, the ON-drivers earn $p^* = 27.14$ USD per ride with $30 + 18.8$ minutes between rides but approximately $57\%$ of all drivers are OFFline and earn nothing.\\

\subsubsection{Simulation}
In our simulation of the system, we make a few additional assumptions for the strategy of the drivers and the behavior of the customers. In particular, we assume that the drivers use a strategy informed by the above mean field calculations: Drivers switch OFF their app if there are enough drivers willing to participate in a surge to reduce the number of idle drivers to at least $N_\mathrm{thres}$. However, drivers switch OFF their app only while there are more than $N_\mathrm{idle}^*$ idle drivers at the airport (i.e. while the price is lower than their optimal price), otherwise they wait to serve requests and collect the higher total fare. 

While the drivers collectively earn the most with this optimal strategy, a single driver who stays OFFline for the benefit of the group would earn very little or even nothing. 
To ensure an equitable distribution of payoff across drivers, we assume that the drivers are not perfectly social and are only willing to remain OFFline for a limited amount of time. Drivers turn their app back ON after at most $t^\mathrm{max}_\mathrm{OFF} = 20\,\mathrm{min}$ and only participate in one surge per ride. This means, if a driver participated in a surge, they will not participate in another surge until after they served a ride and received some payoff. This ensures that each driver earns a similar amount over the course of the simulation. In addition, such a behavior might more realistically reflect actions of drivers who cannot switch their app ON and OFF constantly, for example to avoid being detected by automated algorithms of the service provider.

On the customer side, we assume that customers do not immediately leave the system but instead wait for $t^\mathrm{cust}_\mathrm{wait}  = 10 \, \mathrm{min}$, checking the price every $\Delta t^\mathrm{cust} = 2\,\mathrm{min}$ before finally deciding to leave the system. Note that the maximum price they are willing to pay does not change during this time.\\ 

\begin{figure}[!h]
        \centering
        \includegraphics[width=0.45\linewidth]{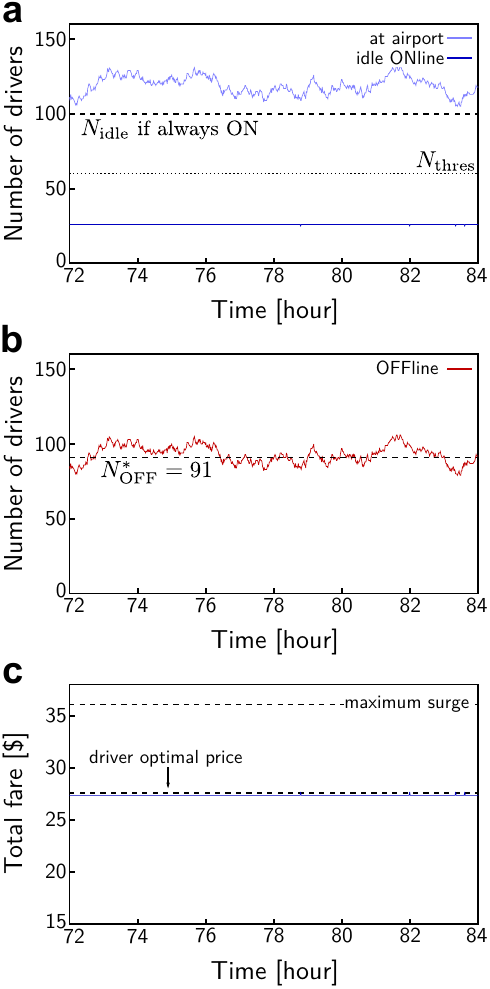}
        \caption{\textbf{Perfectly social drivers reproduce mean-field estimates.} \textbf{a} Total number of drivers at the airport (light blue) and number of idle ONline drivers (dark blue) at the airport. \textbf{b} Number of OFFline drivers. Perfectly social drivers are willing to remain OFFline indefinitely, resulting in the constant (mean-field optimal) number of ONline drivers $N_\mathrm{ON} = N_\mathrm{ON}^* = 26$ in panel \textbf{a}. \textbf{c} The resulting price is constant close to the mean-field optimal price. However, only ONline drivers actually earn this payoff for rides.\\   
        }
        \label{fig:dynamic_game_meanfield_timeseries}
\end{figure}

We aggregate the earnings of all drivers over $T_\mathrm{sim} = 7\,\mathrm{days}$ of continuous requests (roughly equivalent to a month real time with $6$ hours of requests per day). For perfectly social drivers that are willing to remain OFFline indefinitely (but still only participate in one surge per ride) the simulation accurately reproduces the predicted outcome of the mean field calculations, illustrated in Fig.~\ref{fig:dynamic_game_meanfield_timeseries}. The drivers keep the price constant and close to its optimal value. They collectively earn $14.43$ USD per hour on average (standard deviation across drivers $1.12$ USD per hour and minimal earning of an individual driver of $11.23$ USD per hour) in the simulation, agreeing with the predicted $14.39$ USD per hour from mean field calculations. Due to the fluctuations in the demand and the total number of drivers at the airport, the number of drivers required to be OFFline also fluctuates. This creates a sufficient mixing of OFFline and ONline drivers over the course of the simulation to distribute the payoff relatively equally. \\

\begin{figure}[!h]
    \centering
        \includegraphics[width=0.45\linewidth]{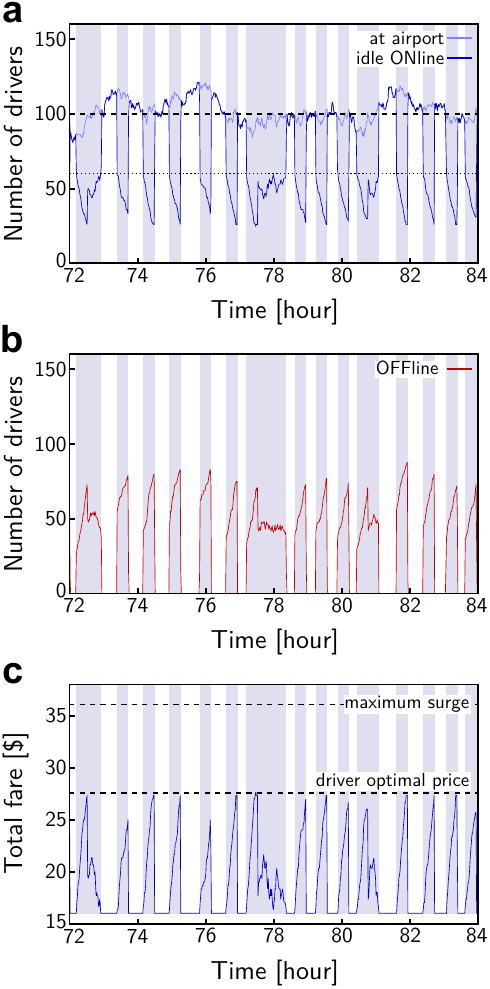}
        \caption{\textbf{Partially social drivers induce repeated price surges.} \textbf{a} Total number of drivers at the airport (light) and number of idle ONline drivers (dark) at the airport. \textbf{b} Number of OFFline drivers. Partially social drivers remain OFFline for at most $t^\mathrm{max}_\mathrm{OFF} = 20$ minutes and only participate in one surge per ride, resulting in repeated periods where a large number of drivers are offline. \textbf{c} The resulting price reflects the anomalous supply dynamics. The drivers' strategy results in repeated price surges with a characteristic amplitude and duration when sufficiently many drivers at the airport are willing to participate in a surge.  
        As drivers rotate being OFFline, all drivers profit equally from these price surges.}
        \label{fig:dynamic_game_spike_timeseries}
\end{figure}

For only partially social drivers, the dynamics changes (Fig.~\ref{fig:dynamic_game_spike_timeseries}). Drivers switch OFF their app but return ONline after $t^\mathrm{max}_\mathrm{off} = 20$ minutes, ending the surge. After the surge, there are not sufficiently many drivers willing to participate in another surge and the price relaxes to its base value $p_\mathrm{base}$ as all drivers are ONline. Only after these drivers have served a customer and returned to the airport, a new surge starts when sufficiently many drivers are willing to participate again. These dynamics determine both the timescale for the duration of the surges (explicitly as $t^\mathrm{max}_\mathrm{off} = 20$) and the time between the surges (implicitly via the driver turn-over rate depending on the request rate $\lambda$). The induced surge peaks typically reach the driver-optimal price for a short time, resulting in a characteristic pattern of repeated short price spikes with (almost) fixed amplitude. In reality, the dynamics would be affected also by the timescale of the dynamic pricing algorithm reacting to the supply and demand changes as well as by additional demand-side incentives such as fewer alternative transport options (resulting in lower price elasticity of customers or higher willingness to pay) late at night and other external influences such as traffic conditions and variable round-trip time (see Fig.~\ref{fig:dynamic_game_variable_ravel_time} or Fig.~2c in the main manuscript). Overall, the drivers earn $13.64$ USD per hour with this strategy (standard deviation $0.24$ USD per hour and minimal earning $12.98$ USD per hour), corresponding to lower total earnings but also much lower risk for extremely low income. \\

In contrast, when drivers remain always ONline, they only earn $12.07$ USD per hour on average (standard deviation $0.04$ USD per hour), agreeing with the $12$ USD per hour expected from the mean field calculation (see above). Thus, drivers collectively increase their profits compared to the baseline always-ON strategy but they do not achieve the optimal payoff. However, in contrast to the case of perfectly social drivers who remain OFFline indefinitely, they keep the distribution of income much more similar among all drivers without the need to explicitly share the profits at the end of the game, reducing the risk for individual drivers to participate. Overall, the optimal strategy for the drivers depends on their risk-aversion preferences and their trust in the other drivers to cooperate (during the current day as well as across longer timescales). \\

\clearpage

\begin{figure*}[h]
    \centering
    \includegraphics[width=1.0\linewidth]{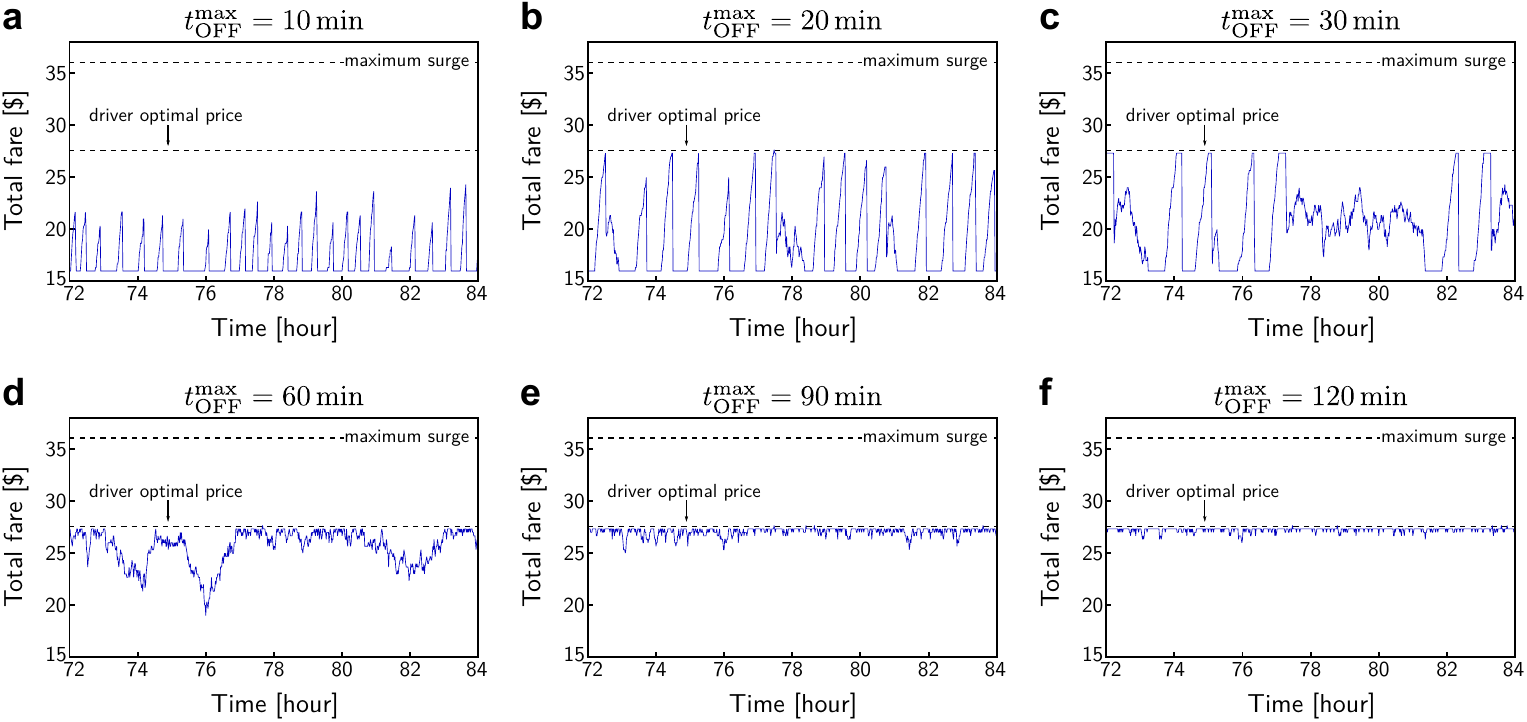}
    \caption{\textbf{Repeated price surges transition into constant surge as drivers become more social.} \textbf{a-c} When drivers are only weakly social and want to ensure a high payoff for themselves, they remain OFFline only for a short amount of time $t^\mathrm{max}_\mathrm{OFF} \le 2 t_d = 30$ minutes. After a surge, most drivers are not willing to participate in another one. Only when sufficiently many drivers have served a ride they initiate the next surge. This leads to repeated short price spikes with characteristic amplitude and duration set by $t^\mathrm{max}_\mathrm{OFF}$. 
    \textbf{d-f} When drivers are more trusting and social (sharing of profits, reciprocation over longer timescales) and $t^\mathrm{max}_\mathrm{OFF} > 2 t_d$, drivers can bridge the time until a driver returns from a ride. This creates a permanently high surge fee, similar to the mean-field limit (compare Fig.~\ref{fig:dynamic_game_meanfield_timeseries}).}
    \label{fig:dynamic_game_switch_off_time_robustness}
\end{figure*}

\paragraph*{Robustness:}
The qualitative results of the model are robust to changes of the parameters, such as the price limits and thresholds. In particular, the qualitative picture of repeated surge peaks occurs for a range of values of $t^\mathrm{max}_\mathrm{OFF}$. This parameter explicitly sets the timescale of surges, describing how social or trusting the drivers are (see Fig.~\ref{fig:dynamic_game_switch_off_time_robustness}a-c). When $t^\mathrm{max}_\mathrm{OFF} > 2 t_d$, drivers are sufficiently social to bridge the time until another driver has serviced a customer and is willing to join the surge again. This leads to a permanently high surge fee since some drivers are OFFline at all times. As $t^\mathrm{max}_\mathrm{OFF}$ increases, the dynamics become more and more similar to the mean-field limit where drivers remain OFFline indefinitely  (compare Fig.~\ref{fig:dynamic_game_switch_off_time_robustness}d-f).

\begin{figure*}[h]
    \centering
    \includegraphics[width=1.0\linewidth]{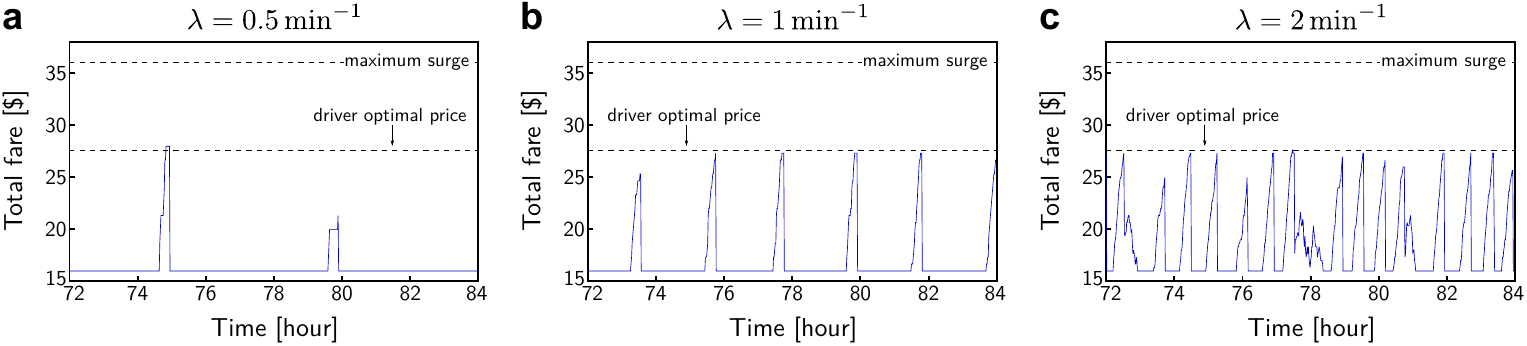}
    \caption{\textbf{Price surges are more common with higher demand.} As the demand increases (request rate $\lambda$ increases from \textbf{a} to \textbf{c}), surges become more frequent. With the request rate, the driver turn over rate increases. As a consequence, drivers are willing to participate in surges more often and the time between surges decreases.}
    \label{fig:dynamic_game_demand_robustness}
\end{figure*}

\begin{figure*}[h]
    \centering
    \includegraphics[width=0.5\linewidth]{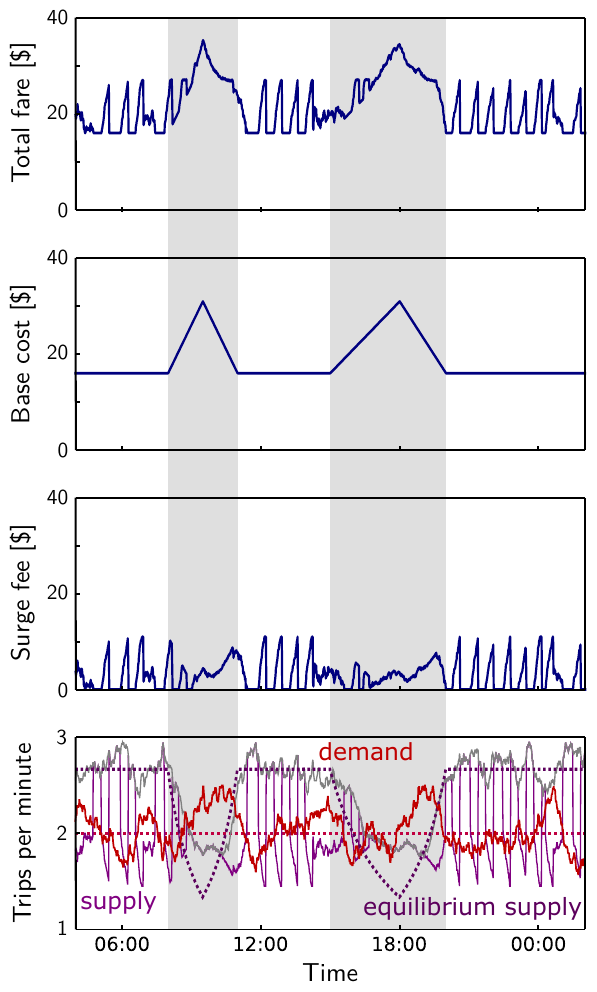}
    \caption{\textbf{Simulation with variable travel time qualitatively reproduces surge dynamics observed at DCA.} The figure shows the total fare, base cost and surge fee of the simulation replicating typical rush hour dynamics with variable travel time during morning and evening commuting hours (compare Fig.~2c in the main manuscript). The last panel shows additional details on the supply and demand evolution. The supply (grey, number of drivers at the airport) fluctuates corresponding to the stochastic demand (60 minute moving average, solid red line). The actual supply of online drivers (solid purple line) shows repeated supply shortages creating price surges. The dashed lines indicate the average request rate (demand, red) and the mean field equilibrium number of drivers expected at the airport at the given conditions and full demand (ignoring reduced demand due to increased price).}
    \label{fig:dynamic_game_variable_ravel_time}
\end{figure*}

The second relevant timescale in the system is the driver turn over rate, influencing the duration between individual surges. This turn over rate is directly related to the request rate $\lambda$ (demand). As the demand decreases, more drivers need to go OFFline to induce a surge since fewer drivers are required to serve the predicted upcoming requests. That means the parameter $N_\mathrm{thres}$ in the price function Eq.~(\ref{eq:dynamic_game_price}) decreases while the total number of drivers remains constant, $N=160$. This makes surges more difficult to coordinate and less frequent, as drivers participating in one surge need longer times to serve a ride and become willing to induce a surge again. Resulting time series for different request rates are illustrated in Fig.~\ref{fig:dynamic_game_demand_robustness}. The results reflect the qualitative intuition from the two previous discussions: a lower demand makes it more difficult to organize surges as the incentive to defect is higher and more trust is required between the players.

\newpage

\clearpage
\paragraph*{Variable travel time}
In the main manuscript, we showed simulation results with variable travel time (variable base cost) to more closely model the daily dynamics including rush hour and periods of heavy traffic. For these simulations, we modulate the round-trip time such that $\left<t_s\right>$ increases to $60$ minutes during rush hour (doubling the standard travel time from $t_d = 15$ to $t_d = 30$, approximately reflecting the average taxi trip duration during rush hour from DCA) to more closely replicate the price dynamics observed at DCA. As the round-trip time changes during the simulation, we adjust the driver strategy based on the optimal mean-field strategy assuming static conditions with the current round-trip time.
We use the price function $p_\mathrm{base} = 1\,\mathrm{USD} + t_d\, \frac{\mathrm{USD}}{\mathrm{minute}}$ to determine the base cost (changing from $16$ USD to $31$ USD) depending on the expected travel time. At the same time, the threshold before surge pricing sets in also changes over time based on the current round-trip time, calculated as described above as the number of drivers required to bridge the time until the first driver returns, $N_\mathrm{thresh} = \lambda \, (2t_d)$.\\

Figure~\ref{fig:dynamic_game_variable_ravel_time} shows the same results as Fig.~2c in the main manuscript with additional detail on the supply and demand in the last panel. The supply (number of drivers at the airport) was converted to units of trips per minute to be comparable to the demand by comparing it to the number of requests that can be served before surge sets in,
\begin{equation}
    \text{supply} = 1 + \frac{\text{number of drivers at airport }}{\lambda \, (2 t_d)} = 1 + \frac{\text{number of drivers at airport }}{N_\mathrm{thresh}} \,,
\end{equation}
effectively assuming $\lambda \, (2 t_d)$ drivers are currently busy. This means surge pricing sets in as soon as the supply is lower than the expected demand $\lambda = 2$ requests per minute. 
The expected equilibrium supply follows as 
\begin{equation}
    \text{equilibrium supply} = \frac{\text{total number of drivers }}{\lambda \, (2 t_d)} = \frac{\text{total number of drivers }}{N_\mathrm{thresh}} \,,
\end{equation}
Due to the stochasticity of the demand (shown as a 60 minute moving average to avaerage the comparatively low number of requests), the number of requests per minute is not constant but fluctuates around the expected value. Correspondingly, the available supply varies stochastically around the mean field equilibrium supply. Note that the sotchastic fluctuations of the demand or supply are not correlated with the induced price surges. 

\clearpage
\subsection{Summary}
The actual mechanism used by Uber to assign ride requests to drivers at airports is unknown. In particular not all drivers that take a trip from the airport will immediately return and not all drivers will have to wait a long time and be able or willing to coordinate. For example, a driver arriving at the airport terminal to drop off a passenger may be preferred to immediately pick up a customer due to lower waiting time since this driver is already at the terminal compared to drivers waiting in a parking lot further away. 
The exact dynamics depend also on the layout of the airport and local regulations (where drivers may park, how long customers take to exit the airport and book a ride etc.). Moreover, while we showed in the main manuscript and in section \ref{sec:demand} of this document that the demand is approximately constant over the day and the price surges are not significantly correlated with the arrivals, fluctuations due to clustered arrivals of airplanes likely still enter in the pricing algorithm. Importantly, clustered arrivals and temporarily increased demand also change the incentive structure, as the artificially induced price surges become easier to organize and more profitable at higher demand as illustrated in all above models. Moreover, dynamic pricing algorithms typically do not react instantaneously to changes. A longer reaction timescale may allow drivers to profit from their own induced surge (as assumed in the discussion of the two-player and multiplayer static model). This may also contribute to easier organization of artificial price surges.

Overall, while the above models do not exactly replicate reality, they qualitatively illustrate the incentives inherent in dynamic pricing mechanisms for drivers to maximize their profit by collectively inducing price surges. In particular, they illustrate that these incentives exist even with the most basic form of dynamic pricing over a range of models, showing that artificial supply shortages are a general problem across dynamic pricing schemes. Specifically, the models illustrate the following:
\begin{itemize}
    \item The two-player model illustrates that the basic incentive exists in the simplest possible setting independent of the parameters of the pricing algorithm and provides a simple model to understand the conditions that promote anomalous supply dynamics.
    \item The multiplayer game extends this setting and illustrates that the effect remains when drivers act in a socially optimal way. This one-time game would describe the dynamics when the demand is strongly localized in time, for example for simultaneous requests from many passengers arriving on the same plane.
    \item The dynamic model serves to illustrate that repeated price surges may emerge in a simple model, even without explicit planning. In particular, it illustrates that these price surges emerge for a broad range of parameters and that the incentives illustrated in the static, one-off games are sufficient to induce repeated price surges.
\end{itemize}
We note that the same arguments made here for supply-induced price surges in principle also hold for demand-induced surge fee manipulation, for example by temporarily reducing demand to lower the price. Naturally, organizing such collective action on the demand side is much harder (as all players are strangers) and the game is typically not repeated with the same group of people (more incentive to defect). Additionally, service providers have little incentive to reduce prices as the price for ride-hailing services is typically already low (due to competition with other service providers, taxi cabs and public transport), leaving not much room for price decreases. Thus, while the mechanism in principle translates to demand-induced down-surges, the conditions make these much harder to execute than supply-induced up-surges.

\clearpage
\section{Data -- methodology and details}
\label{sec:data}

Dynamic pricing mechanisms are increasingly adopted in on-demand mobility services. In ride-hailing, Uber Technologies Inc.~operates one of the largest digital marketplaces where more than 90 million monthly active users from all around the world request rides and are matched with a driver subject to a dynamically adapted price \cite{UberStatistics2019}. In the main manuscript, we illustrated how specific market conditions incentivize collective action among drivers that induce non-stationary price dynamics. We based our findings on Uber price estimate time series collected for trip in different locations all around the globe. We compared the price data against time series of aircraft deplanements collected for Reagan National Airport (DCA) as well as historic taxi trip records in Washington D.C. In this section of the Supplementary Material, we provide detailed information on all sources of data, our data acquisition approach, data content and quality as well as how we analyzed the raw data to obtain our main findings.

\subsection{Data sources and acquisition}

In this subsection, we detail the different data sources used to acquire the data required to conduct our analyses.

\subsubsection{Acquisition of Uber price time series}

\paragraph{Uber price estimates.} Uber's \textit{price estimate} API endpoint provides, amongst others, price estimates for the company's different ride-hailing services on a per trip basis. For a given trip characterized by origin-destination coordinates $(o,d)$, the interface returns a current price estimate vector, whose entries $i=\{1,2,\dots\}$ contain a lower and upper bound $p^i_\mathrm{min}$ and $p^i_\mathrm{max}$ per Uber product $i$ operated in the area, a trip duration estimate $t_d$, as well as a trip distance estimate $l$. Note that trip duration and distance estimates are universal for the different Uber products. We attach a time stamp $t$ per tuple $((o,d),\mathbf{p}_\mathrm{min},\mathbf{p}_\mathrm{max},l,t_d)_t$ and define the ride-hailing price dynamics as the time series $\{((o,d),\mathbf{p}_\mathrm{min},l,t_d)_t, t\in T\}$ where $T$ is the set of all measurement times. In our price estimate time series analysis, we use the lower bound on the price estimate $\mathbf{p}_\mathrm{min}$ as the reference quantity for our calculations.\\

\paragraph{Uber products.} We focus on Uber's standard ride-hailing services. Depending on local availability, we use one the two economy products \textit{UberX} or \textit{UberGO} (India). For comparison we also consider one of the six premium products \textit{Black}, \textit{Berline}, \textit{Exec}, \textit{Lux}, \textit{Premier} or \textit{Select} depending on the availability at the location. 

Uber's \textit{products} API endpoint provides, amongst others, information on fee structure parameters for each of these products. We extract the information on booking fees, price per minute, price per mile, distance unit, minimum fees and the currency code (see first section of this document for details on Uber's price model and Tab.~\ref{tab:UberPriceParameters} for an example of the data) for the standard and premium Uber service operating in the respective area. Together with the price estimate time series we use these parameters to compute the base cost and isolate the surge dynamics not explained by distance and duration of the ride (compare Figs.~\ref{fig:PriceTimeSeries}-\ref{fig:TripDurationDistance}).\\

\paragraph{City and trip characteristics.} We compare the price dynamics for Uber standard and premium rides originating from different points of interests in metropolitan areas around the globe. For four types of trip categories we record and analyze Uber price estimate time series. These categories are:
\begin{itemize}
    \item \textit{Airport}: trip originates from an airport and ends at a point of interest in the nearby city center (e.g. train station, business district, convention center)
    \item \textit{Station}: trip originates from a major train station and ends at a point of interest in the corresponding city (e.g. other train station, business district, convention center)
    \item \textit{Convention}: trip originates from a local convention center and ends at a point of interest in the corresponding city (e.g. train station, business district)
    \item \textit{City}: trip originates and ends within the boundaries of a specific city, not specifying a distinct point of interest (e.g. residential area, business district)
\end{itemize}
Note that the choice of trip origin and the local demand and supply situation (at the origin) are the primary determinants for Uber's dynamic pricing mechanism, while the precise choice of destination within a city does not significantly affect the results. Hence, the chosen trip categories classify the ride-hailing price dynamics at points of interest around the globe where demand for rides is typically highly localized and expected to change periodically during a day.

Tab.~\ref{tab:CityList} shows a full list of all 137 trips that Uber price estimate time series have been recorded for, defined in terms of city, region, as well as specifics on origin-destination attributes. It contains 63 airport (see Fig.~\ref{fig:WorldMap}), 12 station, 23 convention and 39 city trip items. Per item, we provide detailed information on origin-destination coordinates, API request rates, time intervals of data recording, standard and premium Uber products used for the analysis, as well as aggregated measures of the surge dynamics in the spreadsheet \textsc{DataRideHailingPrices.xlsx} available in the data section of the online Supplementary Material. 

We selected the trip sample based on three criteria (in order of priority):
\begin{itemize}
    \item \textit{Uber activity}: Uber operates in the local area.
    \item \textit{Global representation}: Balanced mix of regions across the globe.
    \item \textit{Mixed points of interests}: Balanced mix of airport, station, convention and city trips to investigate effects of the qualitatively different demand structure expected at these locations.
\end{itemize}


\begin{center}
\begin{longtable}{|p{5cm}|p{2.7cm}|p{2.8cm}|p{3.cm}|p{3cm}|}
\caption{\textbf{Uber price estimates have been recorded for airport, station, city and convention trips around the world.} The table provides the location (city and region) as well as trip classification characteristics (origin and destination types). Labels are included for reference to the figures in the first two sections of this document. Further trip details can be found in the spreadsheet \textsc{DataRideHailingPrices.xlsx}  available in the online Supplementary Material.
}
\label{tab:CityList}

\\
\hline
\textbf{Label}         & \textbf{City}   & \textbf{Region} & \textbf{Origin type} & \textbf{Destination type} \\
\hline 
\endfirsthead

\multicolumn{5}{l}%
{{\bfseries \tablename\ \thetable{} -- continued from previous page}} \\ \hline
\textbf{Label}         & \textbf{City}   & \textbf{Region} & \textbf{Origin type} & \textbf{Destination type} \\ \hline \endhead

\hline \multicolumn{5}{|r|}{{Continued on next page\dots\qquad }} \\ \hline\endfoot

\hline \hline
\endlastfoot

ARN & Stockholm & Europe & Airport & Station \\ 
ATL & Atlanta & North America & Airport & City \\ 
BMA & Stockholm & Europe & Airport & Station \\ 
BOG & Bogota & South America & Airport & City \\ 
BOM & Mumbai & Asia & Airport & Station \\ 
BRU & Brussels & Europe & Airport & Station \\ 
BWI & Baltimore & North America & Airport & Station \\ 
BWI Residential Area Trip & Baltimore & North America & Airport & City \\ 
BWI Urban Area Trip & Baltimore & North America & Airport & City \\ 
CAI & Cairo & Africa & Airport & Station \\ 
CDG & Paris & Europe & Airport & Station \\ 
CPT & Cape Town & Africa & Airport & City \\ 
DCA & Washington D.C. & North America & Airport & Station \\ 
DCA Residential Area Trip & Washington D.C. & North America & Airport & City \\ 
DCA Urban Area Trip & Washington D.C. & North America & Airport & City \\ 
DEL & Dheli & Asia & Airport & Station \\ 
EWR & New York City & North America & Airport & Station \\ 
EWR Residential Area Trip & New York City & North America & Airport & City \\ 
EWR Urban Area Trip & New York City & North America & Airport & City \\ 
EZE & Buenos Aires & South America & Airport & City \\ 
GIG & Rio de Janeiro & South America & Airport & Station \\ 
GLA & Glasgow & Europe & Airport & Station \\ 
GVA & Geneva & Europe & Airport & Station \\ 
HEL & Helsinki & Europe & Airport & Station \\ 
HKG & Hong Kong & Asia & Airport & Station \\ 
IAD Residential Area Trip & Washington D.C. & North America & Airport & City \\ 
IAD Urban Area Trip & Washington D.C. & North America & Airport & City \\ 
IAH & Houston & North America & Airport & City \\ 
ICN & Seoul & Asia & Airport & Station \\ 
JFK & New York City & North America & Airport & Station \\ 
JFK Residential Area Trip & New York City & North America & Airport & City \\ 
JFK Urban Area Trip & New York City & North America & Airport & City \\ 
JNB & Johannesburg & Africa & Airport & City \\ 
LAX & Los Angeles & North America & Airport & Station \\ 
LGA & New York City & North America & Airport & Station \\ 
LGA Residential Area Trip & New York City & North America & Airport & City \\ 
LGA Urban Area Trip & New York City & North America & Airport & City \\ 
LHR & London & Europe & Airport & City \\ 
LIS & Lisbon & Europe & Airport & Station \\ 
MAA & Chennai & Asia & Airport & City \\ 
MAN & Manchester & Europe & Airport & Station \\ 
MEL & Melbourne & Oceania & Airport & City \\ 
MIA & Miami & North America & Airport & Station \\ 
MRS & Marseille & Europe & Airport & Station \\ 
MSY & New Orleans & North America & Airport & Station \\ 
MUC & Munich & Europe & Airport & Station \\ 
OAK Residential Area Trip & Oakland & North America & Airport & City \\ 
OAK Urban Area Trip  & Oakland & North America & Airport & City \\ 
ORD & Chicago & North America & Airport & Station \\ 
OTP & Bucharest & Europe & Airport & Station \\ 
PER & Perth & Oceania & Airport & City \\ 
PHL & Philadelphia & North America & Airport & City \\ 
SEA & Seattle & North America & Airport & Station \\ 
SFO & San Francisco & North America & Airport & Station \\ 
SFO Residential Area Trip & San Francisco & North America & Airport & City \\ 
SFO Urban Area Trip & San Francisco & North America & Airport & City \\ 
SJC Residential Area Trip & San Francisco & North America & Airport & City \\ 
SJC Urban Area Trip & San Francisco & North America & Airport & City \\ 
TLL & Tallinn & Europe & Airport & Station \\ 
VNO & Vilnius & Europe & Airport & Station \\ 
WAW & Warsaw & Europe & Airport & Station \\ 
YUL & Montreal & North America & Airport & Station \\ 
YYZ & Toronto & North America & Airport & Station \\ 
Hoboken & Hoboken & North America & City & City \\ 
Meatpacking District (NYC) & New York City & North America & City & City \\ 
New York City Trip 1 & New York City & North America & City & City \\ 
New York City Trip 2 & New York City & North America & City & City \\ 
New York City Trip 3 & New York City & North America & City & City \\ 
New York City Trip 4 & New York City & North America & City & City \\ 
New York City Trip 5 & New York City & North America & City & City \\ 
New York City Trip 6 & New York City & North America & City & City \\ 
New York City Trip 7 & New York City & North America & City & City \\ 
New York City Trip 8 & New York City & North America & City & City \\ 
New York City Trip 9 & New York City & North America & City & City \\ 
New York City Trip 10 & New York City & North America & City & City \\ 
New York City Trip 11 & New York City & North America & City & City \\ 
New York City Trip 12 & New York City & North America & City & City \\ 
New York City Trip 13 & New York City & North America & City & City \\ 
New York City Trip 14 & New York City & North America & City & City \\ 
New York City Trip 15 & New York City & North America & City & City \\ 
New York City Trip 16 & New York City & North America & City & City \\ 
New York City Trip 17 & New York City & North America & City & City \\ 
New York City Trip 18 & New York City & North America & City & City \\ 
New York City Trip 19 & New York City & North America & City & City \\ 
San Francisco City Trip 1 & San Francisco & North America & City & City \\ 
San Francisco City Trip 2 & San Francisco & North America & City & City \\ 
San Francisco City Trip 3 & San Francisco & North America & City & City \\ 
San Francisco City Trip 4 & San Francisco & North America & City & City \\ 
San Francisco City Trip 5 & San Francisco & North America & City & City \\ 
San Francisco City Trip 6 & San Francisco & North America & City & City \\ 
San Francisco City Trip 7 & San Francisco & North America & City & City \\ 
San Francisco City Trip 8 & San Francisco & North America & City & City \\ 
San Francisco City Trip 9 & San Francisco & North America & City & City \\ 
Trump Tower (NYC) & New York City & North America & City & City \\ 
Washington, D.C. City Trip 1 & Washington D.C. & North America & City & City \\ 
Washington, D.C. City Trip 2 & Washington D.C. & North America & City & City \\ 
Washington, D.C. City Trip 3 & Washington D.C. & North America & City & City \\ 
Washington, D.C. City Trip 4 & Washington D.C. & North America & City & City \\ 
Washington, D.C. City Trip 6 & Washington D.C. & North America & City & City \\ 
Washington, D.C. City Trip 7 & Washington D.C. & North America & City & City \\ 
Washington, D.C. City Trip 8 & Washington D.C. & North America & City & City \\ 
Washington, D.C. City Trip 9 & Washington D.C. & North America & City & City \\ 
Americas Center Convention Complex & St. Louis & North America & Convention & Airport \\ 
Austin Convention Center & Austin & North America & Convention & Airport \\ 
Boston Hynes Convention Center & Boston & North America & Convention & Airport \\ 
Colorado Convention Center & Denver & North America & Convention & Airport \\ 
Georgia World Congress Center & Atlanta & North America & Convention & Airport \\ 
Henry B. Gonzalez Convention Center & San Antonio & North America & Convention & Airport \\ 
Houston Convention Center & Houston & North America & Convention & Airport \\ 
Hutchinson Convention Center & Dallas & North America & Convention & Airport \\ 
Jacob K. Javits Convention Center & New York City & North America & Convention & Airport \\ 
Los Angeles Convention Center & Los Angeles & North America & Convention & Airport \\ 
Minneapolis Convention Center & Minneapolis & North America & Convention & Airport \\ 
Moscone Center & San Francisco & North America & Convention & Airport \\ 
Nashville Music City Center & Nashville & North America & Convention & Airport \\ 
New Orleans Morial Center & New Orleans & North America & Convention & Airport \\ 
Orange County Convention Center & Orange County & North America & Convention & Station \\ 
Pennsylvania Convention Center & Philadelphia & North America & Convention & Airport \\ 
Phoenix Convention Center & Phoenix & North America & Convention & Airport \\ 
Riverside Convention Center & Riverside & North America & Convention & Airport \\ 
San Diego Convention Center & San Diego & North America & Convention & Airport \\ 
Seattle Convention Center & Seattle & North America & Convention & Airport \\ 
Tampa Convention Center & Tampa & North America & Convention & Airport \\ 
Walter E. Washington Convention Center & Washington D.C. & North America & Convention & Airport \\ 
Washington, D.C. City Trip 5 & Washington D.C. & North America & Convention & City \\ 
30th Street Station (Philadelphia) & Philadelphia & North America & Station & Airport \\ 
Grand Central Station (NYC) & New York City & North America & Station & City \\ 
Jamaica Station (Queens) & New York City & North America & Station & Airport \\ 
Millenium Station (Chicago) & Chicago & North America & Station & Airport \\ 
Montreal Central Station & Montreal & North America & Station & Airport \\ 
Pennsylvania Station (Newark) & Newark & North America & Station & Airport \\ 
Pennsylvania Station (NYC) & New York City & North America & Station & City \\ 
Union Station (Chicago) & Chicago & North America & Station & Airport \\ 
Union Station (LA) & Los Angeles & North America & Station & Airport \\ 
Union Station (Toronto) & Toronto & North America & Station & City \\ 
Union Station (Washington, D.C.) & Washington D.C. & North America & Station & Airport \\ 
World Trade Center (NYC) & New York City & North America & Station & City \\

\end{longtable}
\end{center}

\subsubsection{Acquisition of aircraft deplanement time series}
Flightradar24 is an online platform that provides real-time flight information on commercial aircraft activity around the world. The service includes flight tracks, origins and destinations, aircraft call signs and types, positions, altitudes, headings and speeds. We use the platform's open API to compile landing time statistics about deplanements at the airports specified in Tab.~\ref{tab:CityList}. 

Per deplanement at airport $x$, we create a landing event at time $t$ provided as real landing time $t_\mathrm{real}$ and attach it with information on the aircraft's official call sign $s$ and its current seat configuration $c_s$. We retrieve the information on current aircraft seat configuration via flightera.net, an online platform that provides detailed aircraft information per call sign. 
The list of tuples $(x,s,c_s)_{t_\mathrm{real}}$ recorded over the time span $T$ defines the aircraft deplanement time series that we use as a proxy to model the time-dependent demand for rides originating from the airports.

\subsubsection{Acquisition of taxi trip records}

The Department of For-Hire Vehicles and the D.C. Office of the Chief Technology Officer provide taxicab trip records for the Washington D.C. region. In our analysis, we focus on the data recorded between 17/01/02 and 17/08/27, including 34 full weeks of data. From this data set, we use a subset of approximately 370000 taxi trip records originating from Reagan National Airport (DCA) to compile spatio-temporarily resolved trip statistics for the average demand for taxi-like mobility services. We use taxi trips' origin-destination timestamps, their zip codes as well as longitudes and latitudes to determine a statistical demand model per day of week.

\subsubsection{Acquisition of foreign currency exchange rates}

Ratesapi.io is a publicly accessible API that provides current and historical foreign exchange rates for different currencies based on the data made available through the European Central Bank. We use the API to convert Uber price estimates in non-USD countries into USD with the help of the exchange rate for the respective day.

\subsubsection{Acquisition of timezone information}
We gathered the time zones for each airport using a JSON database from \url{https://github.com/mwgg}, (accessed on 2019-07-12). 
For city, convention and station trips we entered the time zone information manually. The conversion between the measured time in CEST and the desired timezones was done using python's \textit{pytz} module, which takes care of daylight saving times automatically.

\subsection{Data content and quality}

In this subsection, we detail the data quality per source and the data cleansing methodology applied to obtain the data serving as inputs for our analyses.

\subsubsection{Uber price estimate time series} 

Uber price estimates provide possible fare ranges for the requested ride, upper and lower bounded by the price vectors $\mathbf{p}_\mathrm{max}$ and $\mathbf{p}_\mathrm{min}$. $\mathbf{p}_\mathrm{max}$ and $\mathbf{p}_\mathrm{min}$ are provided as integer values in the local currency by the Uber API.  
Typically, the difference between $\mathbf{p}_\mathrm{max}$ and $\mathbf{p}_\mathrm{min}$ is fixed for most of the time (e.g. 2 USD). For our analyses, we consider the lower bound of the price estimate only. 

Price estimates do not update continuously, but only approximately every two minutes (compare Fig.~\ref{fig:PriceTimeSeries}). Hence, price estimate time series $\{((o,d),\mathbf{p}_\mathrm{min},l,t_d)_t, t\in T\}$ contain identical elements if sampling at higher frequency than two minute intervals. We provide detailed information on the sampling frequency per recorded trip request in the spreadsheet \textsc{DataRideHailingPrices.xlsx}  available in the online Supplementary Material.

We do not clean the price estimate time series and use the raw data as inputs for our analyses as described in more detail in the following subsection.\\

\subsubsection{Aircraft deplanement time series} 

In rare occasions, flightradar24's API does not return a real landing time, or information on the airplane's call sign. In those cases, we assume the flight to be cancelled and exclude the event from the time series that we use for our analyses. 

Furthermore, in few cases there is no information on the current seat configuration available on flightera.net under the provided aircraft call sign. In those cases, we use the aircraft model to estimate the number of seats from data containing call signs. 
First we match all aircraft arrivals with known call signs with the number of seats obtained from flightera.net. We then compute the average number of seats for every aircraft model. 
Arrivals without a call sign entry, or where flightera.net does not provide the seat configuration, are then assigned the average of the number of seats according to the aircraft model.
In cases where there is no number of seats information available for an aircraft model, we assume it to be the average over all aircraft models.

Comparing seat configurations across the aircraft in the deplanement time series reveals a maximum variation of 25 seats for aircraft of the same model (Boeing 757-232, equipped with 180-205 seats). This type of aircraft has landed seven times at DCA within the time frame considered in our time series. For smaller aircraft in the order of 50 seats, the maximum deviation in seat configuration was at most 20\%.\\

\subsubsection{Taxi trip records} 

Selected taxi trip records contain entries where geographical information are not properly decoded, longitude or latitude information indicate locations outside the US, or zip codes contain placeholders. Similarly, a small number of time stamps are not properly specified. 

We filter these entries from the data recorded between 17/01/02-17/08/27, including 34 full weeks of records. Furthermore, our data cleansing procedure excludes data records less than 0.25 miles or more than 1000 miles.\\ 

\subsection{Data processing}

In this final subsection, we detail how we integrated and analyzed the cleaned data to obtain the results presented in the main manuscript and the first three sections of this Supplementary Material.

\subsubsection{Isolation and characterization of Uber surge dynamics}
\label{sec:data_processing:surge_time_series}

For illustrations of the price time series we use the raw data recorded. To study the surge fee time series we subtract the trip fee and pickup fee of the respective product calculated from the recorded trip duration and distance estimates together with the trip fee parameters. This leaves only the surge fee and surcharges. Since data on some surcharges is not available from Uber's API, we assume that surcharges are constant in time.\footnote{This is not accurate in individual cases where airport pickup fees may change over time or tolls may be in effect during the day but not at night.} We estimate these surcharges as the minimum value of the remaining surcharge plus surge fee. Subtracting this constant value results in an estimate of the absolute surge fee time series that attains a minimum value zero (no surge) at least once. Note that due to the rounding to integer values, the price estimate may not reflect all changes of the trip fare, leading to small fluctuations in the isolated surge fee that do not correspond to actual surge activity. 

For better comparison between the different trips, we normalize the absolute surge fee time series by the base cost (sum of the pickup fee, trip fee and estimated surcharges) at that time, resulting in a relative surge factor time series for each trip and Uber product. In general, we observe that premium Uber services have less frequent surge fee contributions and we assume that these are typically caused by global changes such as strong demand-side price surges or network problems influencing the price estimates. 
The resulting data is, for example, shown in Fig.~\ref{fig:PriceTimeSeries} above.

To analyze the changes in the surge factor time series, we first create a uniform representation of the surge factor time series in one minute intervals. Each point in the new time series is calculated as the average in the five minute time window centered around $t$ for the respective combination of trip origin $o$, destination $d$. Each component of the price vector corresponds to an Uber service. Data points include information about approximately three updates of the price estimate. Based on this data, we calculate the relative price changes as the difference between consecutive time steps and discard any changes smaller than $\Delta p^2 < 10^{-7}$. The remaining changes form the basis for the analysis and are illustrated in the histograms in Fig.~\ref{fig:timescales}.

\subsubsection{Computation of cross-correlations between aircraft deplanement and Uber price estimate time series}

The aircraft arrival data has a maximum resolution of one minute, with some minutes without arrivals. Hence, we first aggregate the number of seats for each minute, and set minutes without arrivals to zero. Then we compute the moving average with a window of five minutes, to reduce noise and to remain consistent over the treatment of our time series.

We compute the surge fee component from the Uber price estimate according to Eq. (\ref{eq:uber_price_components}). Because we have a much larger but irregular resolution for the Uber price estimates, we compute the moving average of the surge fee and compute the averages for every minute. This leaves us with the same granularity of the data as the deplanement data.

In the next step, we select a time window from the surge fee time series. We normalize this fragment of the series by subtracting its mean and dividing by the standard deviation. 

We then iterate over various lags $\Delta t$. For each lag, we shift the deplanement data by the corresponding amount and select the overlap with the surge fee window. We compute the Pearson correlation coefficient $\rho$ for these two time series according to Eq.~(\ref{eq:correlation}).

We repeat this procedure for different time windows (8:00 to 14:00,
14:00 to 20:00,
20:00 to 02:00,
08:00 to 20:00,
14:00 to 02:00,
08:00 to 02:00) and data from different days (19/06/04, 19/06/05 and 19/06/06), see Fig.~\ref{fig:AircraftCapacity}.

\subsubsection{Computation of taxi trip statistics}

We organize the taxi trip data from the period of 17/01/02-17/08/27 by weekdays (Monday, Tuesday, etc.). For each weekday we determine the average number of taxi trips starting at Reagan National Airport for every minute of the day.

Because of irregularities in the timestamps of the data we are not able to obtain data for some minutes. Therefore, we set missing values to zero, and correct them by computing a moving average with a window of five minutes.

\end{document}